\documentclass{IEEEtran}
\usepackage[T1]{fontenc}
\usepackage[latin9]{inputenc}
\usepackage{mathrsfs}
\usepackage{amsmath}
\usepackage{amssymb}
\usepackage{stackrel}
\usepackage{graphicx}
\usepackage{esint}
\usepackage[unicode=true,
 bookmarks=true,bookmarksnumbered=true,bookmarksopen=true,bookmarksopenlevel=1,
 breaklinks=false,pdfborder={0 0 0},pdfborderstyle={},backref=false,colorlinks=false]
 {hyperref}
\hypersetup{pdftitle={Your Title},
 pdfauthor={Your Name},
 pdfpagelayout=OneColumn, pdfnewwindow=true, pdfstartview=XYZ, plainpages=false}

\makeatletter

\providecommand{\tabularnewline}{\\}

\makeatother

\begin{document}
\title{A Survey on Fundamental Limits of Integrated Sensing and Communication}
\author{{\normalsize{}An Liu$^{*}$, }\textit{\normalsize{}Senior Member,
IEEE}{\normalsize{}, Zhe Huang, }\textit{\normalsize{}Student Member,
IEEE}{\normalsize{}, Min Li, Yubo Wan, Wenrui Li, Tony Xiao Han$^{*}$,
Chenchen Liu, Rui Du, Danny Tan Kai Pin, Jianmin Lu}\textit{\normalsize{},}{\normalsize{}
Yuan Shen, }\textit{\normalsize{}Senior Member, IEEE}{\normalsize{},
Fabiola Colone,}\textit{\normalsize{} Senior Member, IEEE}{\normalsize{}
and Kevin Chetty}\thanks{This work was supported in part by National Science Foundation of
China (No.62071416), and in part by Huawei Techologies Co., Ltd. (Corresponding
authors: An Liu, Min Li.)

An Liu, Zhe Huang, Min Li, Yubo Wan and Wenrui Li are with the College
of Information Science and Electronic Engineering, Zhejiang University,
Hangzhou 310027, China (email: anliu@zju.edu.cn, min.li@zju.edu.cn).

Tony Xiao Han, Chenchen Liu, Rui Du, Danny Tan Kai Pin and Jianmin
Lu are with Huawei Techologies Co., Ltd. (email: tony.hanxiao@huawei.com).\protect \\
$\text{ }$$\text{ }$$\text{ }$Yuan Shen is with the Department
of Electronic Engineering, Tsinghua University, and Beijing National
Research Center for Information Science and Technology, Beijing 100084,
China (e-mail: shenyuan\_ee@tsinghua.edu.cn).\protect \\
$\text{ }$$\text{ }$$\text{ }$Fabiola Colone is with the Department
of Information Engineering, Electronics and Telecommunications, Sapienza
University of Rome, 00184 Rome, Italy (email: fabiola.colone@uniroma1.it)\protect \\
$\text{ }$$\text{ }$$\text{ }$Kevin Chetty is with the Department
of Security and Crime Science, University College London, WC1H 9EZ
London, U.K. (e-mail: k.chetty@ucl.ac.uk).\protect \\
$\text{ }$$\text{ }$$\text{ }$$^{*}$Co-first author.}}
\maketitle
\begin{abstract}
The integrated sensing and communication (ISAC), in which the sensing
and communication share the same frequency band and hardware, has
emerged as a key technology in future wireless systems due to two
main reasons. First, many important application scenarios in fifth
generation (5G) and beyond, such as autonomous vehicles, Wi-Fi sensing
and extended reality, requires both high-performance sensing and wireless
communications. Second, with millimeter wave and massive multiple-input
multiple-output (MIMO) technologies widely employed in 5G and beyond,
the future communication signals tend to have high-resolution in both
time and angular domain, opening up the possibility for ISAC. As such,
ISAC has attracted tremendous research interest and attentions in
both academia and industry. Early works on ISAC have been focused
on the design, analysis and optimization of practical ISAC technologies
for various ISAC systems. While this line of works are necessary,
it is equally important to study the fundamental limits of ISAC in
order to understand the gap between the current state-of-the-art technologies
and the performance limits, and provide useful insights and guidance
for the development of better ISAC technologies that can approach
the performance limits. In this paper, we aim to provide a comprehensive
survey for the current research progress on the fundamental limits
of ISAC. Particularly, we first propose a systematic classification
method for both traditional radio sensing (such as radar sensing and
wireless localization) and ISAC so that they can be naturally incorporated
into a unified framework. Then we summarize the major performance
metrics and bounds used in sensing, communications and ISAC, respectively.
After that, we present the current research progresses on fundamental
limits of each class of the traditional sensing and ISAC systems.
Finally, the open problems and future research directions are discussed. 
\end{abstract}

\begin{IEEEkeywords}
Integrated sensing and communication, Radar sensing, Localization,
Fundamental limits

\thispagestyle{empty}
\end{IEEEkeywords}

\section{Introduction}

\begin{figure}[tbh]
\begin{centering}
\includegraphics[width=90mm]{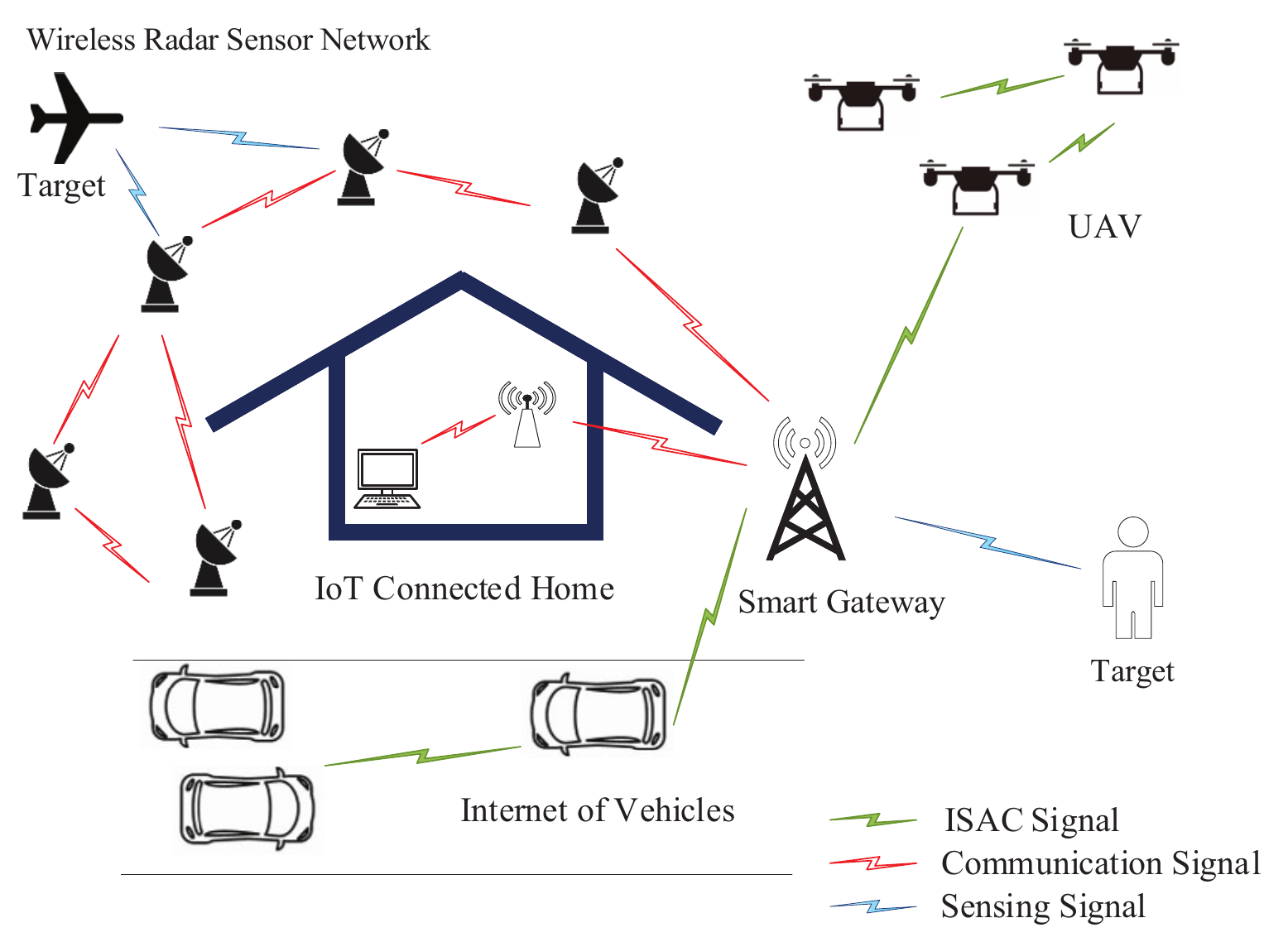}
\par\end{centering}
\caption{\label{fig:Illustration for ISAC-1}Illustration for the applications
of ISAC.}
\end{figure}

Future beyond 5G and sixth generation (6G) wireless systems are expected
to provide various high-accuracy sensing services, such as indoor
localization for robot navigation, Wi-Fi sensing for smart home and
radar sensing for autonomous vehicles. Sensing and communication systems
are usually designed separately and occupy different frequency bands.
However, due to the wide deployment of the millimeter wave and massive
MIMO technologies, communication signals in future wireless systems
tend to have high-resolution in both time and angular domain, making
it possible to enable high-accuracy sensing using communication signals.
As such, it is desirable to jointly design the sensing and communication
systems such that they can share the same frequency band and hardware
to improve the spectrum efficiency and reduce the hardware cost. This
motivates the study of integrated sensing and communication (ISAC).
It is believed that ISAC will become a key technology in future wireless
systems to support many important application scenarios \cite{Bliss,IOR,JRC,JRC_2,Internet_radar}.
For example, in future autonomous vehicle networks, the autonomous
vehicles will obtain a large amount of information from the network,
including ultra-high resolution maps and near real-time information
to help navigate and avoid upcoming traffic congestion \cite{V2V_network}.
In the same scenario, radar sensing in the autonomous vehicles should
be able to provide robust, high-resolution obstacle detection on the
order of a centimeter \cite{AV_radar}. The ISAC technology for autonomous
vehicles provides the potential to achieve both high-data rate communications
and high-resolution obstacle detection using the same hardware and
spectrum resource. Other applications of ISAC include Wi-Fi based
indoor localization and activity recognition, unmanned aerial vehicle
(UAV) communication and sensing, extended reality (XR), joint radar
(target tracking and imaging) and communication systems etc. Each
application has different requirements, limits, and regulatory issues.
Fig. \ref{fig:Illustration for ISAC-1} illustrates some possible
application areas for ISAC.

Under this background, ISAC has attracted tremendous research interest
and attentions in both academia and industry. For example, recently,
there have been an increasing number of academic publications on ISAC,
ranging from transceiver architecture and frame structure \cite{JRC,Index_modulation},
ISAC waveform design \cite{DMMSE_waveform,OFDM_waveform,Range_sidelobe_waveform},
joint coding design \cite{joint_coding_oppermann,joint_coding_complementary,joint_coding_golay},
temporal-spectral-spatial signal processing \cite{JRC_beaforming,Subcar_Power,MU_JRC_beamforming},
to experimental performance demonstrations, prototyping, and field-tests
\cite{SAR_com}. The authors of this paper have also organized IEEE
WTC Special Interest Group (SIG) on ISAC and a workshop on ISAC in
IEEE Global Communications Conference in 2020. Furthermore, in September
and November 2019, IEEE 802.11 formed the WLAN Sensing Topic Interest
Group and Study Group, respectively, and formed a new official Task
Group IEEE 802.11bf in September 2020, with the objective of incorporating
wireless sensing as a new feature for next-generation WiFi systems
(e.g., Wi-Fi 7).

Despite these early research efforts on ISAC, many important problems
about ISAC remain open, such as the unified theoretical frameworks,
the fundamental performance limits, and the optimal ISAC schemes and
signal processing algorithms. In particular, characterizing the fundamental
limits of ISAC, including the distortion bounds for sensing parameters
(such as the direction of arrival (DOA), signal propagation time delay,
Doppler frequency, position, velocity etc.) as well as the channel
capacity and capacity-distortion tradeoff performance, is of great
importance to make breakthrough in ISAC technologies. On one hand,
the fundamental limits provide a performance bound for practical ISAC
technologies, which reveals the potential gap between the current
technologies and the optimal solution. On the other hand, the fundamental
limits analysis also provides useful guidance and insight for the
design and analysis of practical ISAC systems. Recently, a number
of works have been dedicated to studying the fundamental limits of
ISAC, see e.g., \cite{bistatic_MAC,zhang}. However, many important
questions remain open and need further study. In this paper, we conduct
a comprehensive survey on the fundamental limits of various sensing
systems and ISAC systems, and discuss the open problems and potential
research directions. We hope that this survey serves as a starting
point for interested researchers to work on this important and challenging
research area.

Some related works are reviewed below. There are many survey papers
for traditional sensing technologies, including radar sensing \cite{radar_survey},
wireless localization \cite{tempcoop,Wireless_Local,Wireless_Local2,wireless_local3},
WiFi and mobile sensing \cite{Wifi_local1,wifi_sensing2,wifi_sensing3},
among which only a few works have discussed the fundamental limits
of traditional sensing. For example, the fundamental limits of radar
sensing and wireless localization have been surveyed in \cite{radar_survey}
and \cite{tempcoop}, respectively. Recently, several works have also
presented the recent research progress on joint radar and communication
(JRC) system, which can be viewed as a special case of ISAC considered
in this paper. In \cite{Bliss}, the authors presented the applications,
topologies, levels of system integration, the current state of the
art, and outlines of future information-centric JRC systems. In \cite{JRC},
the authors overviewed the application scenarios and research progress
in radar-communication coexistence and dual-functional radar-communication
systems. In \cite{rahman2020enabling}, the author first reviewed
the work on coexisting communication and radar systems, then provided
a brief review for three types of JRC systems and finally reviewed
stimulating research problems and potential solutions. However, previous
works such as \cite{Bliss}, \cite{JRC} and \cite{rahman2020enabling}
mainly focus on the design, analysis and optimization of practical
JRC systems, and there still lacks a comprehensive survey on the fundamental
limits of ISAC. To summarize, a number of contributions differentiate
this paper from existing works:
\begin{itemize}
\item We propose a systematic classification method for both traditional
radio sensing technologies (such as radar sensing and wireless localization)
and ISAC technologies so that they can be naturally incorporated into
a unified framework.
\item Existing survey works on ISAC mainly focus on the joint system design
and integration, but pay little attention to the fundamental limits
of the integrated system. To our best of knowledge, this is the first
work to provide a comprehensive survey on the fundamental limits of
both radio sensing and ISAC systems.
\item We propose several typical ISAC channel topologies as abstracted models
for various ISAC systems, analogous to traditional communication channel
topologies. We point out that the fundamental limits of ISAC channels
cannot be obtained by a trivial combinations of existing performance
bounding techniques in separate sensing and communication systems.
\item We present a list of important open challenges and potential research
directions on ISAC, many of which have not been mentioned in the previous
works.
\end{itemize}
The rest of the paper is organized as follows. Section II describes
the classifications of integrated sensing and communication. Section
III presents some essential performance metrics for radio sensing
as well as integrated sensing and communication. Sections IV - VII
present the current research progress on the fundamental limits of
the device-free sensing, device-based sensing, device-free ISAC, and
device-based ISAC, respectively. Section VIII discusses open problems
and future research directions in ISAC. Finally, we make our conclusions
in Section IX.

\section{Classifications of Integrated Sensing and Communication\label{sec:Classifications-of-Integrated}}

\begin{figure*}[tbh]
\begin{centering}
\includegraphics[width=130mm]{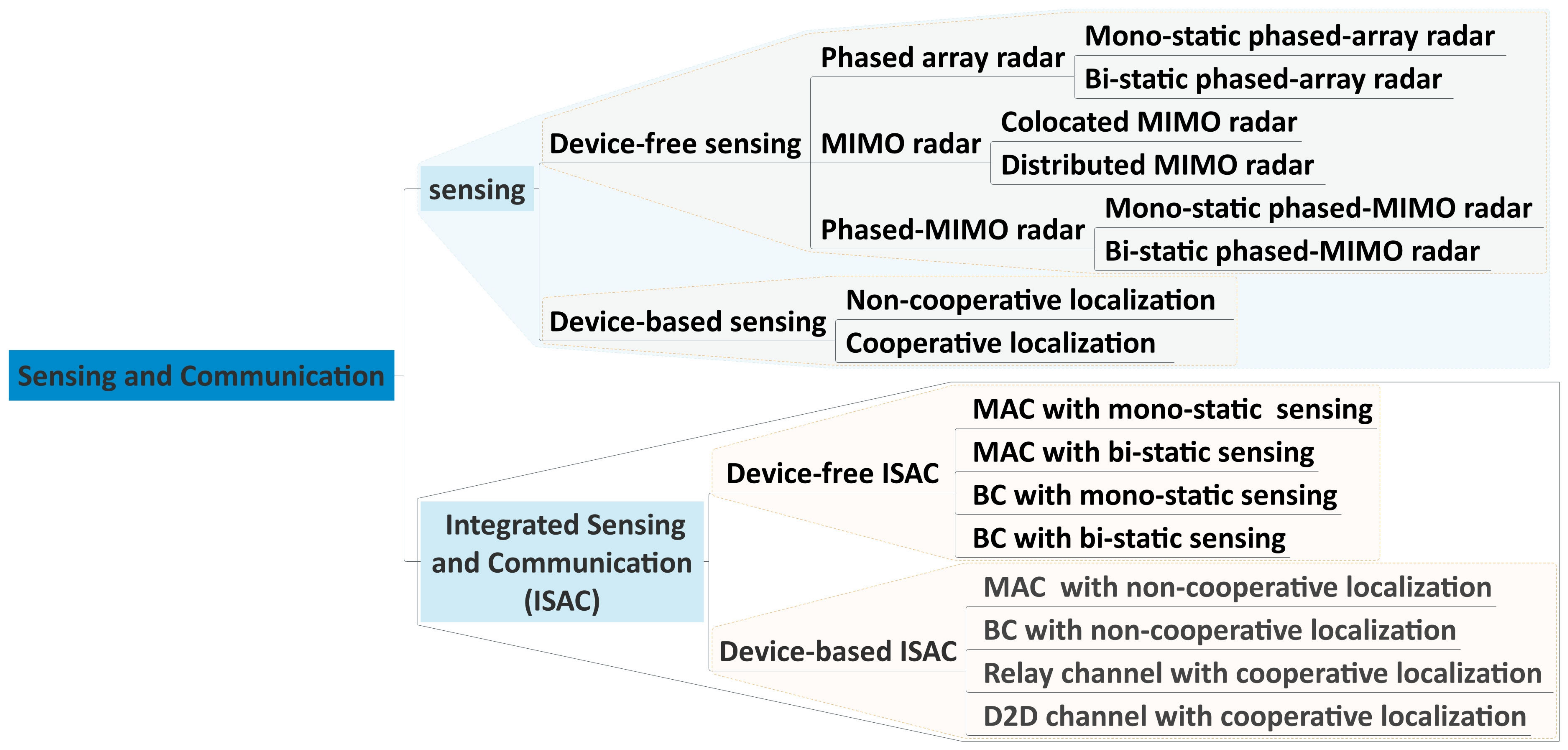}
\par\end{centering}
\centering{}\caption{\label{fig:ISAC-all}An overview of classifications of ISAC.}
\end{figure*}

Traditional radio sensing can be classified into two categories, namely,
the device-free sensing and device-based sensing. 
\begin{itemize}
\item \textit{Device-free sensing} means that the sensing targets (e.g.,
a bird) are not capable of transmitting and/or receiving the sensing
signal, or means that the sensing procedure does not rely on the transmitting
and/or receiving of the sensing target (e.g., a target vehicle). A
typical example for device-free sensing is the radar sensing. 
\item \textit{Device-based sensing} means that the sensing targets are capable
of transmitting and/or receiving the sensing signal, and the sensing
procedure relies on the transmitting and/or receiving of the sensing
target. A typical example is the wireless-based localization to localize
mobile devices. 
\end{itemize}
Naturally, ISAC can also be classified into device-free ISAC and device-based
ISAC as it will be illustrated later. In this section, we first briefly
discuss the history of device-free and device-based sensing/ISAC.
Then we provide a detailed classifications for each category, which
are also summarized in Fig. \ref{fig:ISAC-all}.

In terms of device-free sensing, the earliest radar can be traced
back to 1904 \cite{hulsmeyer1904telemobiloscope}. In 1950, the concept
of phased-array radar first appeared. Through decades of development,
the concept of MIMO radar is introduced in 2004 \cite{MIMO_radar_proposed}
and the concept of phased-MIMO radar was proposed in 2010 \cite{phased-MIMO_radar_performance}.
As an attempt to integrate the radar and communication, the concept
of joint radar-communications (JRC) was proposed in 2006 \cite{joint-radar-and-communication}.
In terms of device-based sensing, Global Navigation Satellite System
(GNSS) has been used to provide location services initially. Owing
to the poor performance of the GNSS in indoor environments, the cellular-based
localization was proposed as a good alternative to GNSS. The first
cellular-based localization system is called E-911 used for providing
emergency services \cite{firstwirelesslocalization}. Starting from
the second generation (2G), wireless localization has been included
as a compulsory feature in the standardization and implementation
of cellular networks, with continuous enhancement on the localization
accuracy over each generation, e.g., from hundreds of meters accuracy
in 2G to tens of meters in the fourth generation (4G). Nowadays, sub
meter-level localization accuracy can even be achieved in 5G by state-of-the-art
techniques, e.g. millimeter wave and massive MIMO. However, the limited
spectrum resource and hardware infrastructure will eventually become
a bottleneck for localization. Furthermore, due to the fact that radio
signals can simultaneously carry data and location-related information
of the transmitters, a unified study on integrated localization and
communication (ILAC) tends to be a natural choice. In this paper,
integrated sensing and communication (ISAC) is proposed as a more
general concept including both the JRC and ILAC as special cases,
since they can be viewed as the device-free ISAC and device-based
ISAC, respectively.

\subsection{Device-free Sensing}

\begin{figure}[tbh]
\begin{centering}
\includegraphics[width=45mm]{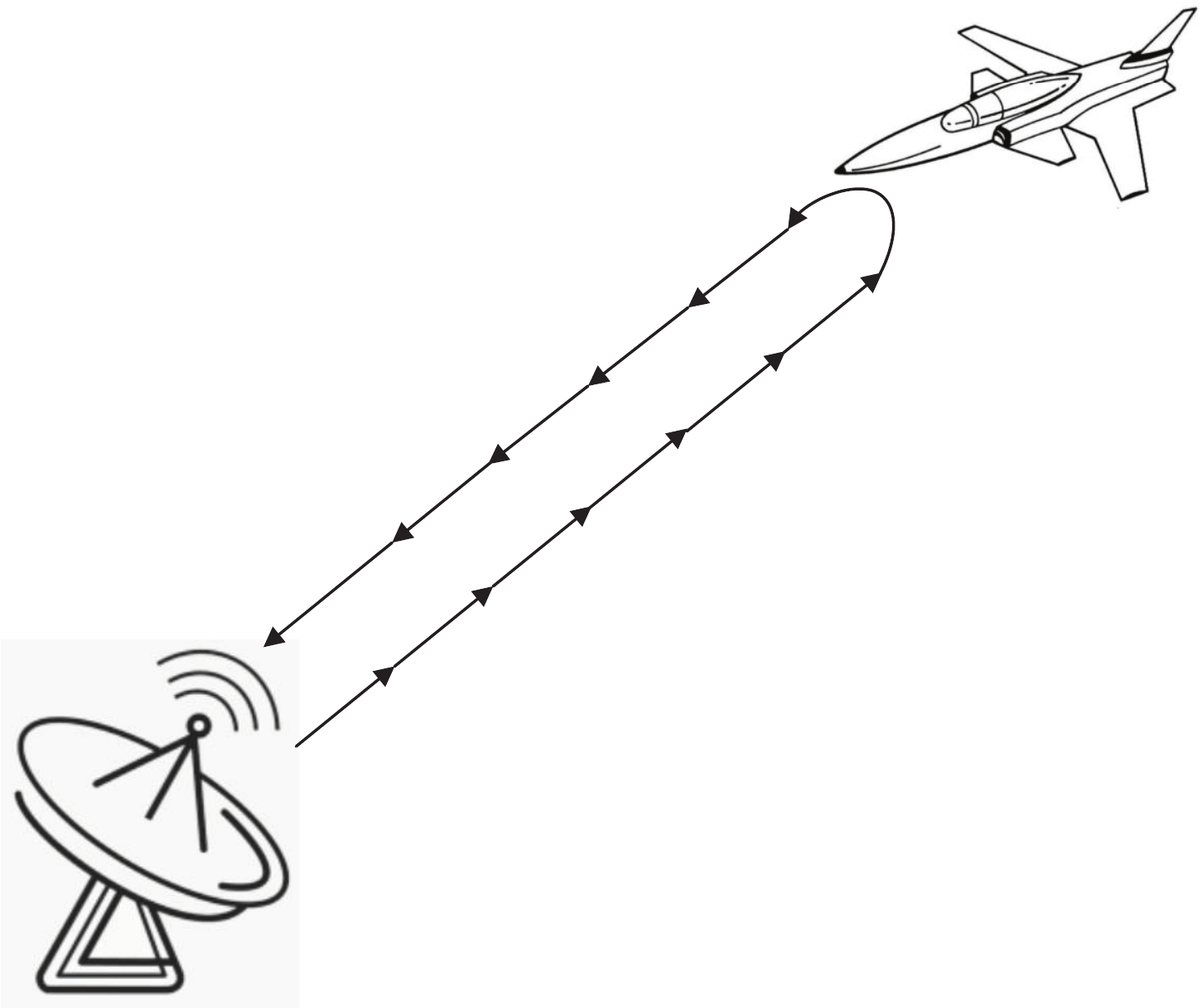}
\par\end{centering}
\centering{}\caption{\label{fig:Remote-Radar-sensing}Remote radar sensing of a single
target.}
\end{figure}

Since the majority of device-free sensing belong to radar sensing,
we will focus on the detailed classifications of radar sensing in
this subsection. As illustrated in Fig. \ref{fig:Remote-Radar-sensing},
radar transmits an omnidirectional or directional probing signal towards
the target. Then the probing signal is reflected by the target and
the radar echo is received by radar. Finally, the target parameters
can be estimated from the received echo.

Generally speaking, there are three radar system architectures: phased-array
radar, MIMO radar and phased-MIMO radar. In this subsection, we will
further divide these three kinds of radar into different classes and
discuss the structure and characteristic of each class.

\subsubsection{Phased-array Radar}

\begin{figure}[tbh]
\begin{centering}
\includegraphics[width=45mm]{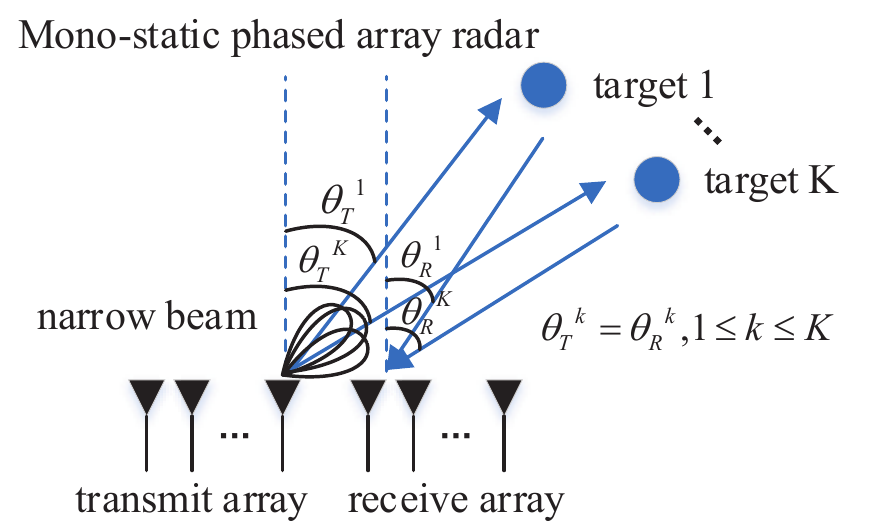}\includegraphics[width=45mm]{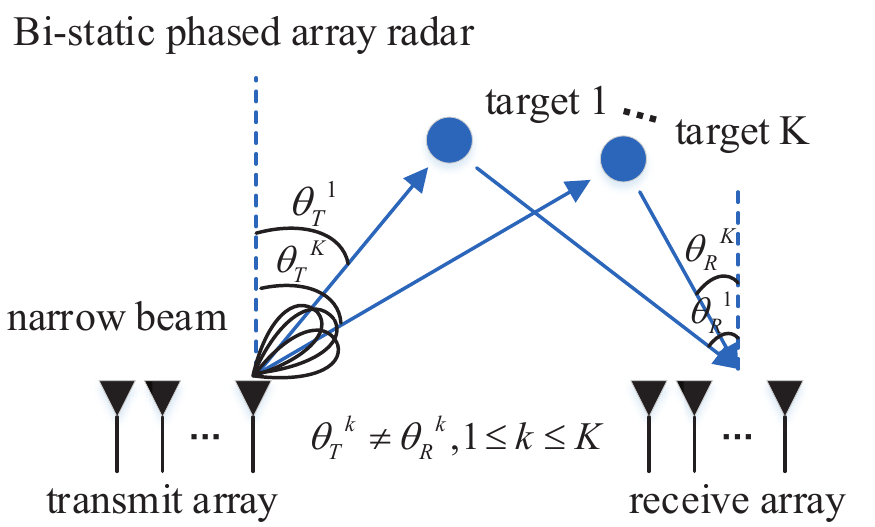}
\par\end{centering}
\centering{}\caption{\label{fig:phased-array radar}An illustration of different classes
of phased-array radar.}
\end{figure}

Phased-array antennas have been an enabling technology for many systems
in support of a variety of radar missions. Phased-array radar employs
many antennas placed together respectively for the transmit and receive
arrays. The spacing between the antennas within an array is set in
the same order of the wavelength of the sensing signal. Each of the
transmit antennas transmits a same baseband signal and transmit beamforming
is employed to steer a high-gain beam in a particular direction \cite{phased_radar_statement}.

As illustrated in Fig. \ref{fig:phased-array radar}, phased-array
radar can be divided into two classes according to whether the transmit
and receive arrays are placed together: mono-static phased-array radar
and bi-static phased-array radar.

Mono-static phased-array radar employs a system in which the transmit
and receive arrays are placed together. In many cases, the same antenna
array is exploited for both transmitting and receiving. In this paper,
we slightly extend this concept to include radar systems where the
transmit and receive antenna arrays are co-located. The advantage
of this kind of placement is that the AoD (angle of departure) and
AoA (angle of arrival) are the same in this case, thus fewer parameters
need to be estimated. However, the interference from the transmit
array to the receive array is non-negligible and needs to be eliminated.
One common method for interference elimination is to use pulsed waveforms
so that the transmit and the receive functions are performed at different
time intervals to avoid interference.

Bi-static phased-array radar employs a system in which the transmit
and receive arrays are placed in different sites. Since the AoD and
AoA are different in this case, more parameters need to be estimated.
However, the interference from the transmit array to the receive array
is smaller due to the larger distance.

\subsubsection{MIMO Radar}

MIMO radar was first proposed in \cite{MIMO_radar_proposed}. Contrary
to the phased-array radar, MIMO radar transmits decorrelated probing
signals from independent transmitters. Since independent signals undergo
independent fading, MIMO radar can overcome target Radar Cross Section
(RCS) scintillations \cite{MIMO_radar_proposed}. Moreover, the received
signal in MIMO radar is a superposition of independently faded signals,
and thus the average SNR of the received signal is more or less constant
\cite{MIMO_radar_proposed}.

\begin{figure}[tbh]
\begin{centering}
\includegraphics[width=75mm]{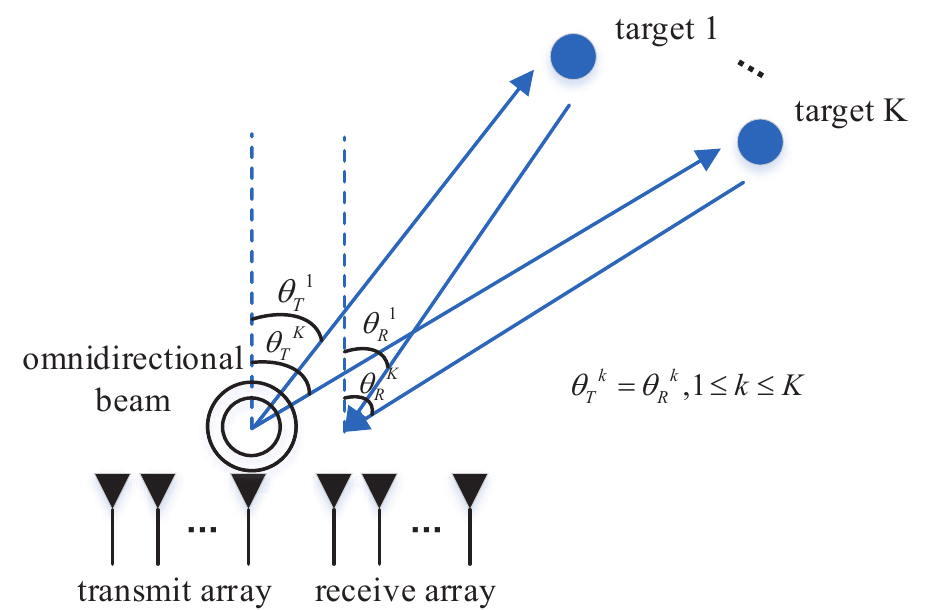}
\par\end{centering}
\centering{}\caption{\label{fig:MIMO-Radar}An illustration of colocated MIMO radar.}
\end{figure}

\begin{figure}[tbh]
\begin{centering}
\includegraphics[width=75mm]{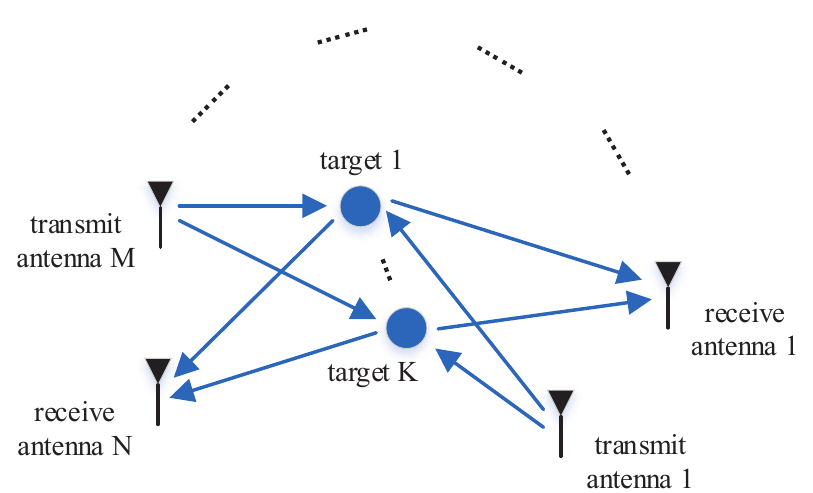}
\par\end{centering}
\centering{}\caption{\label{fig:MIMO-Radar-1}An illustration of distributed MIMO radar.}
\end{figure}

MIMO radar can be divided into two classes: colocated MIMO radar and
distributed MIMO radar \cite{MIMO_radar_proposed}.

In colocated MIMO radar, the antennas in the transmit or receive antenna
array are placed together, and the spacing between the antennas within
an array is set in the same order of the signal wavelength, as illustrated
in Fig. \ref{fig:MIMO-Radar}. Note that although the antenna placement
of colocated MIMO radar is similar to that of the phased-array radar,
the transmit signals are fundamentally different in these two radars,
i.e., independent signals in MIMO radar versus beamformed signals
in phased-array radar, as explained above. With decorrelated signals
transmitted from different transmitters and received by different
receivers placed together, the target has been observed multiple times
from the same direction, and each observation is independent from
each other. In this way, the waveform diversity gain can be achieved
to enhance the radar sensing performance \cite{colocated_MIMO_radar_2011,colocated_MIMO_radar_2013,colocated_MIMO_radar_2014}.

In distributed MIMO radar, the antennas in the transmit or receive
antenna array are widely distributed in different locations, and the
spacing between any two antennas is far larger than the wavelength,
as illustrated in Fig. \ref{fig:MIMO-Radar-1}. With independent signals
transmitted from distributed transmitters and received by distributed
receivers, the target has been observed multiple times from different
directions. Hence, the spatial diversity gain can be achieved to increase
the accuracy of localization \cite{statistical_MIMO_radar1,statistical_MIMO_radar2,statistical_MIMO_radar3}.
Note that there is no mono-static distributed MIMO radar since the
antennas in both transmit and receive arrays are distributed in the
space. However, each node might implement both transmit and receive
functions.

\subsubsection{Phased-MIMO Radar}

\begin{figure}[tbh]
\begin{centering}
\includegraphics[width=45mm]{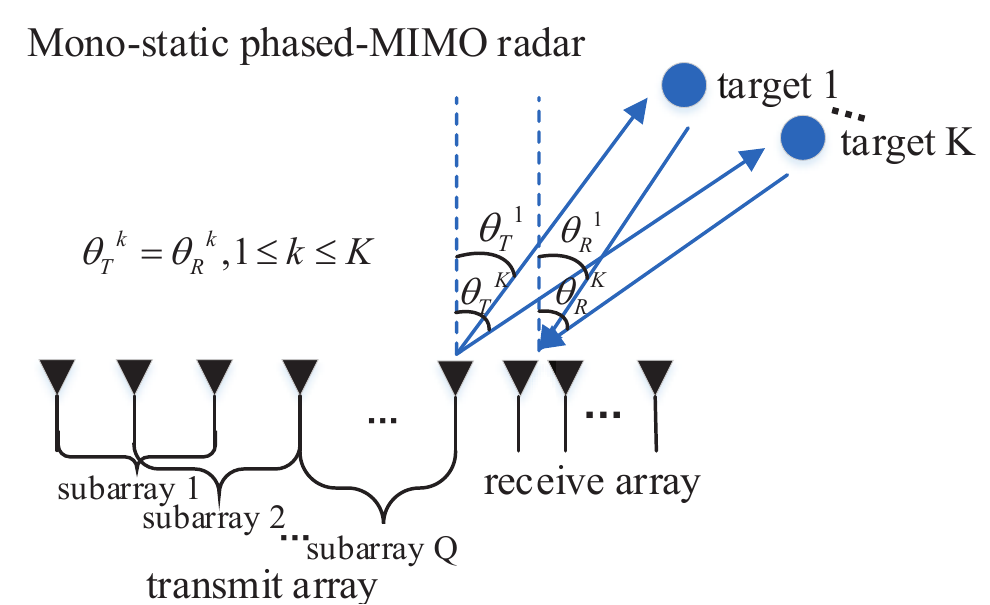}\includegraphics[width=45mm]{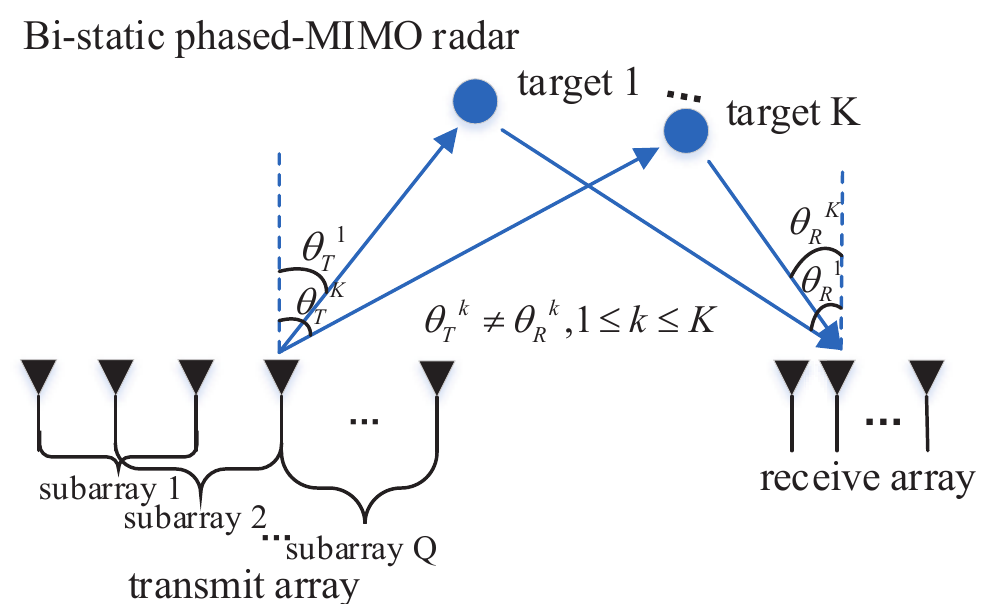}
\par\end{centering}
\centering{}\caption{\label{fig:Phased-MIMO-Radar}Phased-MIMO radar.}
\end{figure}

Phased-MIMO Radar was first proposed in \cite{phased-MIMO_proposed}
and it achieves a tradeoff between phased-array radar (beamforming
gain) and MIMO radar (waveform diversity gain). As illustrated in
Fig. \ref{fig:Phased-MIMO-Radar}, the transmit array of phased-MIMO
radar is divided into different sub-arrays which are allowed to have
overlapping. Each subarray is composed of any number of antennas ranging
from 1 to $M$, and forms a beam towards a certain direction. Different
waveforms are transmitted by different subarrays. Therefore, each
subarray can be regarded as a phased-array radar and all subarrays
can be jointly regarded as a MIMO radar. There is no specific limitation
imposed on the receive array, but a colocated receive array is typically
used \cite{phased-MIMO_proposed}. As illustrated in Fig. \ref{fig:Phased-MIMO-Radar},
phased-MIMO radars can be further divided into mono-static phased-MIMO
radars and bi-static phased-MIMO radars according to whether the transmit
and receive arrays are placed together.

\subsubsection{Other Device-free Sensing Scenarios}

There are some other device-free sensing scenarios that do not necessarily
fall into the above classes. For example, passive radar is another
technique for device-free sensing which has been investigated for
several decades, especially for defence applications \cite{passive-radar1}.
This kind of radar is not intended to send radar probing signals actively.
It instead parasitically exploits the echoes from the targets that
are illuminated by pre-existing transmitters, being intrinsically
bistatic. Various communication transmitters might be employed as
illuminators of opportunity thus enabling different applications.
Radio and television broadcast transmitters are usually preferred
for long range surveillance applications. On the other hand, WiFi
access points might be employed for local area monitoring \cite{passive-radar2,passive-radar3}.
The passive radar can estimate the desired parameters of the target
from the passively received signals. Passive radar has received renewed
interest for surveillance purposes because it allows target detection
and localization with advantages such as low cost, covert operation,
no frequency allocation requirement, etc. However, the sensing performance
of the passive radar is totally subject to the communication component.
Consequently, its performance is very sensitive to the characteristics
of the received waveforms, which may vary significantly over time
depending on the requirements and the characteristics of the communication
signals and channel. Therefore, advanced methodologies and signal
processing techniques have to be implemented to improve the reliability
of the resulting sensor against this time-varying scenario \cite{passive-radar4}.

\subsection{Device-based Sensing}

\begin{figure}[tbh]
\centering{}\includegraphics[width=75mm]{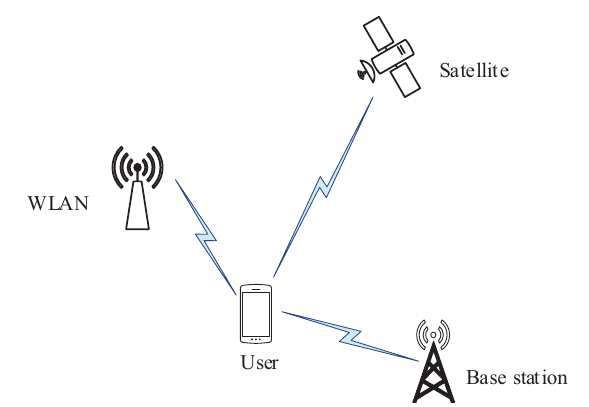}\caption{\label{fig:Wireless-based-localization-syst}Wireless-based localization
systems.}
\end{figure}

For device-based sensing, we will focus on the wireless-based localization.
Though different localization systems exist, e.g., GNSS, localization
systems based on WLAN or cellular networks as shown in Fig. \ref{fig:Wireless-based-localization-syst},
they all aim to estimate the location of the targeting object based
on a set of wireless reference signals propagated between the reference
nodes and the targeting object. The targeting objects with unknown
locations are often referred to as agent nodes, and the reference
nodes with known locations are often called anchor nodes. In most
cases, the agent receives reference signals from the anchors to localizes
itself. However, there are also cases in which the anchors receive
reference signals from the agents to localize the agents. In this
case, if the agent wants to obtain its own position, the anchors will
send the estimated position to the agent via a communication link.
In this paper, unless otherwise specified, we will mostly focus on
the case when the target want to localize itself based on the signals
received from multiple anchors. The localization problems in wireless
networks can be classified into two classes, namely, cooperative localization
and non-cooperative localization. With cooperation among neighboring
agent nodes, higher localization accuracy can be achieved, which reveals
a different fundamental limits compared to the non-cooperative localization,
as will be elaborated below.

\subsubsection{Non-Cooperative Localization}

\begin{figure}[tbh]
\centering{}\includegraphics[width=70mm]{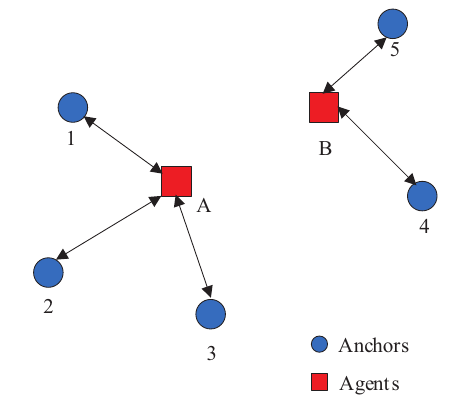}\caption{\label{fig:Non-Cooperative-Localization}Non-Cooperative Localization.}
\end{figure}

Consider a non-cooperative localization network with $N_{a}$ agents
and $N_{b}$ anchors, where each agent localizes itself based on signals
transmitted from neighboring anchors only. Fig. \ref{fig:Non-Cooperative-Localization}
illustrates a special case when $N_{a}=2$, $N_{b}=5$. Agent A receives
reference signals from Anchor 1, 2 and 3, while Agent B receives reference
signals from Anchor 4 and 5. The position of the agent can be inferred
from different metrics of the received signal, including time of arrival
(TOA), angle of arrival (AOA), angle of departure (AOD), time difference
of arrival (TDOA) and received signal strength (RSS), as detailed
below.

TOA or TDOA-based localization method extracts time-based metric from
received signals for localization. Generally speaking, TOA-based method
estimates the distance by multiplying the signal propagation delay
with the light speed. Then, based on trilateration relationship, the
agent position can be estimated. TOA-based method requires time synchronization
between the agent and all the anchors, which is quite difficult to
achieve in practical systems. To overcome this challenge, the TDOA-based
method, which only measures the differences in the TOAs from several
anchors, is proposed to get rid of the requirement on the time synchronization
between the agent and the anchors. In this case, the relative distance
is estimated in contrast to the TOA-based absolute distances estimation.

AOA-based localization is another commonly used approach that uses
the angles (AOA/AOD) between anchors and the agent node to achieve
localization. The angle-based metric can be extracted by an array
of antennas. Based on the AOA measurements, the agent can be localized
by two anchors in a 2D plane theoretically.

The RSS measurements can also be used for localization. RSS-based
localization method neither requires time synchronization among different
nodes nor relies on the LOS signal propagation. However, this method
has a fatal drawback, namely the poor localization accuracy. This
is because the RSS measurements highly rely on the characteristic
of the propagation environment. When the environment is harsh, e.g.,
in destructive shadowing, the localization performance will degrade
severely.

It is also possible to combine the above metrics to further enhance
the localization performance by using a hybrid method, e.g., based
on both TOA and AOA. Nonetheless, in real-life scenario, high-accuracy
localization may not be guaranteed by non-cooperative localization
owning to limited anchor deployment, especially in harsh environments.
For example, some agents may not receive strong signals from a sufficient
number of anchors. In this case, it is important to consider cooperative
localization which also utilizes signals from other agents, as elaborated
below.

\subsubsection{Cooperative Localization}

\begin{figure}[tbh]
\centering{}\includegraphics[width=70mm]{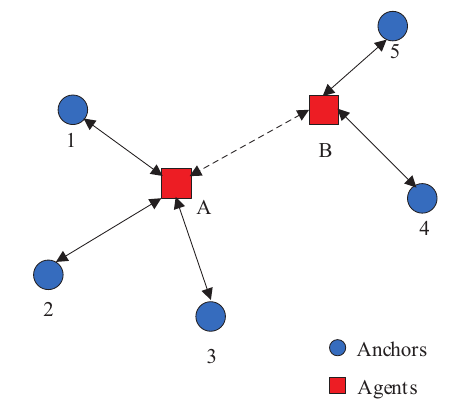}\caption{\label{fig:Cooperative-Localization}Cooperative localization.}
\end{figure}

In cooperative localization networks, each agent localizes itself
based on measurements from both anchors and other agents. Specifically,
as shown in Fig. \ref{fig:Cooperative-Localization}, the agents (A
and B) receive signals from the anchors (1, 2, 3, 4 and 5). Agent
A is not in the ranging range of Anchor 4 and 5, while Agent B is
not in the ranging range of Anchor 1, 2 and 3. Conventionally, each
agent needs at least three anchors to accurately localizes itself
based on range measurements in a 2-D plane. Therefore, Agent B cannot
be well localized if it only receives localization signals from two
neighboring anchors. However, if we allow cooperation between Agent
A and B, it is possible for Agent B to localizes itself by also using
the cooperative localization signals from Agent A. Furthermore, the
spatial cooperation mentioned above can be extended to spatio-temporal
cooperation, where each agent can incorporate localization information
both from other agents (spatial cooperation) and its own localization
result in the previous time slot (temporal cooperation).

Based on cooperative localization, higher coverage and accuracy can
be achieved with the same number of anchors as the non-cooperative
case. The drawback is that in cooperation localization, agents require
for stronger signal processing ability and their location may be exposed
to other agents.

\subsubsection{Other Device-based Sensing Scenarios}

Apart from wireless localization scenarios mentioned above, many other
device-based sensing scenarios have been considered, e.g., fingerprinting-based
localization, proximity-based localization and visible light-based
positioning (VLP) \cite{VLP1,VLP2,fingerprinting1,fingerprinting2,fingerprinting3,fingerprinting4,proximity1}.
In fingerprinting-based localization, unique geotagged signatures,
i.e. fingerprints, are extracted from the data collected by the sensors
firstly. Then the agent can be localized by matching the online signal
measurements against the pre-recorded fingerprints. For fingerprint-based
localization, the fingerprints extracted from the signal measurements
usually correspond to the RSS, because RSS based metric does not rely
on the LOS assumption and performs better in harsh environment. Compared
with geometric-based localization, fingerprinting-based localization
is more robust to clutter environment. However, its offline training
is time-consuming and complex. In proximity-based localization, the
position of the anchor which has the strongest RSS is treated as the
position of the agent node. Obviously, high location accuracy cannot
be guaranteed by this method. VLP is a promising localization method
based on transmitting visible light signals, and it has attracted
increasing attention from industry and academia recently \cite{VLP3,VLP4,VLP5}.
However, VLP has severe performance degradation in NLOS case and heavily
relies on special equipment.

\subsection{Device-free ISAC}

Device-free ISAC means that in the integrated system, the sensing
functionality is achieved by device-free sensing. Device-free ISAC
can be categorized according to different ISAC channel topologies.
In the following, we discuss several typical device-free ISAC channel
topologies, some of which (or simplified versions) have been introduced
in \cite{Bliss}. In all these channels, there are one base station
(BS), $K$ targets and $U$ users.

\subsubsection{Multiple Access Channel with Mono-Static Sensing}

\begin{figure}[tbh]
\begin{centering}
\includegraphics[width=75mm]{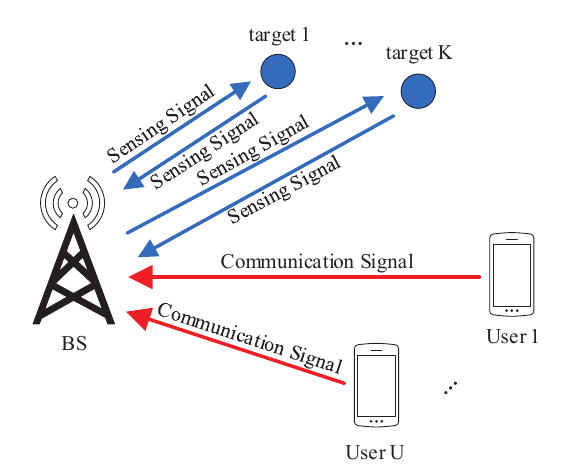}
\par\end{centering}
\centering{}\caption{\label{fig:Mono-static-multiple-access}Multiple access channel with
mono-static BS sensing.}
\end{figure}

The multiple access channel (MAC) with mono-static sensing refers
to a device-free ISAC channel whose communication channel topology
is MAC and radar structure is mono-static (i.e., colocated radar transmitter
and receiver). In general, both the BS and mobile users can act as
the mono-static radar. Two important special cases include the MAC
with mono-static BS sensing in which only the BS acts as a mono-static
radar, and the MAC with mono-static mobile sensing in which only the
mobile users act as mono-static radars. 

Specifically, a MAC with mono-static BS sensing is illustrated in
Fig. \ref{fig:Mono-static-multiple-access}, where the BS acts as
both radar transceiver and communication receiver, while the mobile
user is a communication transmitter. The BS aims to estimate the relevant
parameters of targets and decode the uplink messages from the users.
The challenge is that the uplink signals collide with the probing
radar echoes at the BS, leading to a joint estimation and decoding
problem.

\subsubsection{Multiple Access Channel with Bi-Static Sensing}

\begin{figure}[tbh]
\begin{centering}
\includegraphics[width=75mm]{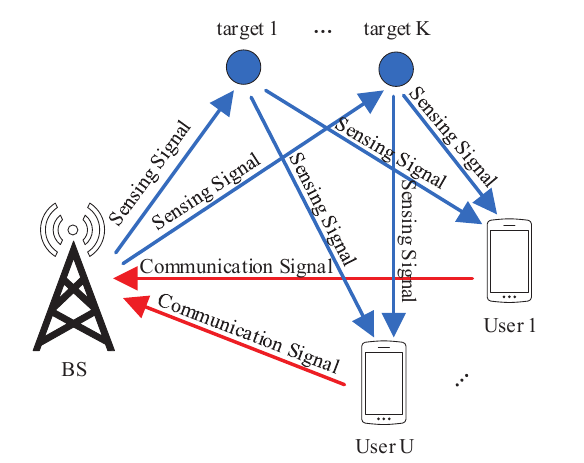}
\par\end{centering}
\centering{}\caption{\label{fig:Bi-static-multiple-access}Multiple access channel with
bi-static mobile sensing.}
\end{figure}

The MAC with bi-static sensing refers to a device-free ISAC channel
whose communication channel topology is MAC and radar structure is
bi-static (i.e., separate radar transmitter and receiver). In general,
both the BS and mobile users can act as the bi-static radar sensor
(radar receiver). Two important special cases include the MAC with
bi-static BS sensing in which only the BS acts as the radar sensor,
and the MAC with bi-static mobile sensing in which only the mobile
users act as bi-static radar sensors. 

Specifically, a MAC with bi-static mobile sensing is illustrated in
Fig. \ref{fig:Bi-static-multiple-access}, where the BS acts as both
radar transmitter and communication receiver, while the user acts
as both radar receiver and communication transmitter. The users aim
to estimate the relevant parameters of the targets while the BS aims
to decode the uplink messages from the users. In this case, the processing
of uplink signals and the probing radar echoes are decoupled if we
do not consider the self-interference at the BS and user sides. A
probable circumstance is that the targets are part of the scatters
for the communication channels. In this case, the user can acquire
partial Channel State information (CSI) from the probing radar echoes.
The challenge is how to use this partial Channel State Information
(CSI) for better uplink communication.

\subsubsection{Broadcast Channel with Mono-Static Sensing}

\begin{figure}[tbh]
\begin{centering}
\includegraphics[width=75mm]{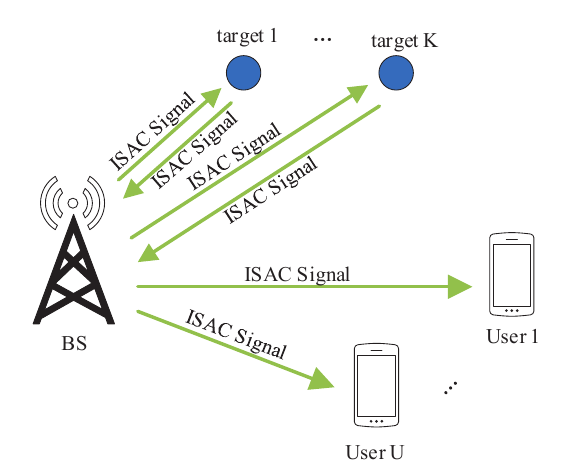}
\par\end{centering}
\centering{}\caption{\label{fig:Mono-static-broadcast-channel}Broadcast channel with mono-static
BS sensing.}
\end{figure}

The broadcast channel (BC) with mono-static sensing refers to a device-free
ISAC channel whose communication channel topology is BC and radar
structure is mono-static. In general, both the BS and mobile users
can act as the mono-static radar. Two important special cases include
the BC with mono-static BS sensing in which only the BS acts as a
mono-static radar, and the BC with mono-static mobile sensing in which
only the mobile users act as mono-static radars. 

Specifically, a BC with mono-static BS sensing is illustrated in Fig.
\ref{fig:Mono-static-broadcast-channel}, where the BS acts as both
radar transceiver and communication transmitter, while each user is
a downlink communication receiver. In general, a joint transmit waveform
can be used for both radar sensing and downlink communications. The
BS aims to estimate the relevant parameters of targets while the users
aim to decode the downlink messages. In this case, the processing
of downlink signals and the probing radar echoes are decoupled since
the BS knows the transmit data. The challenge is the joint design
of the transmit waveform for both the downlink signals and the probing
radar signals at the BS.

\subsubsection{Broadcast Channel with Bi-static Sensing}

\begin{figure}[tbh]
\begin{centering}
\includegraphics[width=75mm]{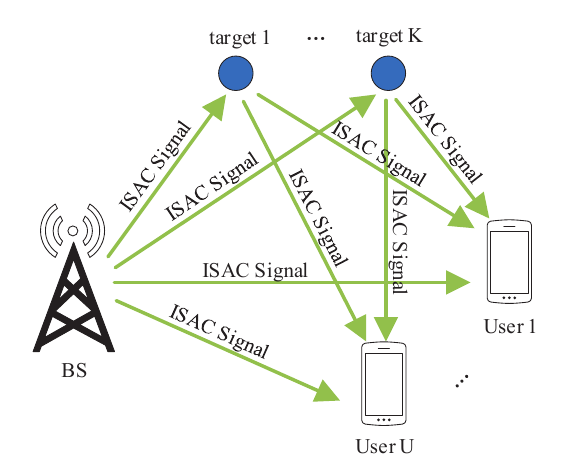}
\par\end{centering}
\centering{}\caption{\label{fig:Bi-static-broadcast-channel}Broadcast channel with Bi-static
mobile sensing.}
\end{figure}

The BC with bi-static sensing refers to a device-free ISAC channel
whose communication channel topology is BC and radar structure is
bi-static. In general, both the BS and mobile users can act as the
bi-static radar sensor (radar receiver). Two important special cases
include the BC with bi-static BS sensing in which only the BS acts
as a bi-static radar sensor, and the BC with bi-static mobile sensing
in which only the mobile users act as bi-static radar sensors. 

Specifically, a BC with bi-static mobile sensing is illustrated in
Fig. \ref{fig:Bi-static-broadcast-channel}, the BS acts as both radar
and communication transmitter, while the user acts as both radar and
communication receiver. The users aim to estimate the relevant parameters
of targets and decode the downlink messages. In this case, the processing
of downlink signals and the probing radar echoes are coupled. The
user needs to jointly estimate the target parameters and decode the
downlink message. The challenges are how to design the joint transmit
waveform at the BS and how to handle the superposition of the downlink
signals and the probing radar signals at the users.

Note that in the above descriptions, we have focused on cellular network
where we call the communication transmitter in the BC or communication
receiver in the MAC as the BS. However, the above device-free ISAC
channel topologies can also be used to model more general ISAC scenarios.
For example, in a general ISAC scenario, we may rename the ``MAC
with mono-static BS sensing'' as ``MAC with mono-static Com-Rx sensing''
since in this case, the communication receiver serves as the mono-static
radar sensor. Similarly, in a general ISAC scenario, we may rename
the ``BC with mono-static BS sensing'' as ``BC with mono-static
Com-Tx sensing'' since in this case, the communication transmitter
serves as a mono-static radar sensor. 

\subsection{Device-based ISAC}

Device-based ISAC means that in the integrated system, the sensing
functionality is achieved by device-based sensing. Device-based ISAC
can also be categorized according to different ISAC channel topologies.
In the following, we discuss several typical device-based ISAC channel
topologies.

\subsubsection{Multiple Access Channel with Non-Cooperative Localization}

\begin{figure}[tbh]
\centering{}\includegraphics[width=75mm]{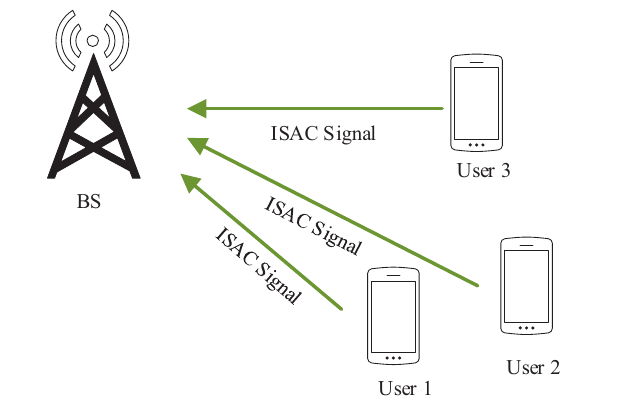}\caption{\label{fig:Non-cooperative-multiple-access}Multiple access channel
with non-cooperative localization.}
\end{figure}

In the multiple access channel with non-cooperative localization illustrated
in Fig. \ref{fig:Non-cooperative-multiple-access}, the users are
going to be localized or communicate with the BS. The BS receives
both communication and localization signals from the users and perform
joint localization and decoding.

\subsubsection{Broadcast Channel with Non-Cooperative Localization}

\begin{figure}[tbh]
\centering{}\includegraphics[width=75mm]{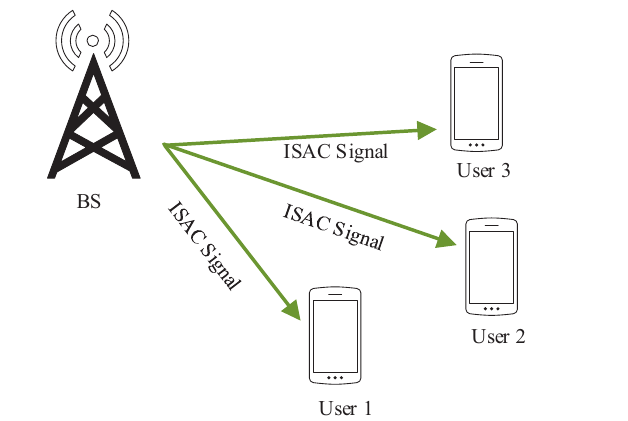}\caption{\label{fig:Broadcast-channel-topology.}Broadcast channel with non-cooperative
localization.}
\end{figure}

In the broadcast channel with non-cooperative localization illustrated
in Fig. \ref{fig:Broadcast-channel-topology.}, the BS transmits a
shared waveform to receivers for both localization and communications.
Each user needs to eliminate interference from others and extract
localization information from the common signals independently. So
the role of joint waveform design at the BS is highlighted, which
has a huge impact on the performance of both localization and communication.

\subsubsection{Relay Channel with Cooperative Localization}

\begin{figure}[tbh]
\centering{}\includegraphics[width=75mm]{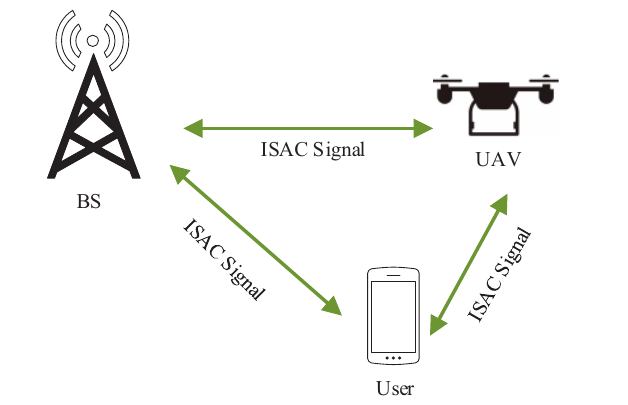}\caption{\label{fig:Relay-channel-topology}Relay channel with cooperative
localization.}
\end{figure}

In the relay channel with cooperative localization illustrated in
Fig. \ref{fig:Relay-channel-topology}, relays such as unmanned aerial
vehicles (UAVs) are used to aid both communications and localization.
For example, the UAV relay can be located by the ground BSs and then
be used as a new anchor node to assist the terrestrial localization.
In the meanwhile, the UAV-aided relaying can also provide communication
services.

\subsubsection{D2D Channel with Cooperative Localization}

\begin{figure}[tbh]
\centering{}\includegraphics[width=75mm]{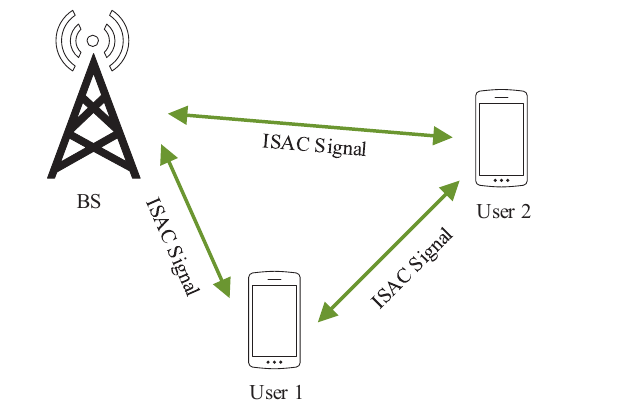}\caption{\label{fig:Cooperative-D2D-channel}D2D channel with cooperative localization.}
\end{figure}

In the D2D channel with cooperative localization illustrated in Fig.
\ref{fig:Cooperative-D2D-channel}, each user receives signals both
from the BS and other neighboring users for communications and localization.
Therefore, from communication perspective, there is a D2D communication
link providing a direct connection between users. From the localization
perspective, there is a cooperative localization link in addition
to the anchor-agent link.

\section{Performance Metrics\label{sec:Performance-Metrics}}

In this section, we present the key performance metrics that are useful
to characterize the fundamental limits of sensing, communication and
ISAC systems. In particular, for sensing systems, estimation-theoretic
metrics are considered, while for communication systems, information-theoretic
framework and metrics are considered. Both estimation-theoretic and
information-theoretic metrics are considered for ISAC systems.

\subsection{Estimation-Theoretic Metrics for Sensing\label{subsec:Metrics:sensing}}

The task of sensing is to obtain awareness of the scene surrounding
the sensor in general, which includes the capability to detect, localize
and track objects, to form images and/or to extract features for recognition/classification
purposes, etc. The focus of this paper is to investigate the performance
bounds in terms of parameter estimation capability, since many important
sensing objectives such as the localization and tracking of objects
can be interpreted as parameter estimation problems, and the capability
to estimate some key parameters, such as the time delay, DOA, and
Doppler frequency, also provides a foundation for more complicated
sensing objectives such as recognition and classification. 

\subsubsection{Mean-Square-Error and Relevant Lower Bounds}

Let $\boldsymbol{\theta}$ be the true parameter vector and ${\bf \hat{\boldsymbol{\theta}}}$
be the estimated vector, both of which are of dimension $K\times1$.
To assess the performance of an estimator, the mean-square error (MSE)
$\epsilon^{2}=\mathbb{E}\left\Vert \boldsymbol{\theta}-\hat{\boldsymbol{\theta}}\right\Vert ^{2}$
is a commonly used metric. Note that this MSE can also be viewed as
the trace of the following error covariance matrix (a.k.a. MSE matrix
in \cite{ZZB}) defined as
\begin{equation}
\mathsf{\mathbf{MSE}}_{\mathbf{\boldsymbol{\theta}}}=\mathbb{E}\left[\left(\mathbf{\boldsymbol{\theta}}-\hat{\boldsymbol{\theta}}\right)\left(\boldsymbol{\theta}-\hat{\boldsymbol{\theta}}\right)^{H}\right],\label{eq:MSE}
\end{equation}
whose diagonal elements quantify the individual MSE for parameters
$\theta_{k}\left(k=1,\cdots,K\right)$. One seeks for the optimal
estimator that minimizes the MSE $\epsilon^{2}$ in general. However,
such an optimal estimator is often difficult to construct and the
minimum MSE (MMSE) is normally hard to characterize.

To gain more insights on the performance limits, a few lower bounds
on $\mathsf{\mathbf{MSE}}_{\mathbf{\boldsymbol{\theta}}}$ have been
proposed in the literature \cite{1945Information}, \cite{ZZB} and
\cite{WWB}. The most famous one is the Cramer-Rao Bound (CRB). This
CRB applies to an unbiased estimator and can be computed as
\begin{equation}
CRB_{\boldsymbol{\theta}}=I^{-1}\left(\boldsymbol{\theta}\right),\label{eq:CRB}
\end{equation}
where $I\left(\boldsymbol{\theta}\right)$ is the Fisher's information
matrix (FIM) with $(i,j)$-th element $\left[I\left(\boldsymbol{\theta}\right)\right]_{ij}=\mathbb{E}_{\boldsymbol{y}}\left[\frac{\partial\ln p\left(\boldsymbol{y};\boldsymbol{\theta}\right)}{\partial\theta_{i}}\frac{\partial\ln p\left(\boldsymbol{y};\boldsymbol{\theta}\right)}{\partial\theta_{j}}\right]$
and $p\left(\boldsymbol{y};\boldsymbol{\theta}\right)$ is the likelihood
function associated with estimating the unknown deterministic parameter
vector $\boldsymbol{\theta}$ from the measurements $\boldsymbol{y}$.
It is known that if an unbiased estimate $\hat{\boldsymbol{\theta}}$
achieves the CRB, then it is the solution to the equation $\frac{\partial\ln p\left(\boldsymbol{y};\boldsymbol{\theta}\right)}{\partial\boldsymbol{\theta}}\mid_{\boldsymbol{\theta}=\hat{\boldsymbol{\theta}}}=0$.
Therefore, the sensitivity of the log-likelihood function $\ln p\left(\boldsymbol{y};\boldsymbol{\theta}\right)$
to changes in $\boldsymbol{\theta}$ determines the minimum achievable
MSE. The steeper the curvature of the log-likelihood function is,
the smaller the CRB is. While the CRB accounts for local errors and
is tight at high SNR, it performs poorly in low SNR regime. This can
be attributed to the lack of the global information of the log-likelihood
function in the CRB, since it is only determined by the local curvature
of the log-likelihood function around the true parameter $\boldsymbol{\theta}$.

The CRB can also be extended to the case when the parameters are random
variables with a known prior distribution \cite{posterior_CRB_NLF,posterior_CRB_target_tracking}.
The CRB with the knowledge of prior distribution is called the posterior
CRB since it serves as an MSE lower bound for the posterior mean estimator
(or equivalently, MMSE estimator). The posterior CRB is given by
\begin{equation}
CRB_{\boldsymbol{\theta}}^{post}=\left(I_{L}+I_{prior}\right)^{-1},
\end{equation}
where $I_{L}=\mathbb{E}_{\boldsymbol{\theta}}\left[I\left(\boldsymbol{\theta}\right)\right]$
is the FIM relevant to measurement and $I_{prior}$ is the FIM relevant
to prior knowledge with $(i,j)$-th element $\left[I_{prior}\right]_{ij}=\mathbb{E}_{\boldsymbol{\theta}}\left[\frac{\partial\ln p\left(\boldsymbol{\theta}\right)}{\partial\theta_{i}}\frac{\partial\ln p\left(\boldsymbol{\theta}\right)}{\partial\theta_{j}}\right]$.
Note that in this case, the FIM contains two components corresponding
to the contributions from the measurements (log-likelihood function)
and the knowledge of prior distribution, respectively. Since the FIM
$I\left(\boldsymbol{\theta}\right)$ for a given parameter vector
$\boldsymbol{\theta}$ still cannot capture the global information
of the log-likelihood function, the posterior CRB is usually loose
in low SNR regime as well.

To improve the tightness, Bayesian lower bounds have been later proposed
by treating the parameters as random variables each with known \textit{a
prior distribution. }Two representatives in this category are the
Weiss-Weinstein and Ziv-Zakai bounds.

In particular, the Weiss-Weinstein bound (WWB) further extended the
CRB by eliminating some regularity conditions on the likelihood function
and introducing free parameters $\boldsymbol{s}\in\left[0,1\right]^{k\times1}$
and $\boldsymbol{H}=\left[\begin{array}{cccc}
\boldsymbol{h}_{1} & \boldsymbol{h}_{2} & \ldots & \boldsymbol{h}_{k}\end{array}\right]\in\mathbb{R}^{K\times k}$ \cite{WWB}, where $k$ is the number of testing points. Specifically,
consider the following equation
\begin{align}
\mathbb{E}\bigg\{\stackrel[i=1]{k}{\sum}a_{i}\left[L^{s_{i}}\left(\boldsymbol{y};\boldsymbol{\theta}+\boldsymbol{h}_{i},\boldsymbol{\theta}\right)-L^{1-s_{i}}\left(\boldsymbol{y};\boldsymbol{\theta}-\boldsymbol{h}_{i},\boldsymbol{\theta}\right)\right] & \times\nonumber \\
\left[f\left(\boldsymbol{\theta}\right)-g\left(\boldsymbol{y}\right)\right]\bigg\} & =\nonumber \\
\stackrel[i=1]{k}{\sum}a_{i}\mathbb{E}\left\{ \left[f\left(\boldsymbol{\theta}-\boldsymbol{h}_{i}\right)-f\left(\boldsymbol{\theta}\right)\right]L^{1-s_{i}}\left(\boldsymbol{y};\boldsymbol{\theta}-\boldsymbol{h}_{i},\boldsymbol{\theta}\right)\right\}  & ,\label{eq:WWB-1}
\end{align}
where $g\left(\boldsymbol{y}\right)$ and $f\left(\boldsymbol{\theta}\right)$
are arbitrary scalar functions of $\boldsymbol{y}$ and $\boldsymbol{\theta}$,
$a_{i}$'s are arbitrary scalars, and $L\left(\boldsymbol{y};\boldsymbol{\theta},\boldsymbol{\theta}^{'}\right)=\frac{p\left(\boldsymbol{y};\boldsymbol{\theta}\right)}{p\left(\boldsymbol{y};\boldsymbol{\theta}^{'}\right)}$.
\begin{flushleft}
Note that $k$ is also a free parameter and as $k$ increases, an
increasingly tighter lower bound is generated. Squaring equation (\ref{eq:WWB-1})
and applying the Schwartz inequality to the left hand side gives
\begin{equation}
\mathbb{E}\left\{ \left[f\left(\boldsymbol{\theta}\right)-g\left(\boldsymbol{y}\right)\right]^{2}\right\} \geq\frac{\left(\boldsymbol{a}^{T}\boldsymbol{w}\right)^{2}}{\boldsymbol{a}^{T}\boldsymbol{Va}},\label{eq:WWB-2}
\end{equation}
where $\boldsymbol{a}=\left[a_{1},a_{2},\ldots,a_{k}\right]^{T}$,
$\boldsymbol{w}$ is a vector with $i$-th element $w_{i}=\mathbb{E}\left\{ \left[f\left(\boldsymbol{\theta}-\boldsymbol{h}_{i}\right)-f\left(\boldsymbol{\theta}\right)\right]L^{1-s_{i}}\left(\boldsymbol{y};\boldsymbol{\theta}-\boldsymbol{h}_{i},\boldsymbol{\theta}\right)\right\} $
and $\boldsymbol{V}$ is a matrix with the $(i,j)$-th element
\begin{align}
V_{ij}= & E\left\{ \left[L^{s_{i}}\left(\boldsymbol{y};\boldsymbol{\theta}+\boldsymbol{h}_{i},\boldsymbol{\theta}\right)-L^{1-s_{i}}\left(\boldsymbol{y};\boldsymbol{\theta}-\boldsymbol{h}_{i},\boldsymbol{\theta}\right)\right]\right.\nonumber \\
\times & \left.\left[L^{s_{j}}\left(\boldsymbol{y};\boldsymbol{\theta}+\boldsymbol{h}_{i},\boldsymbol{\theta}\right)-L^{1-s_{j}}\left(\boldsymbol{y};\boldsymbol{\theta}-\boldsymbol{h}_{i},\boldsymbol{\theta}\right)\right]\right\} .
\end{align}
\par\end{flushleft}

Applying the Schwartz inequality again such that the right hand side
of equation (\ref{eq:WWB-2}) is maximized for the choice $\boldsymbol{a}=\boldsymbol{V}^{-1}\boldsymbol{w}$.
Substitution of $f\left(\boldsymbol{\theta}\right)=\boldsymbol{u}^{T}\boldsymbol{\theta}$
and $g\left(\boldsymbol{y}\right)=\boldsymbol{u}^{T}\hat{\boldsymbol{\theta}}$,
the WWB bound for the MSE matrix is given by
\begin{equation}
\boldsymbol{u}^{T}\mathsf{\mathbf{MSE}}_{\mathbf{\boldsymbol{\theta}}}\boldsymbol{u}\geq\boldsymbol{u}^{T}\boldsymbol{H}Q^{-1}\left(\boldsymbol{s}\right)\boldsymbol{H}{}^{T}\boldsymbol{u},\label{eq:WWB-3}
\end{equation}
where the $(i,j)$-th element of $\boldsymbol{Q}$ is given by
\begin{equation}
Q_{ij}=\frac{V_{ij}}{\mathbb{E}\left\{ L^{1-s_{i}}\left(\boldsymbol{y};\boldsymbol{\theta}-\boldsymbol{h}_{i},\boldsymbol{\theta}\right)\right\} \mathbb{E}\left\{ L^{1-s_{j}}\left(\boldsymbol{y};\boldsymbol{\theta}-\boldsymbol{h}_{i},\boldsymbol{\theta}\right)\right\} }.\label{eq:WWB-4}
\end{equation}

The Ziv-Zakai bound (ZZB) \cite{ZZB} was developed by lower bounding
a quadratic form of the MSE matrix. The derived lower bound starts
from the following identity
\begin{equation}
\boldsymbol{u}^{T}\mathsf{\mathbf{MSE}}_{\mathbf{\boldsymbol{\theta}}}\boldsymbol{u}=\frac{1}{2}\int_{0}^{\infty}Pr\left(\left|\boldsymbol{u}^{T}\left(\boldsymbol{\theta}-\hat{\boldsymbol{\theta}}\right)\right|\geq\frac{h}{2}\right)hdh,\label{eq:ZZB}
\end{equation}
where $\boldsymbol{u}$ is an arbitrarily vector and $Pr\left(\left|\boldsymbol{u}^{T}\left(\boldsymbol{\theta}-\hat{\boldsymbol{\theta}}\right)\right|\geq\frac{h}{2}\right)$
can be lower bounded by
\begin{align}
 & Pr\left(\left|\boldsymbol{u}^{T}\left(\boldsymbol{\theta}-\hat{\boldsymbol{\theta}}\right)\right|\geq\frac{h}{2}\right)\nonumber \\
\geq & \int_{\boldsymbol{\Theta}}\left[p_{\boldsymbol{\theta}}\left(\boldsymbol{\varphi}\right)+p_{\boldsymbol{\theta}}\left(\boldsymbol{\varphi}+\boldsymbol{\delta}\right)\right]P_{min}\left(\boldsymbol{\varphi},\boldsymbol{\delta}\right)d\boldsymbol{\varphi},\label{eq:ZZB-lower-bound}
\end{align}
where $\boldsymbol{\delta}$ can be any vector satisfying $\boldsymbol{u}^{T}\boldsymbol{\delta}=h$,
and
\begin{align*}
 & P_{min}\left(\boldsymbol{\varphi},\boldsymbol{\delta}\right)\\
= & \frac{\int p_{\boldsymbol{y}}\left(\boldsymbol{y}\right)\min\left[p_{\boldsymbol{\theta}\mid\boldsymbol{y}}\left(\boldsymbol{\varphi}\mid\boldsymbol{y}\right),p_{\boldsymbol{\theta}\mid\boldsymbol{y}}\left(\boldsymbol{\varphi+\delta}\mid\boldsymbol{y}\right)\right]d\boldsymbol{y}}{p_{\boldsymbol{\theta}}\left(\boldsymbol{\varphi}\right)+p_{\boldsymbol{\theta}}\left(\boldsymbol{\varphi}+\boldsymbol{\delta}\right)}.
\end{align*}
The lower bound in (\ref{eq:ZZB-lower-bound}) is obtained by relating
the MSE in the estimation problem to the probability of error in a
binary detection problem. Please refer to \cite{ZZB} for the details.
Selecting $\boldsymbol{\delta}$ that maximizes (\ref{eq:ZZB-lower-bound})
leads to a tighter bound
\begin{align}
 & Pr\left(\left|\boldsymbol{u}^{T}\left(\boldsymbol{\theta}-\hat{\boldsymbol{\theta}}\right)\right|\geq\frac{h}{2}\right)\nonumber \\
\geq & \underset{\boldsymbol{\delta}}{\max}\int_{\boldsymbol{\Theta}}\left[p_{\boldsymbol{\theta}}\left(\boldsymbol{\varphi}\right)+p_{\boldsymbol{\theta}}\left(\boldsymbol{\varphi}+\boldsymbol{\delta}\right)\right]P_{min}\left(\boldsymbol{\varphi},\boldsymbol{\delta}\right)d\boldsymbol{\varphi}.
\end{align}
Applying the valley-filling function leads to the ZZB bound as in
(\ref{eq:ZZB-2}) on the top of the next page, where the valley-filling
function is defined as $\mathcal{V}\left\{ p\left(h\right)\right\} \triangleq\underset{\xi\geq0}{\max}p\left(h+\xi\right)$.

\begin{figure*}[tbh]
\begin{equation}
\boldsymbol{u}^{T}\mathsf{\mathbf{MSE}}_{\mathbf{\boldsymbol{\theta}}}\boldsymbol{u}\geq\frac{1}{2}\int_{0}^{\infty}\mathcal{V\left\{ \textrm{\ensuremath{{\displaystyle \underset{\boldsymbol{\delta}}{\max}{\displaystyle \int_{\boldsymbol{\Theta}}\left[p_{\boldsymbol{\theta}}\left(\boldsymbol{\varphi}\right)+p_{\boldsymbol{\theta}}\left(\boldsymbol{\varphi}+\boldsymbol{\delta}\right)\right]P_{min}\left(\boldsymbol{\varphi},\boldsymbol{\delta}\right)d\boldsymbol{\varphi}}}}}\right\} }hdh.\label{eq:ZZB-2}
\end{equation}
\end{figure*}

Both WWB and ZZB improve upon CRB over a wide range of SNRs, however,
they are harder to evaluate in general.

\begin{table*}[tbh]
\caption{\label{tab:MSE's-Lower-Bounds}MSE's Lower Bounds}

\centering{}%
\begin{tabular}{|c|c|c|c|}
\hline 
MSE Bounds & CRB & WWB & ZZB\tabularnewline
\hline 
\hline 
Expression & $\mathsf{\mathbf{MSE}}_{\mathbf{\boldsymbol{\theta}}}\succeq I^{-1}\left(\boldsymbol{\theta}\right)$ & $\boldsymbol{u}^{T}\mathsf{\mathbf{MSE}}_{\mathbf{\boldsymbol{\theta}}}\boldsymbol{u}\geq\boldsymbol{u}^{T}\boldsymbol{H}Q^{-1}\left(\boldsymbol{s}\right)\boldsymbol{H}{}^{T}\boldsymbol{u}$ & (\ref{eq:ZZB-2})\tabularnewline
\hline 
Advantages & Low complexity & A generalization of CRB & More accurate over the full range of SNR\tabularnewline
\hline 
Disadvantages & Inaccurate under low SNR & Free parameters $\boldsymbol{s}$ and $\boldsymbol{H}$ are hard to
choose & Integrals are hard to solve\tabularnewline
\hline 
\end{tabular}
\end{table*}

\subsubsection{Equivalent Fisher's Information Matrix (EFIM)}

In many cases, the unknown parameters can be divided into two subvectors
as $\boldsymbol{\theta}=\left[\begin{array}{cc}
\boldsymbol{\theta}_{1}^{T} & \boldsymbol{\theta}_{2}^{T}\end{array}\right]^{T}\in\mathbb{R}^{K\times1}$, where the first subvector $\boldsymbol{\theta}_{1}\in\mathbb{R}^{m\times1}$
is the parameter of interest and the second subvector is the nuisance
parameter. In this case, the FIM $I\left(\boldsymbol{\theta}\right)$
can be partitioned into submatrices as

\begin{equation}
I\left(\boldsymbol{\theta}\right)=\left[\begin{array}{cc}
I\left(\boldsymbol{\theta}_{1},\boldsymbol{\theta}_{1}\right) & I\left(\boldsymbol{\theta}_{1},\boldsymbol{\theta}_{2}\right)\\
I\left(\boldsymbol{\theta}_{1},\boldsymbol{\theta}_{2}\right)^{T} & I\left(\boldsymbol{\theta}_{2},\boldsymbol{\theta}_{2}\right)
\end{array}\right],
\end{equation}
where $I\left(\boldsymbol{\theta}_{1},\boldsymbol{\theta}_{1}\right)\in\mathbb{R}^{m\times m},$
$I\left(\boldsymbol{\theta}_{1},\boldsymbol{\theta}_{2}\right)\in\mathbb{R}^{m\times\left(K-m\right)}$
and $I\left(\boldsymbol{\theta}_{2},\boldsymbol{\theta}_{2}\right)\in\mathbb{R}^{\left(K-m\right)\times\left(K-m\right)}$.
We only care about the CRB of the first subvector $\boldsymbol{\theta}_{1}$.
One possible solution is to first calculate the inverse of the FIM
of the entire parameter vector as $I^{-1}\left(\boldsymbol{\theta}\right)$
and then obtain the CRB of the first subvector by extracting the submatrix
$\left[I^{-1}\left(\boldsymbol{\theta}\right)\right]_{m\times m}$
at the left-top corner of $I^{-1}\left(\boldsymbol{\theta}\right)$.
A more efficient method is to directly calculate the FIM of the first
subvector by introducing the concept of EFIM. Specifically, the EFIM
for $\boldsymbol{\theta}_{1}$ is defined as
\begin{equation}
I_{e}\left(\boldsymbol{\theta}_{1}\right)=I\left(\boldsymbol{\theta}_{1},\boldsymbol{\theta}_{1}\right)-I\left(\boldsymbol{\theta}_{1},\boldsymbol{\theta}_{2}\right)I\left(\boldsymbol{\theta}_{2},\boldsymbol{\theta}_{2}\right)^{-1}I\left(\boldsymbol{\theta}_{1},\boldsymbol{\theta}_{2}\right)^{T}.
\end{equation}
Note that the EFIM $I_{e}\left(\boldsymbol{\theta}_{1}\right)$ retains
all the necessary information to derive the information inequality
for the parameter $\boldsymbol{\theta}_{1}$, since $\left[I^{-1}\left(\theta\right)\right]_{m\times m}=I_{e}^{-1}\left(\boldsymbol{\theta}_{1}\right)$
and the MSE matrix of $\boldsymbol{\theta}_{1}$ is bounded below
by $I_{e}^{-1}\left(\boldsymbol{\theta}_{1}\right)$.

\subsubsection{Other Performance Metrics}

Other forms of performance criterions have also been considered in
the literature. For instance, in radar sensing, the theory of radar
resolution has been developed to facilitate understanding of the fundamental
resolution limitations of radar systems. A well known theoretical
estimate of radar resolution is $\Delta R=\frac{c}{2B}$, where $c$
is the speed of light and $B$ is the bandwidth \cite{skolnik1980introduction}.
In addition, a normalized cross-ambiguity function was introduced
in \cite{woodward1953probability}, and a multi-dimensional ambiguity
function has been proposed in \cite{GResolution} to characterize
the tradeoff between system parameters and resolution in range, angle
(azimuth and elevation) and Doppler. In particular, the concept of
an ambiguity function has been obtained by introducing a physically
meaningful and mathematically tractable definition of a difference
function between the two sets of signals produced at the elements
of a receiving aperture by two targets differing in range, angle (azimuth
and elevation) or Doppler. Under the narrow band assumption, this
multi-dimensional ambiguity function is factorized as the product
of the range-Doppler ambiguity function and the azimuth-elevation
ambiguity function, where the overall resolution constant depends
upon up the effective area of the aperture \cite{GResolution}. 

There are also performance metrics for target detection. In general,
the task of detection is to decide whether a target exists through
a sequence of measurements. Two important performance metrics for
target detection are detection probability and false alarm probability.
The detection probability indicates the probability of detecting a
target when a target actually exists, while the false alarm probability
indicates the probability of detecting a target when a target does
not exist \cite{richards2005fundamentals}.

In addition, considering each independent resolution cell (e.g., range-angle-Doppler)
as a binary information storage unit, i.e., ``0\textquotedbl{} = target
absent, ``1\textquotedbl{} = target present, J. Guerci \textit{et
al. }introduced the notion of radar capacity (analogous to the capacity
of communication) \cite{radar_capacity} by the Hartley capacity measure
\begin{equation}
C_{R}=\log_{2}N,
\end{equation}
where $N$ is the total number of independent radar resolution cells
given by
\begin{equation}
N\propto\frac{R_{\max}}{\Delta R}\frac{2\pi}{\Delta\theta}\frac{PRF}{\Delta f_{d}},
\end{equation}
where $R_{\max}$ is the maximum range, $\Delta R$ is the range resolution,
$\Delta\theta$ is the bearing resolution, $PRF$ is the pulse repetition
frequency and $\Delta f_{d}$ is the Doppler resolution.

\subsubsection{Summary}

The MSE and its lower bounds have been proposed to investigate fundamental
limits of the parameter estimation problem. The most commonly used
bounds include CRB, WWB and ZZB. The CRB is generally easier to compute,
but it does not adequately characterize the performance in particular
in the low SNR regime. WWB and ZZB are Bayesian bounds and improves
upon CRB but at the expense of heavier computational complexity. Complementary
to these MSE bounds, in the context of radar sensing, the theory of
radar resolution has also been developed to quantify the limit at
which radar is able to separate two targets in the range, angular
or Doppler domain. The comparison of MSE's lower bounds are summarized
in Table \ref{tab:MSE's-Lower-Bounds}. In WWB, if we let $\boldsymbol{H}=hI$,
$k=K$ and $h\rightarrow0$, we will arrive at the expression of CRB
\cite{WWB}.

\subsection{Information-Theoretic Metrics for Communication\label{subsec:Metrics:communication}}

The task of communication is to transmit message from source to destination
as reliably as possible. Channel capacity, originally conceived by
Shannon, is one of the most important notions for assessing the fundamental
limits of a communication system. Shannon capacity measures the maximum
communication rate in bits per transmission such that the probability
of error can be made arbitrarily small when the coding block length
is sufficiently large. In what follows, we first briefly review the
channel capacity of a time-invariant channel and then moves on to
discuss two important capacity definitions tailored to the time-varying
channel.

\subsubsection{Channel capacity of a time-invariant channel}

For a single-user time-invariant channel, the Shannon capacity is
defined as the maximum mutual information $I\left(X;Y\right)$ between
the channel input $X$ and output $Y$, i.e., $C=\max_{p\left(x\right)}I\left(X;Y\right)$
bits per channel use (bcu). When specialized to a Gaussian channel
with additive white Gaussian noise and an average transmit power constraint
$P$ on input $X$, the capacity $C$ corresponds to the well-known
Shannon's formula: $C_{awgn}=\log_{2}\left(1+\frac{P}{\sigma^{2}}\right)$
bcu, where $\sigma^{2}$ is the noise variance. The capacity notion
has also been applied to various multi-user time-invariant channels,
such as multiple-access channels (MAC), broadcast channels, interference
channels and relay channels \cite{el2011network,cover1999elements}.
In particular, the capacity region of discrete memoryless and Gaussian
MAC is fully characterized, while for other channel topologies, achievable
rate regions have been proposed and the capacity region is known for
a limited class of channels.

\subsubsection{Ergodic and outage capacity of a time-varying channel}

\begin{figure}[tbh]
\begin{centering}
\includegraphics[width=85mm]{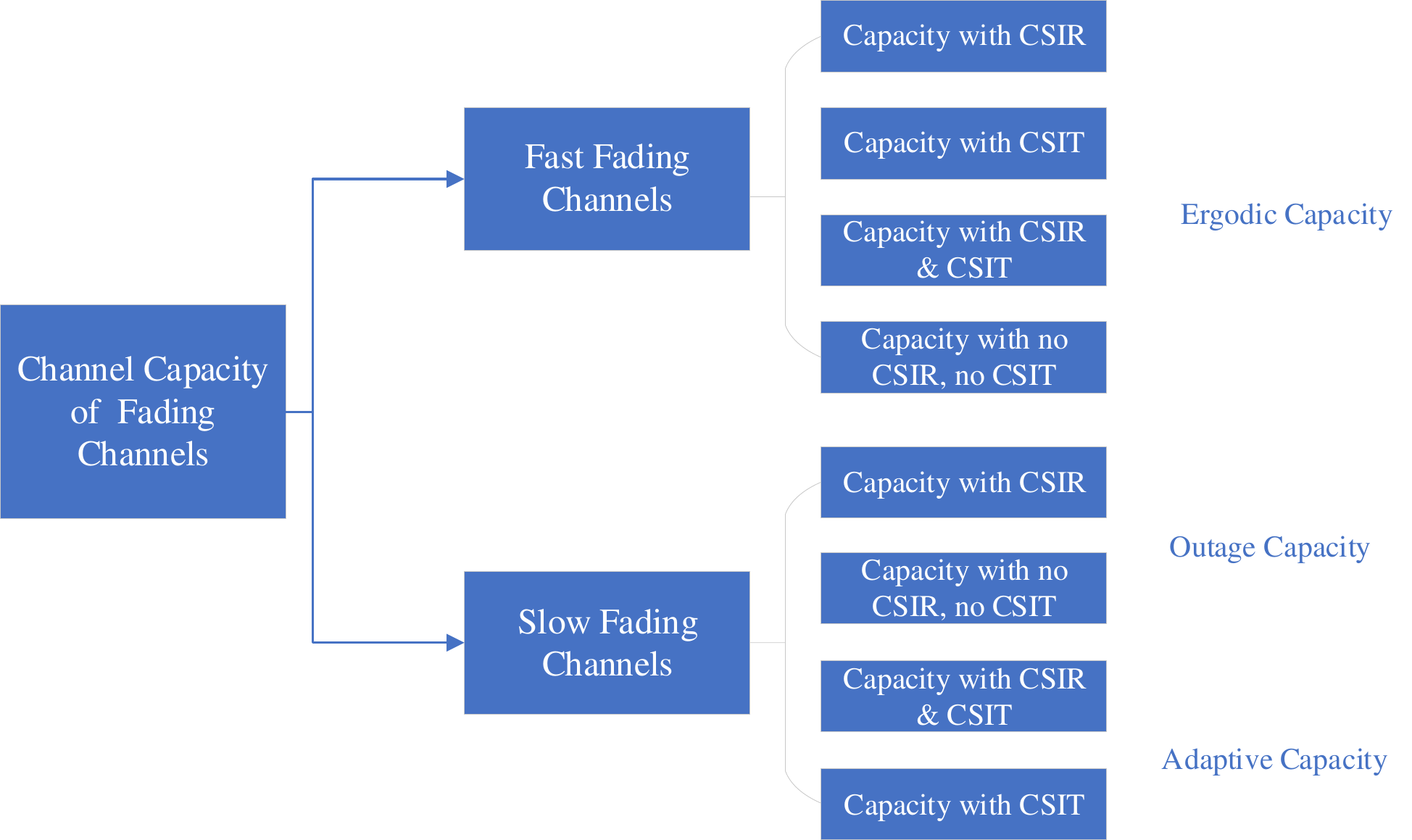}
\par\end{centering}
\centering{}\caption{\label{fig: performance metrics for fading channel}An overview of
performance metrics for fading channels.}
\end{figure}

Considering wireless fading time-varying channels, we can distinguish
fast fading and slow fading and further classify each case into subcases
each with or without Channel State Information at Transmitter (CSIT)
and/or Channel State Information at Receiver (CSIR), see Fig. \ref{fig: performance metrics for fading channel}.
Two capacity definitions are reviewed:
\begin{itemize}
\item Ergodic capacity: In the case of fast fading, the coding block length
spans a large number of channel coherence time intervals. The channel
is thus ergodic (i.e., each codeword seeing all possible fading realizations)
and has a well-defined Shannon ergodic capacity. For a single-user
channel with perfect CSIR and CSIT, the ergodic capacity is given
by
\begin{align}
C_{CSIR/CSIT} & =\underset{p\left(X\mid H\right)}{\max}\mathbb{E}_{H}\left[I\left(X;Y\mid H=h\right)\right]\label{eq:fast fading_CSIR_CSIT}\\
 & \text{s.t. }\mathbb{E}\left[\left|X\right|^{2}\right]\leq P,\nonumber 
\end{align}
which is attained by adapting the transmission power and rate to the
channel state variations, i.e., the input distribution $p\left(X\mid H\right)$
depends on the channel state $H$. On the other hand, if with perfect
CSIR but without CSIT, no adaptive transmission strategy is allowed
and the ergodic capacity reduces to
\begin{align}
C_{CSIR} & =\underset{p\left(X\right)}{\max}\text{ }\mathbb{E}_{H}\left[I\left(X;Y\mid H=h\right)\right]\\
 & \text{s.t. }\mathbb{E}\left[\left|X\right|^{2}\right]\leq P.
\end{align}
In this case, the input distribution $p\left(X\right)$ does not depend
on the channel state $H$ anymore.
\item Outage capacity: In the case of slow fading, the coding block length
is on the order of the channel coherence time interval. The channel
is thus no longer ergodic and Shannon capacity is not well defined
in this case. However, if the system can tolerate a loss of a fraction
$p_{out}$ of the messages on average, reliable communication can
be achieved at any rate lower than an outage capacity. For a single-user
channel with perfect CSIR without CSIT, the outage capacity is given
by
\begin{align}
C_{out} & =\underset{\begin{array}{c}
p\left(X\right):\mathbb{E}\left[\left|X\right|^{2}\right]\leq P\end{array}}{\max}R\\
 & \text{ s.t. }p\left(I\left(X;Y\mid H=h\right)<R\right)\leq p_{out}.
\end{align}
\end{itemize}
The definitions above can also be generalized to the multiuser scenario,
leading to ergodic capacity region and outage capacity region, see,
e.g., \cite{el2011network}, \cite{cover1999elements}. More information-theoretic
modeling and fundamental limits on the state-dependent channels can
also be found in \cite{2008Channel}.

\subsubsection{Summary}

Shannon capacity serves as an ultimate limit that a communication
system can achieve. In addition to its theoretical importance, the
establishment of the capacity can also guide the design of capacity-achieving
structured codes (such as LDPC and polar codes) and drive innovative
transmission strategy, such as adaptive power and rate transmission
in the ergodic fading case. However, Shannon capacity is not always
well defined in particular when the channel is time-varying and the
transmitter or receiver may or may not access to all realizations
or the full statistics of the channel state. Alternative approaches,
such as outage capacity here, or adaptive capacity and broadcast capacity
\cite{el2011network}, \cite{cover1999elements} can be useful in
characterizing the performance bounds of such a communication system.

\subsection{Performance Metrics for ISAC}

In the above two subsections, we have presented some performance metrics
for sensing and communication functionalities, respectively. ISAC
systems aim to integrate both functionalities in a synergetic manner
and therefore fundamental communication-sensing performance tradeoff
should be fully understood. Towards this end, a unified capacity-distortion
performance metric is considered, where the capacity measures the
communication performance as presented in Subsection \ref{subsec:Metrics:communication},
while the distortion notion slightly generalizes the MSE as in Subsection
\ref{subsec:Metrics:sensing} to account for estimation of parameters
of finite alphabet and to accommodate arbitrary estimation cost function.
In the following, we review three approaches for representing the
capacity-distortion tradeoff in the literature. We would like to point
out that the existing performance metrics for ISAC are still primeval
and thus deserve further study.

\subsubsection{Estimation-Information-Rate Induced Approach}

The estimation information rate was introduced by \cite{Information_estimation_rate}
and represents an approximate mutual information between the observation
$Y$ and the true parameter $\theta$. Specifically, consider $\theta$
is Gaussian distributed with variance $P$ and it is estimated as
$\hat{\theta}$ with MSE distortion $D$. It is standard to establish
the following inequality chain
\begin{equation}
I\left(\theta;Y\right)\geq I\left(\theta;\hat{\theta}\right)\geq\frac{1}{2}\log\left(\frac{P}{D}\right),
\end{equation}
where the first inequality uses the Markov chain $\theta-Y-\hat{\theta}$
and follows by the data processing inequality, while the second inequality
holds because
\begin{equation}
\begin{array}{l}
I\left(\theta;\hat{\theta}\right)=h\left(\theta\right)-h\left(\theta\mid\hat{\theta}\right)\\
\geq h\left(\theta\right)-h\left(\theta-\hat{\theta}\right)\\
\geq h\left(\theta\right)-\frac{1}{2}\log\left(2\pi e\mathbb{E}\left[\left(\theta-\hat{\theta}\right)^{2}\right]\right).
\end{array}
\end{equation}
This lower bound therefore converts the MSE distortion to an estimation
information rate for sensing. Hence one can examine the tradeoff between
the communication information rate and the estimation information
rate both in the same unit for ISAC systems.

\subsubsection{Equivalent-MSE Induced Approach}

Instead of deriving equivalent estimation information rate for sensing,
\cite{DMMSE} proposed to derive the equivalent of communication information
rate to the MSE metric. In particular, consider a Gaussian channel
$Y=\sqrt{snr}X+Z$, where $X,Z\sim\mathcal{CN}\left(0,1\right)$.
Then the MMSE of estimating input $X$ from output $Y$ is given by:
$D\left(snr\right)=1/\left(1+snr\right)$. Therefore one can convert
a given communication capacity $C\left(snr\right)=\log\left(1+snr\right)$
to a MSE metric by $D_{\textrm{Equivalent}}=2^{-C}$ . In this way,
one can examine the tradeoff between the communication equivalent-MSE
and the estimation MSE both in the same unit for ISAC systems.

\subsubsection{Capacity-Distortion Function Induced Approach}

Different from traditional channel capacity, the capacity-distortion
function $C\left(D\right)$ is the channel capacity under certain
distortion constraints $D$ since we need to send message while simultaneously
estimating the channel state $S$ \cite{P2P},\cite{zhang}. Specifically,
a general capacity-distortion function is given by
\begin{equation}
C\left(D\right)=\underset{p\left(X\right)}{\max}\text{ }I\left(X;Y\mid S\right),\text{ s.t. }\mathbb{E}\left[d\left(S,\hat{S}\right)\right]\leq D,
\end{equation}
where $X,Y$ are input and output symbol respectively, $\hat{S}$
is the estimated sensing state and $\mathbb{E}\left[d\left(S,\hat{S}\right)\right]$
is the average distortion of an estimator. More details can be found
in Section \ref{sec:device-free isac} and Section \ref{sec:device-based isac}.

\subsubsection{Summary}

The first two approaches above represent very preliminary attempts
at constructing a unified capacity-distortion performance metric for
ISAC systems. Each has its own obvious limitations. The first approach
assumes Gaussian distributed sensing parameters and estimation errors
and requires to know the MSE of an estimator, while the second approach
also works only in a simple linear Gaussian channel modeling. The
third approach seems to be a more natural way to unify the analysis
of the fundamental limits of ISAC under the information-theoretic
framework. However, the current information-theoretic models considered
in \cite{P2P},\cite{zhang} are oversimple and cannot cover many
important ISAC scenarios. As such, new frameworks and more general
approaches are called upon for better characterizing the performance
limits of ISAC.

\section{Fundamental Limits of Device-free Sensing}

In this section, we will discuss the current research progress on
the fundamental limits of device-free sensing. In particular, we will
focus on the fundamental limits for different classes of radar sensing
as classified in Section \ref{sec:Classifications-of-Integrated}.
For each class, we will highlight several important works, and present
the system model, performance bounds and key insights learned from
the analysis of the fundamental limits.

\subsection{Fundamental Limits of Phased-array Radar}

A few works have investigated the fundamental limits of phased-array
radar. In \cite{richards2005fundamentals}, the author studied the
performance limits of the mono-static phased-array radar system with
a single transmit antenna and $N$ receive antennas. Assuming that
the target is quasi-static and the Doppler effect can be ignored,
the $N$-dimensional received signal for one radar pulse is given
by
\begin{equation}
\boldsymbol{Y}(t)=\alpha\boldsymbol{a}_{R}(\theta)s\left(t-\tau\right)+\boldsymbol{Z}(t),
\end{equation}
where $\alpha$ is the reflection coefficient of the target, $\tau$
is the delay of the target, $\boldsymbol{a}_{R}(\theta)=[e^{j\frac{2\pi}{\lambda}R_{1}\sin\theta},e^{j\frac{2\pi}{\lambda}R_{2}\sin\theta},\cdots,e^{j\frac{2\pi}{\lambda}R_{N}\sin\theta}]^{T}$
is the receive steering vector with $R_{n}$ denoting the locations
of the $n$-th antennas, $s\left(t\right)$ is the transmit waveform
with normalized energy and $\boldsymbol{Z}(t)$ is the noise matrix,
including the interfering echoes from the clutters and the background
noise. The noise matrix has i.i.d complex Gaussian entries of zero
mean and variance $\sigma_{z}^{2}$. Under these assumptions, the
CRBs of delay $\tau$ and direction of arrival (DOA) $\theta$ are
given by
\begin{equation}
CRB_{\tau}=\frac{1}{8\pi^{2}\textrm{SNR}N\beta^{2}},\label{eq:CRBtau_siso}
\end{equation}

\begin{equation}
CRB_{\theta}=\frac{6}{\left(2\pi\right)^{2}\frac{d^{2}}{\lambda^{2}}\cos^{2}\theta\textrm{SNR }N\left(N^{2}-1\right)},\label{eq:CRBtheta_siso}
\end{equation}
where $\beta^{2}=\frac{\int_{-\infty}^{\infty}f^{2}\left|S(f)\right|^{2}df-\left(\int_{-\infty}^{\infty}f\left|S\left(f\right)\right|^{2}df\right)^{2}}{\int_{-\infty}^{\infty}\left|S(f)\right|^{2}df}$
is the squared effective bandwidth, $S(f)$ is the Fourier transformation
of transmitted baseband signal $s(t)$, $\text{SNR}=\frac{\left|\alpha\right|^{2}}{\sigma_{z}^{2}}$
is the received SNR, $\lambda$ is the signal wavelength, and $d$
is antenna spacing. Note that if $\left|S\left(f\right)\right|$ is
symmetric with respect to zero, the right integral representation
$\int_{-\infty}^{\infty}f\left|S\left(f\right)\right|^{2}df$ will
become zero.

From (\ref{eq:CRBtau_siso}) and (\ref{eq:CRBtheta_siso}), we conclude
that the estimation performance of both delay $\tau$ and DOA $\theta$
improves with the increase of SNR and the number of receive antennas
$N$. In addition, the estimation performance of $\tau$ also improves
with the increase of the squared effective bandwidth $\beta^{2}$,
while the estimation performance of $\theta$ improves with the increase
of the normalized antenna spacing $d/\lambda$.

\begin{figure}[tbh]
\begin{centering}
\includegraphics[width=75mm]{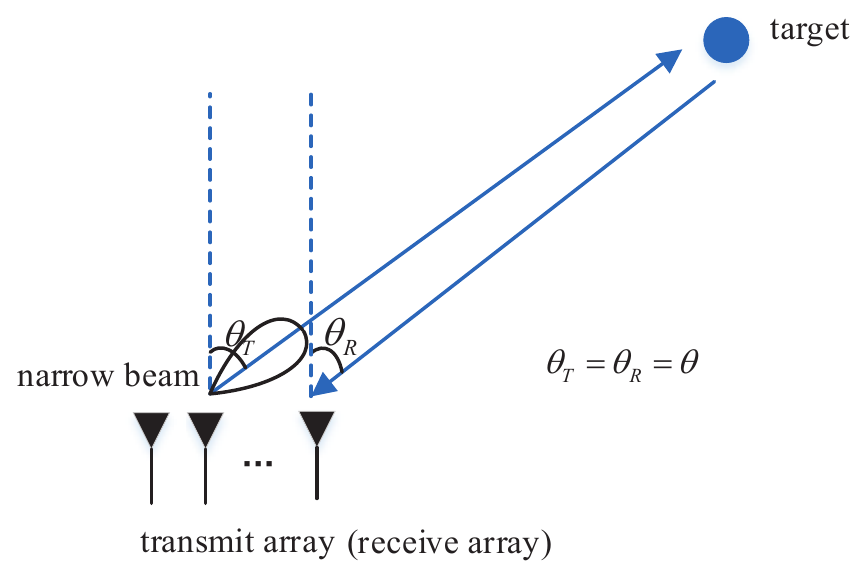}
\par\end{centering}
\centering{}\caption{\label{fig:Single-target-phased-array-radar}Single-target sensing
in the mono-static phased-array radar system.}
\end{figure}

The performance limits of the mono-static phased-array radar system
with multi-antenna transmit and receive arrays was further studied
in \cite{phased_radar_bound}, in terms of CRB. The system model is
illustrated in Fig. \ref{fig:Single-target-phased-array-radar}. Uniform
linear array (ULA) is adopted as the transmit/receive array and the
spacing $d$ between two adjacent antennas is assumed to be half of
the signal wavelength $\lambda$. Both transmit and receive antenna
arrays are assumed to have $M$ antennas. Additionally, the target
is assumed to be static (i.e., there is no Doppler shift). In this
case, the target parameters are range $r$ and DOA $\theta$. The
range $r$ is estimated from the time delay $\tau$ according to the
relationship $r=\tau c/2$, while the DOA $\theta$ is estimated directly
based on the received radar echo. Specifically, the $M\times1$ received
signal for one radar pulse is given by
\begin{equation}
\boldsymbol{Y}(t)=\alpha\boldsymbol{a}_{R}(\theta)\boldsymbol{a}_{T}^{T}(\theta)\boldsymbol{w}s\left(t-\tau\right)+\boldsymbol{Z}(t),
\end{equation}
where $\alpha$ is the reflection coefficient of the target, $\boldsymbol{a}_{T}(\theta)$
is the transmit steering vector, and $\boldsymbol{w}=\left[w_{1},\ldots,w_{M}\right]^{T}$
is the beamforming vector. To facilitate the analysis, the beamforming
vector is assumed to be $\boldsymbol{w}=\boldsymbol{a}_{T}(\theta)$
in \cite{phased_radar_bound} to obtain the highest possible processing
gain at the actual DOA $\theta$.

Under the above assumptions, the CRB of $r$ and $\theta$ is given
by \cite{phased_radar_bound}
\begin{align}
CRB_{r} & =\frac{3}{2\pi^{2}\text{SNR}M^{3}\beta^{2}},\label{eq:CRB_range}\\
CRB_{\theta} & =\frac{1}{2\text{SNR}M^{3}\xi^{2}},\label{eq:CRBrtheta}
\end{align}
where $\beta^{2}$ is the squared effective bandwidth and
\begin{equation}
\xi^{2}=\frac{\pi^{2}d^{2}\cos^{2}\theta(M^{2}-1)}{3\lambda^{2}}\label{eq:rmsaofBF}
\end{equation}
is the root mean square aperture width of the beampattern.

From (\ref{eq:CRBrtheta}) and (\ref{eq:CRB_range}), we can make
similar conclusion as that for the case of single transmit antenna.
The main difference is that the CRB in (\ref{eq:CRBrtheta}) for the
case of $M$ transmit antennas has an additional factor of $1/M^{2}$,
which is contributed by the transmit beamforming gain and that the
total transmit power increases with the number of transmit antennas
$M$ when the transmit power of each transmit antenna is fixed.

\subsection{Fundamental Limits of MIMO Radar}

\subsubsection{Colocated MIMO Radar for Single-Target Sensing}

\begin{figure}[tbh]
\begin{centering}
\includegraphics[width=75mm]{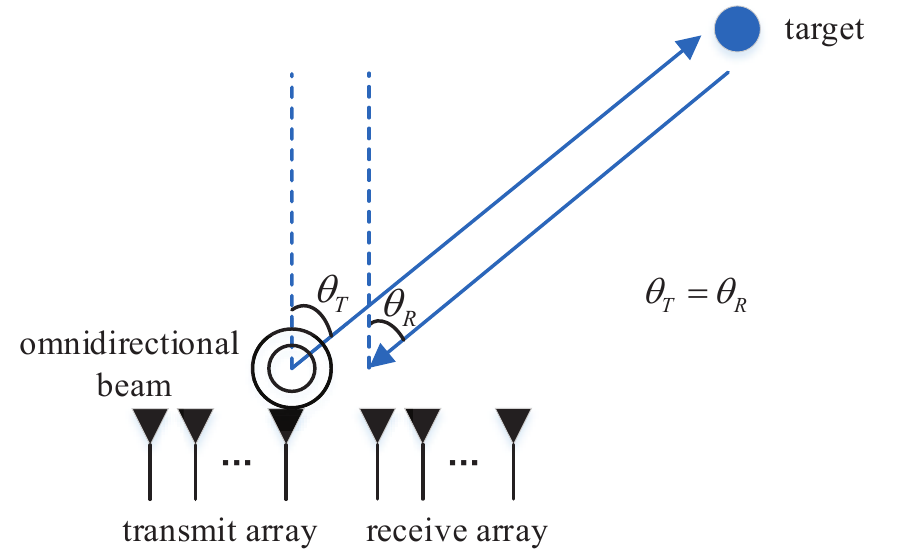}
\par\end{centering}
\centering{}\caption{\label{fig:Single-target-colocated-MIMO}Single-target sensing via
colocated MIMO radar.}
\end{figure}

The CRB of the sensing performance via colocated MIMO radar has been
studied in \cite{colocated_MIMO_radar_2011} for single-target sensing.
The system model is illustrated in Fig. \ref{fig:Single-target-colocated-MIMO}.
In the system model, a colocated MIMO radar formed by $M$ transmit
antennas and $N$ receive antennas is used to detect a moving target.
The MIMO radar is assumed to be moving with the velocity $v_{S}$
and $L$ radar pulses are transmitted in a coherent processing interval
(CPI) for target sensing. At the radar receiver, a matched filter
bank is used to estimate the time delay first, and then the signals
after the matched filter bank are assumed to be sampled at the perfect
timing without any delay estimation error. Finally, the discrete samples
after matched filtering are used to estimate the DoA $\theta$ and
velocity $v$ of the target. Specifically, after the matched filter
bank, the $N\times M$ received signal for the $l$-th radar pulse
is given by
\begin{equation}
\boldsymbol{Y}(l)=\alpha\boldsymbol{a}_{R}(\theta)\boldsymbol{a}_{T}^{T}(\theta)e^{j2\pi f_{D}l}+\boldsymbol{Z}(l),
\end{equation}
where $\alpha$ is the reflection coefficient of the target, $\boldsymbol{a}_{T}(\theta)=[e^{j\frac{2\pi}{\lambda}T_{1}\sin\theta},e^{j\frac{2\pi}{\lambda}T_{2}\sin\theta},\cdots,e^{j\frac{2\pi}{\lambda}T_{M}\sin\theta}]^{T}$
is the transmit steering vector and $\boldsymbol{a}_{R}(\theta)=[e^{j\frac{2\pi}{\lambda}R_{1}\sin\theta},e^{j\frac{2\pi}{\lambda}R_{2}\sin\theta},\cdots,e^{j\frac{2\pi}{\lambda}R_{N}\sin\theta}]^{T}$
is the receive steering vector, $T_{m}$ and $R_{n}$ are the locations
of the $m$-th and $n$-th sensors for the transmit and receive antennas
respectively, $f_{D}=2T_{P}(v_{S}sin(\theta)+v)/\lambda$ is the normalized
Doppler frequency and $T_{P}$ is the radar pulse period, and $\boldsymbol{Z}(l)$
is the noise matrix with i.i.d complex Gaussian entries of zero mean
and variance $\sigma_{z}^{2}$.

Assuming that the linear array is used, the CRBs of $\theta$ and
$v$ are given by \cite{colocated_MIMO_radar_2011}
\begin{equation}
CRB_{\theta}=\frac{1}{2\text{SNR}\pi^{2}\cos^{2}\theta MNL(\sigma_{R}^{2}+\sigma_{T}^{2})},\label{eq:CRBthetaMIMO}
\end{equation}
\begin{equation}
CRB_{v}=\frac{1}{8\text{SNR}\pi^{2}MNL}\left(\frac{3\lambda^{2}}{(L^{2}-1)T_{P}^{2}}+\frac{4v_{S}^{2}}{\sigma_{R}^{2}+\sigma_{T}^{2}}\right),\label{eq:CRBv}
\end{equation}
where $\text{SNR}=\frac{\left|\alpha\right|^{2}}{\sigma_{z}^{2}}$
is the received SNR, $L$ is the number of radar pulses in a CPI,
$\sigma_{T}^{2}$ and $\sigma_{R}^{2}$ are the sample-variances of
the transmit and receive antenna positions, which are defined as
\begin{align*}
\sigma_{R}^{2} & =\frac{4}{N\lambda^{2}}\left(\tilde{\kappa_{R}}-\frac{\kappa_{R}^{2}}{N}\right),\\
\sigma_{T}^{2} & =\frac{4}{M\lambda^{2}}\left(\tilde{\kappa_{T}}-\frac{\kappa_{T}^{2}}{M}\right),
\end{align*}
where $\tilde{\kappa_{R}}=\sum_{n=0}^{N-1}R_{n}^{2}$, $\tilde{\kappa_{T}}=\sum_{m=0}^{M-1}T_{m}^{2}$,
$\kappa_{R}=\sum_{n=0}^{N-1}R_{n}$ and $\kappa_{T}=\sum_{m=0}^{M-1}T_{m}$.

Note that the sample-variances of the transmit and receive antenna
positions $\sigma_{R}^{2}$ and $\sigma_{T}^{2}$ are related to the
root mean square aperture width of the beampattern $\xi^{2}$ in (\ref{eq:rmsaofBF}).
Specifically, if both the transmit and receive antenna arrays are
uniform linear arrays (ULAs) with $M$ antennas and antenna spacing
$d$, we have
\[
\pi^{2}\cos^{2}\theta(\sigma_{R}^{2}+\sigma_{T}^{2})=O\left(\frac{\pi^{2}d^{2}\cos^{2}\theta M^{2}}{\lambda^{2}}\right),
\]
which has the same order as $\xi^{2}$. In this case, if $L=1$, the
order of the CRB of $\theta$ in (\ref{eq:CRBthetaMIMO}) is given
by
\[
CRB_{\theta}=O\left(\frac{1}{\text{SNR}M^{2}\xi^{2}}\right).
\]
Compared to the order of the CRB of $\theta$ for the phased-array
radar in (\ref{eq:CRBrtheta}), i.e., $O\left(\frac{1}{\text{SNR}M^{3}\xi^{2}}\right)$,
the CRB of $\theta$ for the colocated MIMO radar decreases with $M$
at the order of $1/M^{2}$ instead of $1/M^{3}$. This is not surprising
since the phased-array radar can focus its transmit energy on the
direction of the target to achieve a beamforming gain of $O\left(M\right)$,
while the colocated MIMO radar cannot enjoy such beamforming gain
since it transmits independent waveforms from different antennas.
However, the advantage of MIMO radar is that its transmit signal can
cover the whole angular space and thus the initial search time for
a target can be reduced.

From (\ref{eq:CRBthetaMIMO}) and (\ref{eq:CRBv}), it can be observed
that the the estimation performance of the DOA $\theta$ and velocity
$v$ is positively relative to SNR, the number of pulses $L$ in a
CPI, the product of the transmit and receive antennas $MN$ and the
sample-variances of the antenna positions $\sigma_{R}^{2},\sigma_{T}^{2}$.
The estimation performance of the velocity also improves with the
increase of the radar pulse period $T_{P}$ and the decrease of the
radar velocity $v_{S}$ and signal wavelength $\lambda$. However,
the movement of the radar has no impact on the estimation performance
of the DOA of the target.

In \cite{colocated_MIMO_radar_2013}, the CRB of colocated MIMO radar
using time multiplexing is analysed. The obtained CRB shows that the
accuracy of the DOA estimators decreases in a MIMO radar if the target
moves with relative radial velocity, because the motion causes an
unknown phase rotation of the baseband signal due to the Doppler effect.

\subsubsection{Colocated MIMO Radar for Multi-Target Sensing}

\begin{figure}[tbh]
\begin{centering}
\includegraphics[width=75mm]{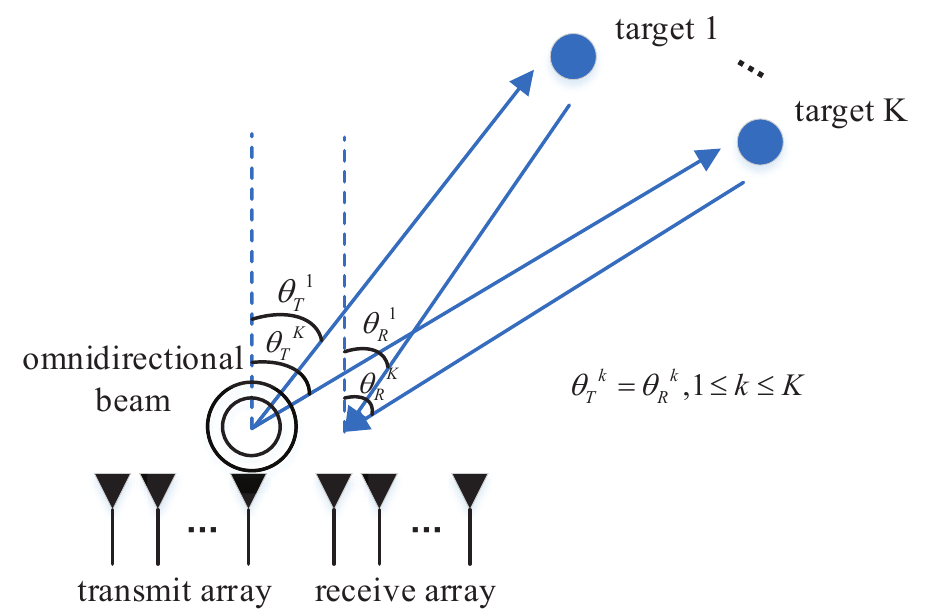}
\par\end{centering}
\centering{}\caption{\label{fig:Multi-target-colocated-MIMO}Multi-target sensing via colocated
MIMO radar.}
\end{figure}

In \cite{colocated_two_target}, the CRB analysis of the colocated
MIMO radar is extended to the multi-target case, as illustrated in
Fig. \ref{fig:Multi-target-colocated-MIMO}. There are $K$ targets
and the DOA of the $k$-th target is $\theta^{k}$. The transmit signal
is narrowband and thus the time delay is ignored in the system model.
After the matched filter bank, the $N\times M$ received signal for
the $l$-th radar pulse is given by
\begin{equation}
\boldsymbol{Y}(l)=\sum_{k=1}^{K}\alpha^{k}\boldsymbol{a}_{R}(\theta^{k})\boldsymbol{a}_{T}^{T}(\theta^{k})e^{j2\pi f_{D}^{k}l}+\boldsymbol{Z}(l),
\end{equation}
where $\alpha^{k}$ is the reflection coefficient of the $k$-th target,
$\boldsymbol{a}_{T}(\theta^{k})$ and $\boldsymbol{a}_{R}(\theta^{k})$
are the transmit and receive steering vectors respectively, $f_{D}^{k}$
is the Doppler shift associated with the $k$-th target, and $\boldsymbol{Z}(l)$
is the noise matrix.

The target parameters are the DOAs $\theta^{k}$'s and velocities
$v^{k}$'s of the targets, where the velocity $v^{k}$ is estimated
from the Doppler shift $f_{D}^{k}$. In \cite{colocated_two_target},
the special case of two targets is studied in details. To facilitate
analysis, intermediate target parameters $\vartheta^{1}$, $\vartheta^{2}$,
$f_{D}^{1}$ and $f_{D}^{2}$ are adopted, where $\vartheta^{1}=sin(\theta^{1})$
and $\vartheta^{2}=sin(\theta^{2})$. CRB of these intermediate parameters
is deduced. Since the expression of the CRB is very complicated, we
do not give the exact expression. The main conclusion is that the
CRB of the DOAs and velocities of the two targets only depends on
the differences of their DOAs and Doppler frequencies, i.e.,
\begin{equation}
CRB_{\vartheta^{1},\vartheta^{2},f_{D}^{1},f_{D}^{2}}=CRB_{\vartheta^{1},\vartheta^{2},f_{D}^{1},f_{D}^{2}}(\bigtriangleup\vartheta,\bigtriangleup f_{D})
\end{equation}
where $\bigtriangleup\vartheta=\vartheta^{1}-\vartheta^{2}$ and $\bigtriangleup f_{D}=f_{D}^{1}-f_{D}^{2}$.
The estimation performance is better if the differences $\bigtriangleup\vartheta,\bigtriangleup f_{D}$
between the parameters of the two targets are larger. When $\bigtriangleup\vartheta$
and $\bigtriangleup f_{D}$ are sufficiently large, the estimation
performance for two targets will approach that for a single target.

\subsubsection{Distributed MIMO Radar for Single-Target Sensing}

\begin{figure}[tbh]
\begin{centering}
\includegraphics[width=75mm]{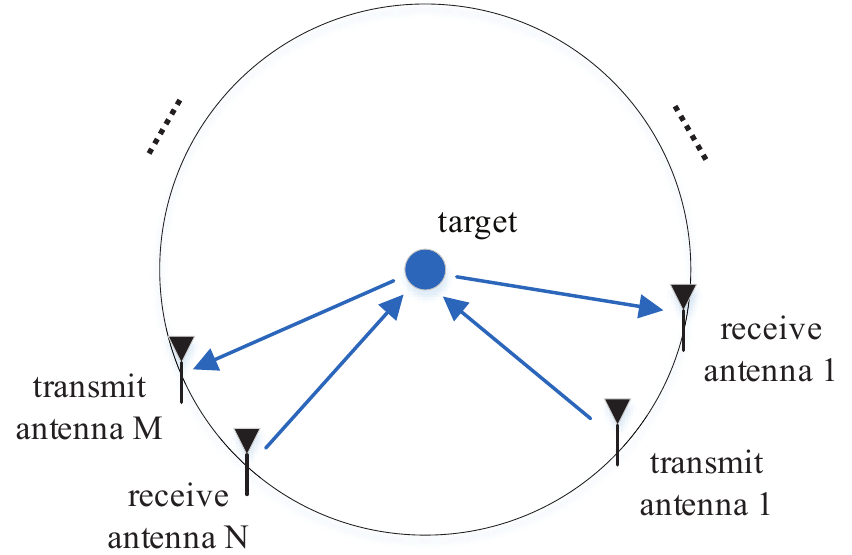}
\par\end{centering}
\centering{}\caption{\label{fig:Single-target-distributed-MIMO}Single-target sensing via
distributed MIMO radar.}
\end{figure}

The CRB of the sensing performance via the distributed MIMO radar
has been studied in \cite{statistical_MIMO_radar1} for single-target
sensing. As illustrated in Fig. \ref{fig:Single-target-distributed-MIMO},
the transmit and receive antennas are placed symmetrically around
the target so that the sensing performance can be improved \cite{statistical_MIMO_radar1}.
The lowpass equivalent of the signal transmitted from the $m$-th
transmitter is $s_{m}(t)$, and the energy of the waveform $s_{m}(t)$
is normalized to be one. Assume that the transmitted signals $s_{m}(t)$'s
from different transmit antennas are approximately orthogonal and
they maintain approximate orthogonality for time delays and Doppler
shifts of interest. Under these assumptions, the received signal model
at receiver $n$ due to the signal transmitted from transmitter $m$
is
\begin{equation}
y_{n,m}(t)=\alpha_{n,m}s_{m}(t-\tau_{n,m})e^{j2\pi f_{n,m}t}+z_{n,m}(t),
\end{equation}
where $\tau_{n,m}$, $f_{n,m}$ and $\alpha_{n,m}$ represent the
time delay, Doppler shift and reflection coefficients, respectively,
corresponding to the path between the $m$-th transmitter and the
$n$-th receiver, and $z_{n,m}(t)$ is noise.

The parameters of interest are the location and velocity of the target,
which are expressed in the form of coordinates in rectangular coordinate
systems as $(x,y)$ and $(v_{x},v_{y})$. These parameters are estimated
from the time delays $\tau_{n,m}$ and Doppler shifts $f_{n,m},1\leq n\leq N,1\leq m\leq M$,
which can be regarded as intermediate parameters.

Since the number of the intermediate parameters is large, it is difficult
to obtain a closed-form expression of CRB. Nonetheless, we can analyse
the order of the CRB for the time delay and Doppler shift of the path
between transmitter $m$ and receiver $n$, as given by
\begin{equation}
CRB_{\tau_{n,m}}=O\left(\frac{1}{\textrm{SNR}_{n,m}\beta_{m}^{2}}\right),\label{eq:CRMtaodist}
\end{equation}
\begin{equation}
CRB_{f_{n,m}}=O\left(\frac{1}{\textrm{SNR}_{n,m}\gamma_{n,m}^{2}}\right),\label{eq:CRMfdist}
\end{equation}
where $\beta_{m}^{2}$ is the squared effective bandwidth of $s_{m}(t)$,
$\textrm{SNR}_{n,m}=\frac{\left|\alpha_{n,m}\right|^{2}}{\sigma_{z}^{2}}$
and
\begin{align*}
 & \gamma_{n,m}^{2}\\
= & \frac{\int_{-\infty}^{\infty}t^{2}\left|s_{m}(t-\tau_{n,m})\right|^{2}dt-\left|\int_{-\infty}^{\infty}t\left|s_{m}(t-\tau_{n,m})\right|^{2}dt\right|^{2}}{\int_{-\infty}^{\infty}\left|s_{m}\left(t\right)\right|^{2}dt},
\end{align*}
are the received SNR and squared effective pulse length for the $(n,m)$-th
receive-transmit pair, respectively. Furthermore, assume that $\textrm{SNR}_{n,m}=O\left(\text{SNR}\right),\gamma_{n,m}^{2}=O\left(\gamma^{2}\right),\forall n,m$
and $\beta_{m}^{2}=O\left(\beta^{2}\right),\forall m$, where $\text{SNR}=\frac{1}{NM}\stackrel[n=1]{N}{\sum}\stackrel[m=1]{M}{\sum}\textrm{SNR}_{n,m},$
$\gamma^{2}=\frac{1}{NM}\stackrel[n=1]{N}{\sum}\stackrel[m=1]{M}{\sum}\gamma_{n,m}^{2}$
and $\beta^{2}=\frac{1}{M}\stackrel[m=1]{M}{\sum}\beta_{m}^{2}$ are
the average received SNR, average squared effective pulse length and
average squared effective bandwidth, respectively. Then it can be
shown that the orders of the CRB for the position and velocity are
respectively given by \cite{statistical_MIMO_radar1}
\begin{equation}
CRB_{(x,y)}=O\left(\frac{1}{\textrm{SNR}MN\left(\beta^{2}+\gamma^{2}\right)}\right),
\end{equation}
\begin{equation}
CRB_{(v_{x},v_{y})}=O\left(\frac{1}{\textrm{SNR}MN\gamma^{2}}\right).
\end{equation}
The key insights revealed from the CRB analysis in \cite{statistical_MIMO_radar1}
is that the estimation performance improves with the number of antennas
$M,N$ and SNR. Moreover, the estimation performance of the position
$(x,y)$ improves with both the squared effective bandwidth and squared
effective pulse length, while the estimation performance of the velocity
$(v_{x},v_{y})$ improves with the squared effective pulse length.

In \cite{statistical_MIMO_radar2}, the CRB is also analyzed for directly
estimating the velocity of the target only. It is concluded that the
estimation performance of the velocity is also positively relative
to the SNR and the squared effective radar pulse length.

\subsubsection{Distributed MIMO Radar for Multi-Target Sensing}

In \cite{statistical_MIMO_radar3}, the CRB analysis of the distributed
MIMO radar is extended to the multi-target case. Under similar assumptions
as in \cite{statistical_MIMO_radar1}, the received signal model at
receiver $n$ due to the signal transmitted from transmitter $m$
and the reflection of all the $K$ targets is
\begin{equation}
y_{n,m}(t)=\sum_{k=1}^{K}\alpha_{n,m}^{k}s_{m}(t-\tau_{n,m}^{k})e^{j2\pi f_{n,m}^{k}t}+z_{n,m}(t),
\end{equation}
where $\tau_{n,m}^{k}$, $f_{n,m}^{k}$ and $\alpha_{n,m}^{k}$ represent
the time delay, Doppler shift and reflection coefficients, respectively,
corresponding to the path from the $m$-th transmitter to the $k$-th
target and then reflected to the $n$-th receiver.

The target parameters are the locations and velocities of the targets,
which are expressed in the form of coordinates in rectangular coordinate
systems as $(x^{k},y^{k})$ and $(v_{x}^{k},v_{y}^{k}),1\leq k\leq K$.
These parameters are estimated from time delay $\tau_{m,n}^{k}$ and
Doppler shifts $f_{m,n}^{k},1\leq m\leq M,1\leq n\leq N,1\leq k\leq K$
, which can be regarded as intermediate parameters.

The key insights revealed from the CRB analysis in \cite{statistical_MIMO_radar3}
is that if the distances between the targets are large enough, the
interactions between the multiple targets can be ignored and the performance
of the multiple-target sensing can approach that of the single-target
sensing.

\subsection{Fundamental Limits of Phased-MIMO Radar}

\begin{figure}[tbh]
\begin{centering}
\includegraphics[width=75mm]{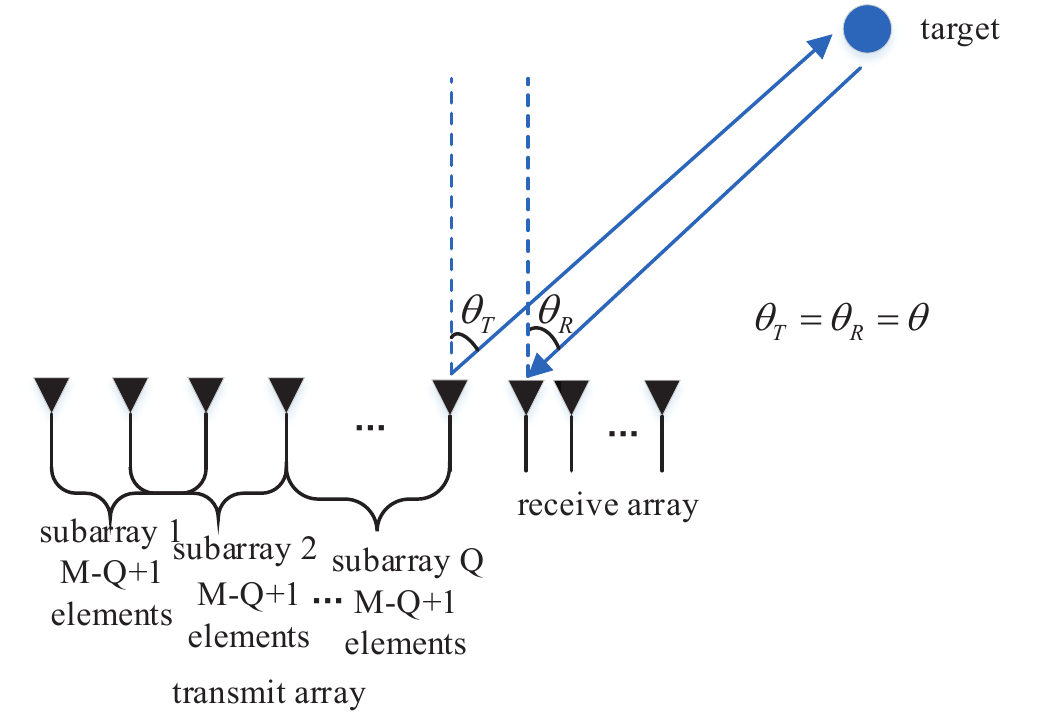}
\par\end{centering}
\centering{}\caption{\label{fig:Single-target-phased-MIMO-radar}Single-target sensing
via phased-MIMO radar.}
\end{figure}

The existing works have been focusing on investigating the performance
limits of single-target sensing in the mono-static phased-MIMO radar
system. In \cite{phased-MIMO_radar_performance}, ambiguity function
(AF) is adopted to analyse the performance of phased-MIMO radar. The
system model is illustrated in Fig. \ref{fig:Single-target-phased-MIMO-radar}.
In phased-MIMO radar, the transmit array is divided into $Q$ subarrays,
while each subarray contains $P=M-Q+1$ adjacent antennas. Meanwhile,
the spacings between two adjacent antennas of the transmit and receive
arrays are assumed to be $d_{T}$ and $d_{R}$, respectively. Additionally,
the reflection coefficient is assumed to be 1 and noise is ignored
to simplify the analysis of AF. Under these assumptions, the $N\times1$
received signal is given by
\begin{align}
\boldsymbol{y}(t,\tau,f_{D},\theta) & =\boldsymbol{a}_{R}(\theta)\sum_{q=1}^{Q}\boldsymbol{a}_{q}^{T}(\theta)\boldsymbol{w}_{q}e^{-j2\pi f_{c}\tau_{q}\text{(\ensuremath{\theta})}}\\
 & \times s_{q}(t-\tau)e^{j2\pi f_{D}t}e^{-j2\pi(f_{c}+f_{D})\tau},
\end{align}
where $\tau$ is the round-trip delay for a target in the $\theta$
direction, $f_{D}$ is the Doppler shift, $f_{c}$ is the carrier
frequency, $\tau_{q}\text{(\ensuremath{\theta})}=qd_{T}\sin\theta/c$
is the relative delay of the zeroth element of the $q$-th subarray
with respect to the zeroth element of the zeroth subarray, $\boldsymbol{a}_{R}(\theta)$
is the receive steering vector, $\boldsymbol{a}_{q}(\theta)$, $s_{q}(t)$
and $\boldsymbol{w}_{q}$ are the transmit steering vector, transmit
waveform and transmit beamforming vector for the $q$-th subarray,
respectively. Note that we have explicitly express the received signal
as a function of $\tau,f_{D},\theta$.

\begin{figure*}[!tp]
\begin{equation}
\intop_{-\infty}^{\infty}\boldsymbol{y}^{H}(t,\tau',f_{D}',\theta')\boldsymbol{y}(t,\tau,f_{D},\theta)dt=\boldsymbol{a}_{\boldsymbol{R}}^{H}(\theta')\boldsymbol{a}_{\boldsymbol{R}}(\theta)\intop_{-\infty}^{\infty}\sum_{q=1}^{Q}(\boldsymbol{w}_{q}^{H}\boldsymbol{a}_{q}^{*}(\theta)s_{q}(t-\tau'))\sum_{q=1}^{Q}(\boldsymbol{a}_{q}^{T}(\theta)\boldsymbol{w}_{q}s_{q}(t-\tau))e^{j2\pi(f_{D}-f_{D}')t}dt.\label{Xlongeq}
\end{equation}
\end{figure*}

\begin{table*}[tbh]
\caption{\label{tab:Order}Comparison of different classes of radars.}

\centering{}%
\begin{tabular}{|c|c|c|c|}
\hline 
Types of Radar & Phased-array Radar & Distributed MIMO Radar & Colocated MIMO Radar\tabularnewline
\hline 
\hline 
CRB order for $\tau$ & $O\left(\frac{1}{\text{SNR}NM^{2}L\beta^{2}}\right)$ & $O\left(\frac{1}{\text{SNR}\beta^{2}L}\right)$ & $O\left(\frac{1}{\text{SNR}NM\beta^{2}L}\right)$\tabularnewline
\hline 
CRB order for $\theta$ & $O\left(\frac{1}{\text{SNR}NM^{2}L\cos^{2}\left(\theta\right)\left(\sigma_{R}^{2}+\sigma_{T}^{2}\right)}\right)$ & N/A & $O\left(\frac{1}{\text{SNR}NML\cos^{2}\left(\theta\right)\left(\sigma_{R}^{2}+\sigma_{T}^{2}\right)}\right)$\tabularnewline
\hline 
CRB order for $f_{D}$ & $O\left(\frac{1}{\textrm{SNR}NM^{2}L\gamma^{2}}\right),$ & $O\left(\frac{1}{\textrm{SNR}L\gamma^{2}}\right)$ & $O\left(\frac{1}{\textrm{SNR}NML\gamma^{2}}\right)$\tabularnewline
\hline 
Advantages & Beamforming Gain & Spatial Diversity Gain & Waveform Diversity Gain\tabularnewline
\hline 
Disadvantages & Long scanning time & High synchronization requirements & SNR degradation\tabularnewline
\hline 
\end{tabular}
\end{table*}

If the matched filters at the receivers are matched to the received
signal with a different set of parameters $\tau',f_{D}',\theta'$,
then the output of the matched filters combined together can be expressed
as in (\ref{Xlongeq}) on the top of the next page. The first term
on the right-hand side of (\ref{Xlongeq}), i.e., $\boldsymbol{a}_{\boldsymbol{R}}^{H}(\theta')\boldsymbol{a}_{\boldsymbol{R}}(\theta)$,
represents the spatial processing in the receiver and is independent
of the transmit waveforms $s_{q}(t),1\leq q\leq Q$. The second term
on the right-hand side of (\ref{Xlongeq}) is defined as the AF \cite{phased-MIMO_radar_performance},
which shows the sensitivity of the output of the matched filter to
the error of the estimation of the parameters. The maximum of the
AF is achieved when $\tau'=\tau$, $f_{D}'=f_{D}$ and $\theta'=\theta$.
The narrower the curve of the AF is, the better the estimator is expected
to be.

In contrast to MIMO radar systems in which the ambiguity function
is fixed, we can adapt the ambiguity function by changing the size
of subarrays and the number of subarrays in the case of phased-MIMO
radar \cite{phased-MIMO_radar_performance}. Meanwhile, adopting the
linear frequency modulation (LFM) waveform can improve the delay resolution
but it is accompanied by the penalty of delay\textendash Doppler coupling
\cite{phased-MIMO_radar_performance}.

\subsection{Summary and Insights}

In existing works, CRB and AF have been used as the performance metrics
for device-free sensing, among which CRB is the most widely used performance
metric. The target parameters to be estimated usually include the
time delay $\tau$, the DOA $\theta$ and the Doppler frequency $f_{D}$.
The other target parameters such as its location and velocity can
be inferred from these intermediate parameters. For all classes of
radars, the estimation performance of all target parameters improves
with the increase of SNR (power resource), the number of antennas
(spatial resource) and the number of pulses in a CPI (time resource),
since the increase of these system resources increases the effective
SNR and the number of observations for parameter estimation.

Specifically, the order of the CRB for the estimation of the time
delay $\tau$ can be expressed in a unified expression as
\begin{equation}
CRB_{\tau}=O\left(\frac{1}{\text{SNR}NM^{a}L\beta^{2}}\right),
\end{equation}
where $M$ and $N$ are the number of transmit and receive antennas
respectively, $L$ is the number of pulses in a CPI, $\beta^{2}$
is the squared effective bandwidth and the exponent $a$ depends on
the type of radar. For example, $a=1$ for MIMO radar and $a=2$ for
phased-array radar due to the additional transmit beamforming gain.
Clearly, the estimation performance of the time delay $\tau$ also
improves with the squared effective bandwidth $\beta^{2}$. 

The order of the CRB for the estimation of the DOA $\theta$ for colocated
antennas can be expressed in a unified expression as
\begin{equation}
CRB_{\theta}=O\left(\frac{1}{\text{SNR}NM^{a}L\cos^{2}\left(\theta\right)\left(\sigma_{R}^{2}+\sigma_{T}^{2}\right)}\right),
\end{equation}
where $\sigma_{T}^{2}$ and $\sigma_{R}^{2}$ are the sample-variances
of the transmit and receive antenna positions and $a$ depends on
the type of radar. For example, $a=1$ for MIMO radar and $a=2$ for
phased-array radar. Clearly, the estimation performance of the DOA
$\theta$ also improves with the sample-variances of the transmit
and receive antenna positions $\sigma_{T}^{2}$ and $\sigma_{R}^{2}$.

The order of the CRB for the estimation of the Doppler frequency $f_{D}$
can be expressed in a unified expression as
\begin{equation}
CRB_{f_{D}}=O\left(\frac{1}{\textrm{SNR}NM^{a}L\gamma^{2}}\right),
\end{equation}
where $\gamma^{2}$ is the squared effective pulse length and $a$
depends on the type of radar. For example, $a=1$ for MIMO radar and
$a=2$ for phased-array radar. Clearly, the estimation performance
of the Doppler frequency $f_{D}$ also improves with the squared effective
pulse length $\gamma^{2}$.

From the above unified expressions, we can conclude that the SNR,
the number of transmit (receive) antennas $M$ ($N$), and the number
of pulses in a CPI are the common influence factors on the estimation
of time delay $\tau$, the DOA $\theta$ and the Doppler frequency
$f_{D}$, while the estimation of $\tau$, $\theta$ and $f_{D}$
are also determined by $\beta$ (effective bandwidth), $\sigma_{R}^{2}$
and $\sigma_{T}^{2}$ (antenna geometry), and $\gamma$ (effective
pulse length), respectively. To summarize, the comparison of order-wise
performances of different classes of radars and their pros and cons
are listed in Table \ref{tab:Order}. There are two additional comments
to the CRB order in Table \ref{tab:Order}. First, for the distributed
MIMO radar, the order of the CRB is given for the intermediate parameters
associated with the path between one transmit and receive antenna
pair. However, the final estimation for the position and velocity
of the target is obtained from the estimates of intermediate parameters
of the paths between all transmit and receive antenna pairs. It can
be shown that the estimation performance of the position and velocity
in the distributed MIMO radar actually has the same order as that
in the colocated MIMO radar \cite{colocated_MIMO_radar_2011}. Second,
in Table \ref{tab:Order}, we follow the convention in the literature
on the fundamental limits of radar sensing and assume a per-antenna
power constraint where the transmit power of each antenna is fixed.
In this case, the total transmit power increases with the number of
transmit antennas $M$. If a total power constraint is assumed, the
CRB order for the phased-array radar and colocated MIMO radar should
be multiplied by a factor of $M$.

For multi-target MIMO radar, the CRB can be improved if the distances
between the targets are larger. In particular, if the targets are
sufficiently far away from each other, the parameters of different
targets can be estimated independently and the performance of the
multiple-target sensing can approach that of the single-target sensing.

\section{Fundamental Limits of Device-based Sensing}

In this section, we will discuss the current research progress on
the fundamental limits of device-based sensing. In particular, we
will focus on the fundamental limits for different classes of wireless-based
localization as classified in Section \ref{sec:Classifications-of-Integrated}.
For each class, we will highlight several important works, and present
the system model, performance bounds and key insights learned from
the analysis of the fundamental limits.

\subsection{Non-cooperative Wireless Localization}

\subsubsection{TOA-based Localization}

The TOA-based localization is the most widely studied wireless localization
method. In the following, we first give a brief historical review
of the key works on the fundamental limits of the TOA-based localization.
Then we discuss the signal model of the TOA-based localization, the
fundamental limits and the associated key insights. In \cite{Qi1},
Qi et al. first derived the CRB of the TOA-based localization in the
presence of non-line-of-sight (NLOS) environment where a single path
propagation (either a single LOS or NLOS path) is assumed \cite{Qi1}.
The authors concluded that NLOS signals do not contribute to the localization
performance when no prior NLOS statistics are available. Furthermore,
the CRB is inversely proportional to the square of effective bandwidth
and depends on geometric configuration of agent/anchor nodes. When
prior information of NLOS signals is attained, the NLOS signals can
also provide useful information for localization and the localization
accuracy can be improved \cite{Qi-new}. In \cite{Qi3}, the authors
further extended their previous work from the single path propagation
case to the multi-path propagation case. Later, further analysis was
developed by Shen et al. \cite{Shen_part1} when the prior knowledge
of the agent's position is available in addition to the NLOS statistics.
In \cite{Shen_part1}, the concepts of equivalent FIM (EFIM) and squared
position error bound (SPEB) were introduced to develop a general framework
for the analysis of the fundamental limits of device-based localization.
Besides, map information of the environment can be regard as a special
form of prior information to the agent, which helps to improve the
estimation accuracy by exploiting some features of the map (e.g.,
its shape and area) \cite{Mapaware}. Furthermore, dynamic scenarios
with moving agents are also investigated in \cite{dynamic_11,dynamic_14}.
In \cite{dynamic_11}, the performance limit is derived in both static
and dynamic scenarios. In the dynamic scenario, the Doppler shift
contributes additional direction information with intensity determined
by the speed of the agent and the root mean squared time duration
of the transmitted signal. In \cite{dynamic_14}, Li proposed a posterior
CRB (P-CRB) for the fundamental limit analysis with dynamic sensor
networks.

\begin{figure}[tbh]
\centering{}\includegraphics[width=75mm]{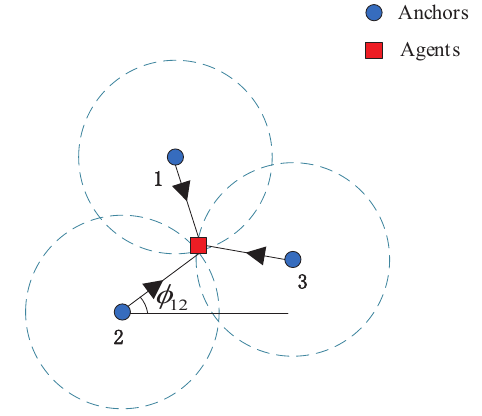}\caption{\label{fig:TOA localization}An illustration of TOA-based localization.}
\end{figure}

To reveal more insights into fundamental limits of the TOA-based localization,
we consider a general TOA-based localization scenario studied in \cite{Shen_part1,tempcoop}.
In this case, consider a multi-path environment which commonly exists
in wireless network and the wireless network consists of $N_{a}$
agent nodes and $N_{b}$ anchor nodes in a 2-D plane, as illustrated
in Fig. \ref{fig:TOA localization}. We define $\mathcal{N}_{a}=\left\{ 1,...,N_{a}\right\} ,\mathcal{N}_{b}=\left\{ 1,...,N_{b}\right\} $
as the set of agent nodes and anchor nodes, respectively. The signal
transmitted from anchor $j\in\mathcal{N}_{b}$ and received by agent
$k\in\mathcal{N}_{a}$ can be written as
\begin{equation}
\ensuremath{y_{k,j}\left(t\right)=\sum_{l=1}^{L_{k,j}}\alpha_{k,j}^{\left(l\right)}s\left(t-\tau_{k,j}^{(l)}\right)+z_{k,j}\left(t\right)},\label{eq:toa}
\end{equation}
where $L_{k,j}$ is the number of multipath components, $\alpha_{k,j}^{\left(l\right)}$
and $\tau_{k,j}^{(l)}$ are the complex gain and delay of $l$-th
path, $s(t)$ is a known waveform whose Fourier transform is denoted
by $S(f)$, $z_{k,j}\left(t\right)$ represents the observation noise
modeled as additive white Gaussian processes with variance $\sigma_{z}^{2}$.
The relationship between the delays and the agent position can be
expressed as
\begin{equation}
\tau_{k,j}^{\left(l\right)}=\frac{1}{c}\left[\left\Vert {\bf p}_{k}-{\bf p}_{j}\right\Vert +b_{k,j}^{\left(l\right)}\right],\label{eq:pdrel}
\end{equation}
where $c$ is the propagation speed of the signal, ${\bf p}_{k}{\bf \triangleq\left[\mathit{x_{k}\textrm{ }y_{k}}\right]^{\mathrm{T}}}$
is the node position, and the range bias $b_{k,j}^{\left(l\right)}>0$
for NLOS propagation while $b_{k,j}^{\left(l\right)}=0$ for LOS propagation.

Since the estimation of individual agent's location is independent,
the analysis can be focused on one agent briefly, e.g., ${\bf p}_{1}$.
Define the range information (RI) from an anchor at direction $\phi$
as $\lambda\mathbf{J}_{\text{r}}(\phi)$, where $\lambda$ is a non-negative
number called the range information intensity (RII), and ${\bf J}_{\text{r}}\left(\phi\right)$
is a 2 $\times$ 2 matrix named the ranging direction matrix (RDM)
with angle $\phi$, given by
\begin{equation}
\mathbf{J}_{\text{r}}(\phi)\triangleq\left[\begin{array}{cc}
\cos^{2}\phi & \cos\phi\sin\phi\\
\cos\phi\sin\phi & \sin^{2}\phi
\end{array}\right].\label{eq:RDM}
\end{equation}
When the prior knowledge of the agent position and range biases $b_{k,j}^{\left(l\right)}$'s
are unavailable, the EFIM for the agent 1's position is
\begin{equation}
\mathbf{J}_{\text{e}}\left(\mathbf{p}_{1}\right)=\sum_{j\in\mathcal{N}_{\mathrm{b},\mathrm{LOS}}}\lambda_{1,j}\mathbf{J}_{\text{r}}\left(\phi_{1,j}\right),\label{eq:EFIMP1}
\end{equation}
where $\phi_{1,j}=\tan^{-1}\frac{y_{1}-y_{j}}{x_{1}-x_{j}}$ is the
angle from anchor $j$ to agent 1 , $\lambda_{1,j}=\frac{8\pi^{2}\beta^{2}}{c^{2}}\left(1-\chi_{1,j}\right)\text{SNR}_{1,j}^{(1)}$
is the RII from anchor $j$, $\mathcal{N_{\mathrm{b,LOS}}}$ denotes
the set of LOS links in $\mathcal{N}_{b}$, $\chi_{1,j}\in[0,1)$
is called path-overlap coefficient (POC), $\beta$ is the effective
bandwidth given by $\beta\triangleq\left(\frac{\int_{-\infty}^{+\infty}f^{2}|S(f)|^{2}df}{\int_{-\infty}^{+\infty}|S(f)|^{2}df}\right)^{1/2}$,
and $\text{SNR}_{1,j}^{(1)}$ is the receive SNR of the agent 1.

\begin{figure}[tbh]
\centering{}\includegraphics[width=60mm]{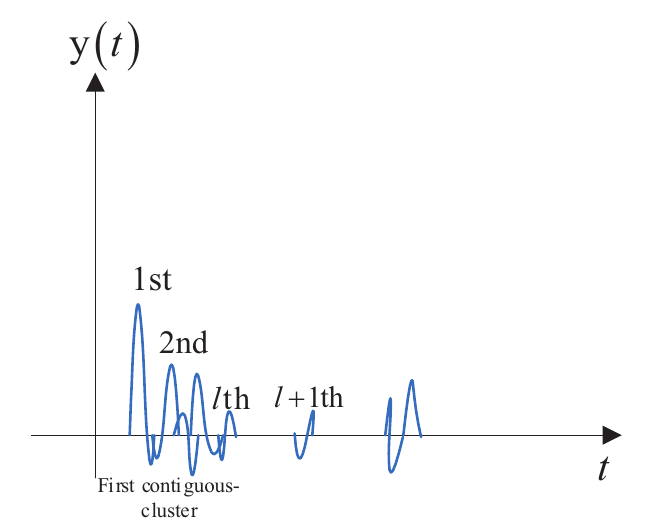}\caption{\label{fig:first cluster}An illustration of the first contiguous
cluster (containing paths) in a LOS signal. The first contiguous cluster
is defined to be the set of paths $\left\{ 1,2,...,l\right\} $ such
that $\left|\tau_{i}-\tau_{i+1}\right|<T_{s}$ for $i=1,2,...,l-1$,
and $\left|\tau_{l}-\tau_{l+1}\right|>T_{s}$ where $T_{s}$ is the
duration of $s\left(t\right)$.}
\end{figure}

The CRB of the position ${\bf p}_{1}$ can be obtained by the matrix
inverse $\mathbf{J}_{\text{e}}^{-1}\left(\mathbf{p}_{1}\right)$.
Therefore, the EFIM in (\ref{eq:EFIMP1}) reveals significant insights
into the fundamental limits of wireless network localization. Specifically,
the performance of localization relies on the NLOS condition, multipath
propagation, network topology and signal bandwidth, as elaborated
below.
\begin{itemize}
\item When no prior knowledge of range biases is available, NLOS signals
make no contribution to the EFIM for the agent position. This is because
the relation between delay and the agent position is affected by the
unknown range bias as shown in (\ref{eq:pdrel}).
\item The POC $\chi_{1,j}$ characterizes the effect of multipath propagation
for localization, which is determined only by the waveform $s(t)$
and the NLOS biases of the multi-path components (MPCs) in the first
contiguous cluster \cite{Shen_part1}, as illustrated in Fig. \ref{fig:first cluster}.
Obviously, the multipath propagation degrades the localization accuracy
compared to single-path propagation, since MPCs interfere the estimation
of the arrival time of the first path. Moreover, $\chi_{1,j}$ is
independent of all the amplitudes. Specially, when the first path
is resolvable from the rest paths, $\chi_{1,j}=0$ and the RII reduces
to that in the single-path propagation.
\item The RII $\lambda_{1,j}$ is proportional to the SNR of the first path
in the receiver (agent 1) and the squared effective bandwidth $\beta^{2}$.
Moreover, due to the connection with POC $\chi_{1,j}$, larger bandwidth
also improves the resolvability of the MPCs.
\item The EFIM is the canonical form of a weighted sum of the RDM from individual
anchors. Anchor $j$ can provide only 1-D RI at the direction $\varphi_{1,j}$
with intensity $\lambda_{1,j}$ for agent 1. Therefore, the localization
performance not only depends on the RII $\lambda_{1,j}$'s but also
the ranging direction $\varphi_{1,j}$'s from the anchors. When there
are anchors contributing RI from diverse ranging directions $\varphi_{1,j}$'s,
the localization performance tends to be better. This phenomenon can
be characterized using the notation of information eclipse introduced
in \cite{Shen_part2}. Please refer to \cite{Shen_part2} for more
detailed discussions.
\end{itemize}
When prior knowledge of the range biases $b_{k,j}^{\left(l\right)}$'s
are available, the EFIM for the agent's position can also be written
as a weighted sum of RDMs from individual anchors, given by
\begin{equation}
\tilde{\mathbf{J}}_{\mathrm{e}}\left(\mathbf{p}_{1}\right)=\sum_{j\in\mathcal{N}_{\mathrm{b},\mathrm{LOS}}}\tilde{\lambda}_{1,j}\mathbf{J}_{\mathrm{r}}\left(\phi_{1,j}\right)+\sum_{j\in\mathcal{N}_{\mathrm{b},\mathrm{NLOS}}}\tilde{\lambda}_{1,j}\mathbf{J}_{\mathrm{r}}\left(\phi_{1,j}\right),
\end{equation}
where the first and second term on the right-hand side (RHS) indicates
the RI of the LOS and NLOS signals, respectively. Moreover, the prior
knowledge increases the RII of both LOS and NLOS signals. The RII
of NLOS signals can be strictly positive and thus contributes to EFIM
in contrast to the case without prior knowledge.

Furthermore, when both prior knowledge of the range biases and the
agent's position is available, the EFIM for the agent's position is
given by
\begin{equation}
\bar{\mathbf{J}}_{\mathrm{e}}\left(\mathbf{p}_{1}\right)=\mathbb{E}_{\mathbf{p}_{1}}\left\{ \tilde{\mathbf{J}}_{\mathrm{e}}\left(\mathbf{p}_{1}\right)\right\} +\mathbf{J}_{\mathrm{p}}\left(\mathbf{p}_{1}\right),
\end{equation}
where $J_{\mathrm{p}}\left(\mathbf{p}_{1}\right)$ denotes the additional
information from the prior position knowledge \cite{Shen_part1}.

\subsubsection{AOA-based Localization}

In the AOA-based localization, the position of the agent is inferred
from the AOAs of the LOS paths, which are extracted from the signals
from the anchors. The main method of the AOA estimation via antenna
arrays is that the differences between the incident signal's arrival
times at different antenna elements contain the angle information.
The basic AOA estimation models can be classified into narrowband
and wideband models. In the narrowband model, the signal bandwidth
$W$ is much less than the center frequency $f_{c}$, and time differences
among different antennas can be represented as phase shifts. Hence,
phased arrays can be applied to the beam-steering process. In the
wideband model, the signal bandwidth $W$ is much larger than the
center frequency $f_{c}$. In this case, since a unique phase value
cannot represent a time delay for a wideband signal, time delayed
lines (timed arrays) are used for the beam-steering, which is the
process to form a beam in a specific direction by assigning specific
weights at each array antenna elements \cite{5G_10}. Then, some typical
scenarios will be discussed in details below.

First, consider the AOA-based localization under narrowband assumption
\cite{AOA-narrowband}, which can be written as
\begin{equation}
\mathbf{y}\left(t\right)=\mathbf{A}\left(\mathbf{\mathbf{\theta}}\right)\mathbf{s}\left(t\right)+\mathbf{z}\left(t\right),
\end{equation}
where $t=1,2,\cdots,L$ is the snapshot index, $\mathbf{A}(\theta)=\left[\mathbf{a}\left(\theta_{1}\right),\mathbf{a}\left(\theta_{2}\right),\ldots,\mathbf{a}\left(\theta_{K}\right)\right]\in\mathbb{C}^{N\times K}$
with $\mathbf{a}\left(\theta_{k}\right)$ denoting the steering vector
associated with AOA $\theta_{k}$ from the $k$-th source (anchor),
$\mathbf{y}\left(t\right)\in\mathbb{C}^{N\times1}$ is the samples
of the received signals, $\mathbf{s}\left(t\right)\in\mathbb{C}^{K\times1}$
is the source signals, and $\mathbf{z}\left(t\right)\in\mathbb{C}^{N\times1}$
denotes the additive noise vector with covariance matrix $\sigma_{z}^{2}\boldsymbol{I}$.
When $K<N$, the CRB is given by
\begin{equation}
\mathrm{CRB}_{\theta}=\frac{\sigma_{z}^{2}}{2L}\left\{ Re\left[\{\mathbf{D}^{H}[\boldsymbol{I}-\mathbf{A}(\mathbf{A}^{H}\mathbf{A})^{-1}\mathbf{A}^{H}]\mathbf{D}\}\right]\right\} ^{-1},
\end{equation}
where $\mathbf{D}=\left[\frac{d\mathbf{a}\left(\theta_{1}\right)}{d\theta_{1}}\cdots\frac{d\mathbf{a}\left(\theta_{K}\right)}{d\theta_{K}}\right]$
denotes the derivative of the steering vectors. For getting more insights
into the CRB of AOA, assume $N$, $L$ are sufficiently large and
the receiver array being a uniform linear array with $\mathbf{a}(\theta_{k})=\left[\begin{array}{lllc}
1 & e^{i\omega\left(\theta_{k}\right)} & \cdots & e^{i(N-1)\omega\left(\theta_{k}\right)}\end{array}\right]^{T}$, where $\omega\left(\theta_{k}\right)=2\pi d\sin\theta_{k}/\lambda$
is a function of $\theta_{k}$. Then, the CRB for $\omega\left(\theta_{k}\right),k=1,...,K$
can be briefly given by
\begin{equation}
\mathrm{CRB}_{\omega}=\frac{6}{N^{3}L}\left[\begin{array}{ccc}
1/\mathrm{SNR}_{1} &  & 0\\
 & \ddots\\
0 &  & 1/\mathrm{SNR}_{K}
\end{array}\right],\label{eq:narrowband-AOA}
\end{equation}
where $\mathrm{SNR}_{k}=1/\sigma_{z}^{2}$ is the receive SNR associated
with the $k$-th source signal (note that both the transmit power
of each anchor and channel gain are normalized to be one in \cite{AOA-narrowband}).
From (\ref{eq:narrowband-AOA}), we can observe that the CRB for the
AOA $\theta_{k}$ is on the order of
\begin{equation}
\mathrm{CRB}_{\theta_{k}}=O\left(\frac{\lambda^{2}}{\text{SNR}_{k}N^{3}Ld^{2}\cos^{2}\theta_{k}}\right).\label{eq:CRBthetaNar}
\end{equation}

Then, wideband signal model is also studied in \cite{AOA-UWB,tempcoop}.
The anchor has a single antenna with normalized transmit power and
the agent has $N$ antennas. The AOA estimation is based on time delay
difference between inter-neighboring-element and can be viewed as
a particular version of TDOA under far-field assumption. The channel
between the anchor and agent is assumed to have a single LOS path.
In this case, the CRB of AOA is given by
\begin{equation}
\mathrm{CRB}_{\theta}=\frac{3c^{2}}{2\pi^{2}d^{2}\left(N-1\right)N\left(2N-1\right)\beta^{2}\text{SNR}\cos^{2}\theta},\label{eq:CRBthetawide}
\end{equation}
where $\beta$ is the effective bandwidth, and $\text{SNR}$ is the
receive SNR. Note that in the limit when $\frac{B}{f_{c}}\rightarrow0$,
the CRB of AOA becomes
\[
\mathrm{CRB}_{\theta}=O\left(\frac{\lambda^{2}}{\text{SNR}N^{3}d^{2}\cos^{2}\theta}\right),
\]
which is consistent with the CRB for the narrowband case in (\ref{eq:CRBthetaNar}).

In \cite{tempcoop}, the authors unified the analysis of the narrowband
and wideband array localization systems where the agent equips $N$
antennas. Specifically, the authors derived the EFIM of AOA-based
localization, which is the form of a weighted sum of the RDM with
direction information intensity (DII) from individual anchors \cite{tempcoop}.
The EFIM can be approximated as
\begin{equation}
{\bf J}_{{\rm e}}\left({\bf p}_{1}\right)\approx N\left(\sum\limits _{j\in{\cal N}_{{\rm b},{\rm LOS}}}\lambda_{1,j}{\bf J}_{{\rm r}}\left(\phi_{1,j}\right)+\mu_{1,j}{\bf J}_{{\rm r}}\left(\phi_{1,j}+\frac{\pi}{2}\right)\right),\label{AOA-shen}
\end{equation}
where $\lambda_{1,j}$ and $\mu_{1,j}$ are the RII and DII from anchor
$j$, respectively. Before getting more insights in (\ref{AOA-shen}),
we note that the squared effective bandwidth of the transmitted RF
signal $s\left(t\right)$ can be decomposed as $\beta^{2}=\beta_{0}^{2}+f_{c}^{2}$,
which contributes to the RI and direction information (DI), respectively.
The first term in the summation (\ref{AOA-shen}) represents the TOA
information (or RI) from the received signals with the RII proportional
to the effective bandwidth of the baseband signal $\beta_{0}^{2}$.
So only the baseband signal contributes to such information. The second
term in the summation (\ref{AOA-shen}) represents the AOA information
(DI) from the received signals with the DII proportional to $f_{c}^{2}$.
Furthermore, in wideband array localization systems, the carrier phases
cannot be used for measuring the TOA information due to an unknown
initial phase in the phase measurements, hence only the baseband part
makes sense, i.e., $\mu_{1,j}\rightarrow0,\forall j$ as $\frac{B}{f_{c}}\rightarrow\infty$.
Conversely, in narrowband array localization systems, the phase differences
between the signals received at the array antennas can eliminate the
unknown initial phase and consequently the carrier part provides extra
AOA information.

\begin{figure}[tbh]
\centering{}\includegraphics[width=85mm]{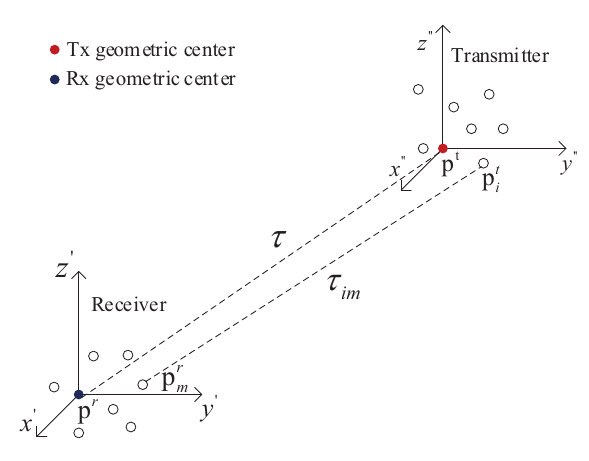}\caption{Localization in a massive MIMO system \cite{5G_10}\label{fig:5G-10}.}
\end{figure}

Recently, millimeter wave and massive MIMO, which are both significant
features for 5G communication networks, are also enabling technologies
for more accurate AOA-based localization and device orientation estimation
\cite{5G_10,5G_15,5Gnew}. In \cite{5Gnew}, the authors studied the
fundamental limits of localization in a narrowband millimeter wave
MIMO system, where only the LOS path was considered. In contrast,
the effect of the delays of different paths was considered in \cite{5G_15}
for localization in a millimeter wave MIMO system. In \cite{5G_10},
a 3-D localization scenario is considered and both the transmitter
and receiver employ massive antenna array with $M$ and $N$ antennas,
respectively, as shown in Fig. \ref{fig:5G-10}. Based on the derivation
of CRB, the author studied the effect of array structure, bandwidth
and synchronization error on the localization accuracy. Specifically,
an example for a planar timed array configuration was considered when
both arrays lying in the XZ-plane and being located one in front to
the other with poisitions $\mathbf{p}^{\mathrm{t}}=[0,0,0]^{T}$ and
$\mathbf{p}^{\mathrm{r}}=[x=0,y,z=0]^{T}$. The CRB for the receiver
position $\mathbf{p}^{\mathrm{r}}$ is given by
\begin{equation}
\begin{array}{l}
\mathrm{CRB}_{x}=\mathrm{CRB}_{z}=\frac{12\kappa_{0}}{SM\left(N-1\right)},\\
\mathrm{CRB}_{y}=\kappa_{0}\frac{1}{MN},
\end{array}\label{eq:AOAcrb}
\end{equation}
where $\kappa_{0}=\frac{c^{2}}{8\pi^{2}\text{SNR}\left(\beta^{2}+f_{c}^{2}\right)}$
denotes the CRB using a single antenna, which is determined by the
receive $\mathrm{SNR}$, effective bandwidth $\beta$, and carrier
frequency $f_{c}$, $S=A_{\mathrm{rx}}/y^{2}$ denotes the ratio between
the receiver array area $A_{\mathrm{rx}}$ and the squared TX-RX distance
$y^{2}$. From (\ref{eq:AOAcrb}), we observe a huge gain obtained
from massive arrays. Compared to the CRB with a single antenna $\kappa_{0}$,
the CRB in (\ref{eq:AOAcrb}) is reduced by a factor of $MN$, where
$M$ accounts for the SNR enhancement due to the beamsteering process
while $N$ accounts for the number of independent measurements available
at the receiver.

\subsubsection{RSS-based Localization}

RSS is also a popular signal metric for localization, especially in
fingerprinting-based and proximity-based localization schemes \cite{RSS1,RSS2}.
For narrowband signals, attenuation of the signal strength through
a mobile radio channel is caused by three nearly independent factors:
path loss, log-normal shadowing, and multipath fading (or fast fading).
In RSS-based localization, time-averaging is commonly used to estimate
the mean received signal strength. For wideband signals, the mean
signal strength is evaluated by summing powers of multipath in a power
delay profile. The mean signal strength is conventionally modeled
in dB scale as
\begin{equation}
P=P_{0}-10\alpha\log_{10}d_{n}+e_{RSS,n},
\end{equation}
\begin{equation}
d_{n}=\sqrt{\left(x_{n}-x\right)^{2}+\left(y_{n}-y\right)^{2}},n=1,2,...,N_{{\rm b}},
\end{equation}
where $\left(x_{n},y_{n}\right)$ denotes the known position of anchor
$n$, $\left(x,y\right)$ denotes the agent's position, $e_{RSS,n}\sim\mathcal{N}\left(0,\eta_{n}^{2}\right)$
is a Gaussian random variable representing the log-normal fading,
and $\alpha$ is the path loss exponent.

The squared position error bound (SPEB) of the RSS-based localization
is expressed as
\begin{equation}
\left(\frac{\ln10}{10\alpha}\right)^{2}\frac{\sum_{i=1}^{N_{\mathrm{b}}}\eta_{i}^{-2}d_{i}^{-2}}{\sum_{i=1}^{N_{\mathrm{b}}}\sum_{j=1}^{i-1}\eta_{i}^{-2}\eta_{j}^{-2}d_{i}^{-2}d_{j}^{-2}\sin^{2}\left(\phi_{i}-\phi_{j}\right)},\label{eq:RSS}
\end{equation}
where $\phi_{i}=\tan^{-1}\frac{y-y_{i}}{x-x_{i}}$ is the angle determined
by the positions of $i$-th anchor and the agent. As can be observed
from (\ref{eq:RSS}), the accuracy of RSS-based localization depends
heavily on the channel parameters, namely the path loss exponent $\alpha$
and the shadowing variances $\eta_{n}^{2}$, wherein the SPEB is proportional
to $\eta_{n}^{2}$ and inversely proportional to the square of $\alpha$.
Furthermore, the effects of NLOS propagation are implicitly included
in the RSS signal model.

\subsubsection{Hybrid Scheme}

Besides the schemes using a single signal metric for localization,
many hybrid schemes have been proposed for localization. In \cite{RSS-TOA},
the author derived the CRB based on hybrid RSS-TOA measurements. \cite{TOA-DOA}
derived the CRB based on a hybrid DOA-TOA localization scheme. Moreover,
for the purpose of eliminating the dependence of estimation accuracy
to the specific anchors' geolocation, the anchor locations were modeled
as Poisson Point Processes (PPP) in \cite{random_12} to study the
average localization performance over the spatial PPP distribution.
The derived average CRB bound acts on the expectation of the MSE with
respect to the random anchor locations depending on the network statistics.

In the following, we elaborate the CRB analysis in \cite{TOA-DOA}
for the hybrid DOA-TOA localization scheme. Consider a far-field scenario
with single-path LOS propagation where the agent is equipped with
an uniform linear array (ULA) of $N$ elements for receiving the reference
signal from an anchor with a single transmit antenna. First, the CRB
of TOA and DOA estimates are derived respectively as follows:
\begin{equation}
\mathrm{CRB}_{\tau}=\frac{1}{8\pi^{2}N\textrm{SNR}\beta^{2}},
\end{equation}
\begin{equation}
\mathrm{CRB}_{\theta}=\frac{3\lambda^{2}}{4\pi^{2}d^{2}\textrm{SNR}\cos^{2}\theta(N-1)N(2N-1)},
\end{equation}
where $\beta^{2}$ denotes the squared effective bandwidth of the
unit-energy transmitted signal $s\left(t\right)$, $d$ is the antenna
element separation, $\theta$ is the DOA and $\lambda$ is the wavelength
of the planewave signal. Apparently, the CRB of TOA is dominantly
determined by the effective bandwidth $\beta$, while that of DOA
is mainly affected by array configuration parameters $N$ and $d$.

Then, the CRB of the location estimate is derived based on that of
the TOA and DOA estimates according to the chain rule. The relation
can be written as
\begin{equation}
\begin{array}{l}
\mathrm{CRB}_{x}=c^{2}\tau^{2}\cos^{2}\theta\mathrm{CRB}_{\theta}+c^{2}\sin^{2}\theta\mathrm{CRB}_{\tau}\\
\mathrm{CRB}_{y}=c^{2}\tau^{2}\sin^{2}\theta\mathrm{CRB}_{\theta}+c^{2}\cos^{2}\theta\mathrm{CRB}_{\tau}
\end{array},
\end{equation}
where $\mathrm{CRB}_{x}$ and $\mathrm{CRB}_{y}$ denote the CRB of
the agent position in x-axis and y-axis of the 2-D plane. The MMSE
of the location estimate is given by
\begin{equation}
\mathrm{MMSE}=\mathrm{CRB}_{x}+\mathrm{CRB}_{y}=c^{2}\tau^{2}\mathrm{CRB}_{\theta}+c^{2}\mathrm{CRB}_{\tau}.
\end{equation}

From the final result, we can conclude that the location accuracy
depends on the SNR, antenna element number, antenna element separation,
squared effective bandwidth, and the relative position between the
anchor and the agent. When the relative distance $c\tau$ is large,
the destructive effect of the estimation errors of DOA for localization
will be magnified.

\subsection{Cooperative Wireless Localization}

\subsubsection{Spatial cooperation}

In spatial cooperation, the agents also transmit reference signals
to aid the localization of the neighbor agents. In this case, the
localization accuracy of all agents can be potentially enhanced. A
few works have studied the fundamental limits of spatial cooperation.
In \cite{cooperative2}, the authors derived the performance bound
based on the signal metrics (e.g., TOA and DOA) extracted from the
received signals. However, such performance bound may not be the fundamental
limit since signal metrics may not contain all useful information
for localization. Hence in \cite{Shen_part2}, the authors extended
their prior work \cite{Shen_part1} to the spatial cooperation scenario
and derived more general fundamental limits of cooperative wireless
localization based on received waveforms rather than signal metrics.
Furthermore, in \cite{cooperative_13}, the authors considered an
anchorless cooperative localization scenario for diminishing the effect
of network topology on the performance bound. Under this assumption,
the localization performance is mainly determined by the number of
nodes in the network and the signal metric used. Also, in \cite{AOA-coop},
the authors considered an AOA-based cooperative localization scheme.

For the scenario of spatial cooperation, the EFIM and RI method introduced
in \cite{Shen_part1} can also be applied similar to the non-cooperative
scenario for whatever the signal metric used, e.g., TOA, TDOA, AOA,
RSS or received waveform itself. Hence, for getting more insights
into the cooperative localization, we next present the fundamental
limit analysis in \cite{Shen_part2}.

In the spatial cooperation considered in \cite{Shen_part2}, the signal
model is the same as (\ref{eq:toa}). The only difference is that
the agents also transmit reference signals, i.e., agent $k$ receives
localization signals from all the other nodes, including both anchors
and the other agents. Suppose there are a set of $N_{a}$ agents denoted
as $\mathcal{N}_{a}$ and a set of $N_{b}$ anchors denoted as $\mathcal{N}_{b}$.
The EFIM for the agent positions $\mathbf{P}=\left[\begin{array}{llll}
\mathbf{p}_{1}^{\mathrm{T}} & \mathbf{p}_{2}^{\mathrm{T}} & \cdots & \mathbf{p}_{N_{\mathrm{a}}}^{\mathrm{T}}\end{array}\right]^{\mathrm{T}}$in this case is a $2N_{\mathrm{a}}\times2N_{\mathrm{a}}$ matrix,
which can be written as
\begin{equation}
\mathbf{J}_{\mathrm{e}}(\mathbf{P})=\mathbf{J}_{\mathrm{e}}^{\mathrm{A}}(\mathbf{P})+\mathbf{J}_{\mathrm{e}}^{\mathrm{C}}(\mathbf{P}),
\end{equation}
where $\mathbf{J}_{\mathrm{e}}^{\mathrm{A}}(\mathbf{P})$ and $\mathbf{J}_{\mathrm{e}}^{\mathrm{C}}(\mathbf{P})$
denote the information from anchors and agent spatial cooperation
respectively. 
\begin{figure*}[tbh]
\begin{equation}
\mathbf{J}_{\mathbf{e}}(\mathbf{P})=\left[\begin{array}{cccc}
\mathbf{J}_{\mathrm{e}}^{\mathrm{A}}\left(\mathbf{p}_{1}\right)+\sum_{j\in\mathcal{N}_{\mathrm{a}}\backslash\{1\}}\mathbf{C}_{1,j} & -\mathbf{C}_{1,2} & \cdots & -\mathbf{C}_{1,N_{\mathrm{a}}}\\
-\mathbf{C}_{1,2} & \mathbf{J}_{\mathrm{e}}^{\mathrm{A}}\left(\mathbf{p}_{2}\right)+\sum_{j\in\mathcal{N}_{\mathbf{a}}\backslash\{2\}}\mathbf{C}_{2,j} & \cdots & -\mathbf{C}_{2,N_{\mathrm{a}}}\\
\vdots & \vdots & \ddots & \vdots\\
-\mathbf{C}_{1,N_{\mathrm{a}}} & -\mathbf{C}_{2,N_{\mathrm{a}}} & ... & \mathbf{J}_{\mathrm{e}}^{\mathrm{A}}\left(\mathbf{p}_{N_{\mathrm{a}}}\right)+\sum_{j\in\mathcal{N}_{\mathrm{a}}\backslash\left\{ N_{\mathrm{a}}\right\} }\mathbf{C}_{N_{\mathrm{a}},j}
\end{array}\right]\label{cooper1}
\end{equation}
\end{figure*}
The detailed formulation is given by (\ref{cooper1}) on the top of
the next page, where
\begin{equation}
\begin{array}{l}
\mathbf{J}_{\mathrm{e}}^{\mathrm{A}}\left(\mathbf{p}_{k}\right)=\sum_{j\in\mathcal{N}_{\mathrm{b}}}\lambda_{k,j}\mathbf{J}_{\mathrm{r}}\left(\phi_{k,j}\right),\\
\mathbf{C}_{k,j}=\mathbf{C}_{j,k}=\left(\lambda_{k,j}+\lambda_{j,k}\right)\mathbf{J}_{\mathbf{r}}\left(\phi_{k,j}\right),j\in\mathcal{N}_{\mathrm{a}}\backslash\{k\},
\end{array}
\end{equation}
where $\lambda_{k,j}$ is RII corresponding to the reference signal
from node $j$ to agent node $k$, $\phi_{k,j}$ is the angle from
node $j$ to agent $k$, and ${\bf J}_{\text{r}}\left(\phi\right)$
is the RDM defined in (\ref{eq:RDM}).

Note that the RI $\lambda_{k,j}\mathbf{J}_{\mathrm{r}}\left(\phi_{k,j}\right)$
from an anchor node $j$ has the same form of that in the non-cooperation
localization, but the RI $\left(\lambda_{k,j}+\lambda_{j,k}\right)\mathbf{J}_{\mathbf{r}}\left(\phi_{k,j}\right)$
from an agent node $j$ is slightly different because the RI between
two agent nodes is obtained by measuring the TDOA instead of TOA,
since it is difficult to achieve time synchronization between two
agents. Nevertheless, the RI is still determined by the SNR and effective
bandwidth of the received waveform, and the POC. Also, each RI corresponds
to an individual received waveform and is a basic building block of
the EFIM. From (\ref{cooper1}), we can observe that $\mathbf{J}_{\mathrm{e}}^{\mathrm{A}}\left(\mathbf{p}_{k}\right)$
is the sub-matrix in the block-diagonal, which indicates that the
localization information from anchors is not interrelated among agents.
Besides, $\mathbf{J}_{\mathrm{e}}^{\mathrm{C}}(\mathbf{P})$ is a
non block-diagonal matrix, implying that the localization information
from agents\textquoteright{} cooperation is highly interrelated. This
is expected since the effectiveness of the localization information
provided by a particular agent depends on its position error.

\subsubsection{Spatio-Temporal Cooperation}

In \cite{spital_temp}, the spatial cooperation is further extended
into spatio-temporal cooperation, which incorporates the intra-node
measurements to further enhance the localization performance by exploiting
the temporal correlation of the localization parameters. In this case,
the EFIM for the positions can be decomposed into two parts, i.e.,
the information obtained from spatial cooperation and temporal cooperation.
Specifically, the EFIM of spatial cooperation is a block-diagonal
matrix of which each block has the same structure as (\ref{cooper1}).
However, the EFIM of temporal cooperation is a non-block-diagonal
matrix because the intra-node measurements are relevant to the agent
positions at two consecutive instants. A more detailed description
can be found in \cite{spital_temp}.

\subsection{Summary and Insights}

In existing works, CRB and EFIM have been used as the performance
metrics for device-based sensing. The two performance metrics are
closely related to each other. Specifically, the CRB can be obtained
by the inverse of the EFIM. As such, it is in general more difficult
to obtain a closed-form expression for the CRB. In this sense, EFIM
can better reveal insights about the structure of the information
that contributes to the sensing/localization performance. All existing
wireless localization schemes explicitly/implicitly utilize three
signal metrics for localization, namely, the TOA/TDOA, the AOA and
the RSS of the reference signals transmitted over the wireless channel.
In all cases, the localization performance improves with the increase
of SNR (power resource) and the number of antennas (spatial resource),
since the increase of these system resources increases the effective
SNR and the number of observations for parameter estimation.

Specifically, in TOA-based schemes, the localization accuracy is affected
by the network topology, multipath environment, the signal effective
bandwidth, the SNR at the receiver, the NLOS condition, and the prior
information about channel parameters or the agent's position. When
there are $M$ transmit antennas at the anchor with a fixed per antenna
transmit power, $N$ receive antennas at the agent, and only two paths
with delays $\tau_{1}$ and $\tau_{2}$ in the multipath environment,
the order of the CRB for the intermediate TOA parameter $\tau_{1}$
associated with each anchor-agent link is given by
\begin{equation}
\mathrm{CRB}_{\tau_{1}}=O\left(\frac{1}{\left(1-\rho\left(\left|\tau_{1}-\tau_{2}\right|\right)\right)MN\beta^{2}{\rm \mathit{\textrm{SNR}}}}\right),\label{eq:summary-tao}
\end{equation}
where $\rho\left(\left|\tau_{1}-\tau_{2}\right|\right)$ denotes a
function of relative delay $\left|\tau_{1}-\tau_{2}\right|$, $\beta$
denotes the effective bandwidth, SNR denotes the receive SNR of the
first path. Compared to the case of single-target device-free sensing
using MIMO radar, there is a slight difference since the CRB of delay
for different paths is coupled as reflected in the term $\rho\left(\left|\tau_{1}-\tau_{2}\right|\right)$.
A similar coupling term also exists in the multi-target device-free
sensing case between the parameters of different targets.

In AOA-based schemes and under narrowband assumption, the localization
accuracy is affected by the the sample-variances of the transmit and
receive antenna positions $\sigma_{T}^{2}$ and $\sigma_{R}^{2}$,
the SNR at the receiver, and the antenna element number. When the
anchor is equipped with a ULA of $M$ transmit antennas and the agent
is equipped with a ULA of $N$ receive antennas, the order of the
CRB for the intermediate AOA parameter $\theta$ associated with each
anchor-agent link is given by
\begin{equation}
\mathrm{CRB}_{\theta}=O\left(\frac{1}{MN\cos^{2}\theta{\rm \mathit{\textrm{SNR}}}\left(\sigma_{T}^{2}+\sigma_{R}^{2}\right)}\right).\label{eq:summary-AOA}
\end{equation}
From (\ref{eq:summary-AOA}), we can observe that AOA estimation accuracy
depends on geometric relation among the agents and the anchors. When
the antenna array and the target happen to lie on a straight line,
e.g., $\theta=\frac{\pi}{2}$, the CRB will become infinitely large.
Note that the CRB order of the AOA parameter is the same as that of
the MIMO radar with a single radar pulse (i.e., $L=1$). In (\ref{eq:summary-tao})
and (\ref{eq:summary-AOA}), we have assumed that the anchor has no
prior information about the agent position and thus there is no transmit
beamforming gain tawards the agent. With perfect transmit beamforming
tawards the agent, the CRB in (\ref{eq:summary-tao}) and (\ref{eq:summary-AOA})
can be reduced by an additional factor of $M$.

In RSS-based schemes, the localization accuracy is affected by the
network topology, propagation environment, e.g., path loss exponent
and shadowing effects, and the distance between the agent and the
anchor.

When multiple signal metrics are used for localization, the estimation
accuracy for the agent position will be improved. Furthermore, cooperation
among agents can significantly improve localization accuracy and reduce
localization outage probabilities, and the localization information
from agents' cooperation is highly interrelated. Agent can treat the
information coming from anchors and other cooperating agents in a
unified way, since anchors can be seen as an agent with infinite prior
position knowledge. Moreover, no matter what kind of cooperation is,
the EFIM of the agent position can be expressed as a weighted sum
of RDMs, and the weight is called the RII, which measures the strength
of the information from a node at a specific direction.

\section{Information-Theoretic Limits of ISAC \label{sec:device-free isac}}

In this section, we present information-theoretic modeling and capacity-distortion
tradeoff limits for several important ISAC building blocks. We start
with a few studies on device-free ISAC and then move on to the studies
on device-based ISAC.

\subsection{Capacity-distortion tradeoff for device-free ISAC memoryless channels\label{subsec:memoryless channel device-free}}

In the device-free ISAC systems, the sensing node, equipped with a
mono-static/bi-static radar, wishes to transmit/receive message to/from
its intended receivers/transmitters and simultaneously estimate the
state parameters of interest upon observing the echo signal. By viewing
this echo signal as generalized feedback, References \cite{P2P,bistatic_MAC,Ahmadipour2020JointSA},
have proposed information-theoretic models for several device-free
ISAC building blocks and investigated the corresponding capacity-distortion
tradeoff.

\subsubsection{Device-free ISAC over Memoryless Point-to-Point Channels with Mono-static
Sensing}

\begin{figure}[tbh]
\begin{centering}
\includegraphics[width=85mm]{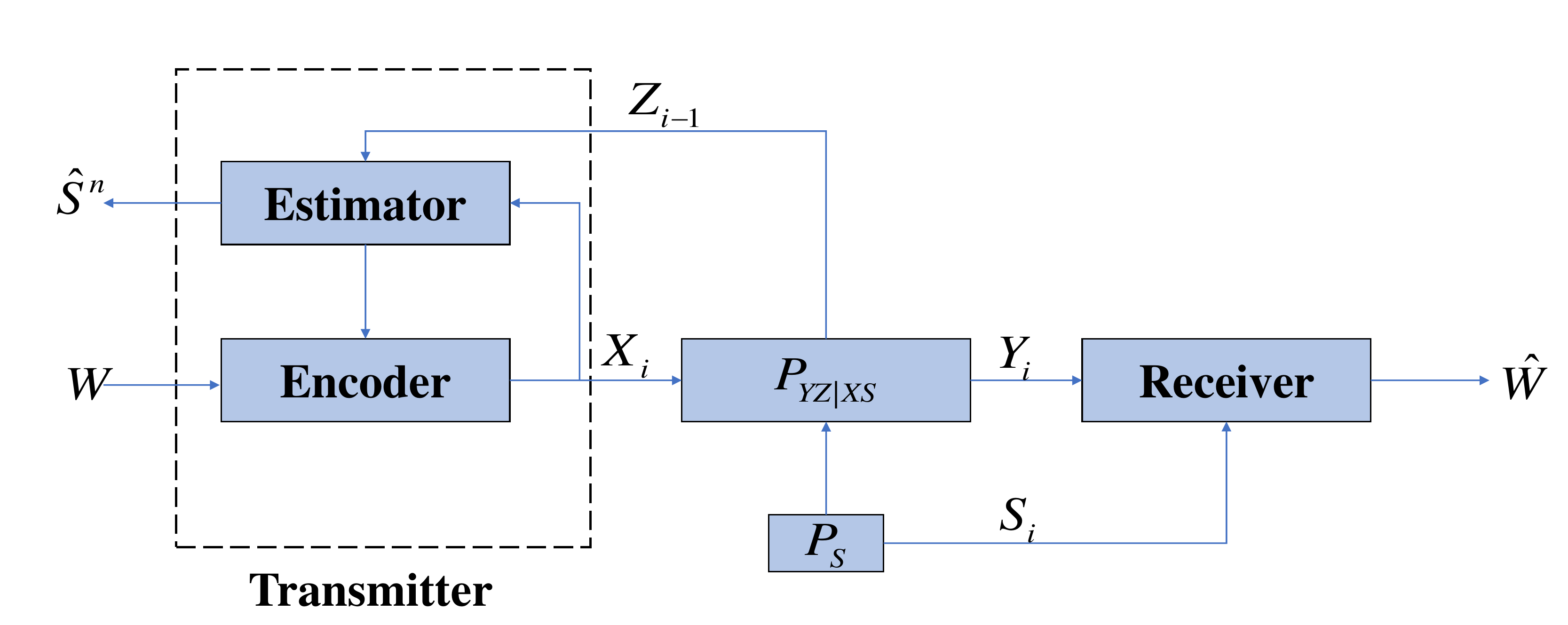}
\par\end{centering}
\caption{\label{fig:P2P}The system model of point-to-point channel with mono-static
sensing (generalized feedback).}
\end{figure}

Reference \cite{P2P} first considered a point-to-point channel with
mono-static sensing, where the ISAC transmitter wishes to send message
$W$ to the receiver while simultaneously estimating the channel state
via output feedback, as illustrated in Fig. \ref{fig:P2P}. The output
feedback can be used to model the radar echo signals in practice.
Assume that the channel is memoryless with i.i.d. state sequence $s^{n}$
(the $i$-th element $s_{i}\in{\mathcal{S}}$) in a coding block of
length $n$, and the receiver knows the state sequence. With input/output
alphabet ${\mathcal{X}},{\mathcal{Y}}$ and ${\mathcal{Z}}$, given
input $X_{i}=x\in{\mathcal{X}}$, the channel produces outputs $(Y_{i},Z_{i})\in{\mathcal{Y}}\times{\mathcal{Z}}$
according to a given transition law $P_{YZ|XS}(.,.|x,s)$ for each
time instance $i$.

For this channel model, the capacity-distortion tradeoff has been
established in \cite[Theorem 1]{P2P}:
\begin{equation}
C(D)=\max_{P_{X}:\frac{1}{n}\sum_{i}\mathbb{E}\left[\left|X_{i}\right|^{2}\right]\leq P}I(X;Y|S),\text{ s.t. }\mathbb{E}[d(S,\hat{S})]\le D
\end{equation}
where the maximum is taken over all input distributions $P_{X}$ satisfying
the average distortion $\mathbb{E}[d(S,\hat{S})]\le D$, and the joint
distribution of ${SXYZ\hat{S}}$ is factorized as
\begin{align}
 & P_{SXYZ\hat{S}}(s,x,y,z,\hat{s})\nonumber \\
= & P_{S}(s)P_{X}(x)P_{YZ|XS}(y,z|x,s)P_{\hat{S}|XZ}(\hat{s}|x,z)
\end{align}

The estimator $P_{\hat{S}|XZ}(\hat{s}|x,z)$ for $S$ is chosen such
that $\mathbb{E}[d(S,\hat{S})]$ is minimized for given input distribution
$P_{X}$. It can be seen that the optimal input is constrained by
the estimation distortion required. In the case of unconstrained distortion
(i.e., $D=\infty$), the result above reduces to the capacity for
a memoryless channel with i.i.d. random states where the state is
available only at the receiver.

To further illustrate the capacity-distortion tradeoff, consider the
following fading channel:
\begin{equation}
Y_{i}=S_{i}X_{i}+N_{i},i=1,\cdots,n,
\end{equation}
where $X_{i}$ is the input satisfying average power constraint $\frac{1}{n}\sum_{i}\mathbb{E}\left[\left|X_{i}\right|^{2}\right]\leq P$
and both $S_{i}$ and $N_{i}$ are i.i.d. Gaussian with zero mean
and unit variance. The generalized feedback is assumed to be perfect,
i.e, $Z=Y$.

\begin{figure}[tbh]
\begin{centering}
\includegraphics[width=75mm]{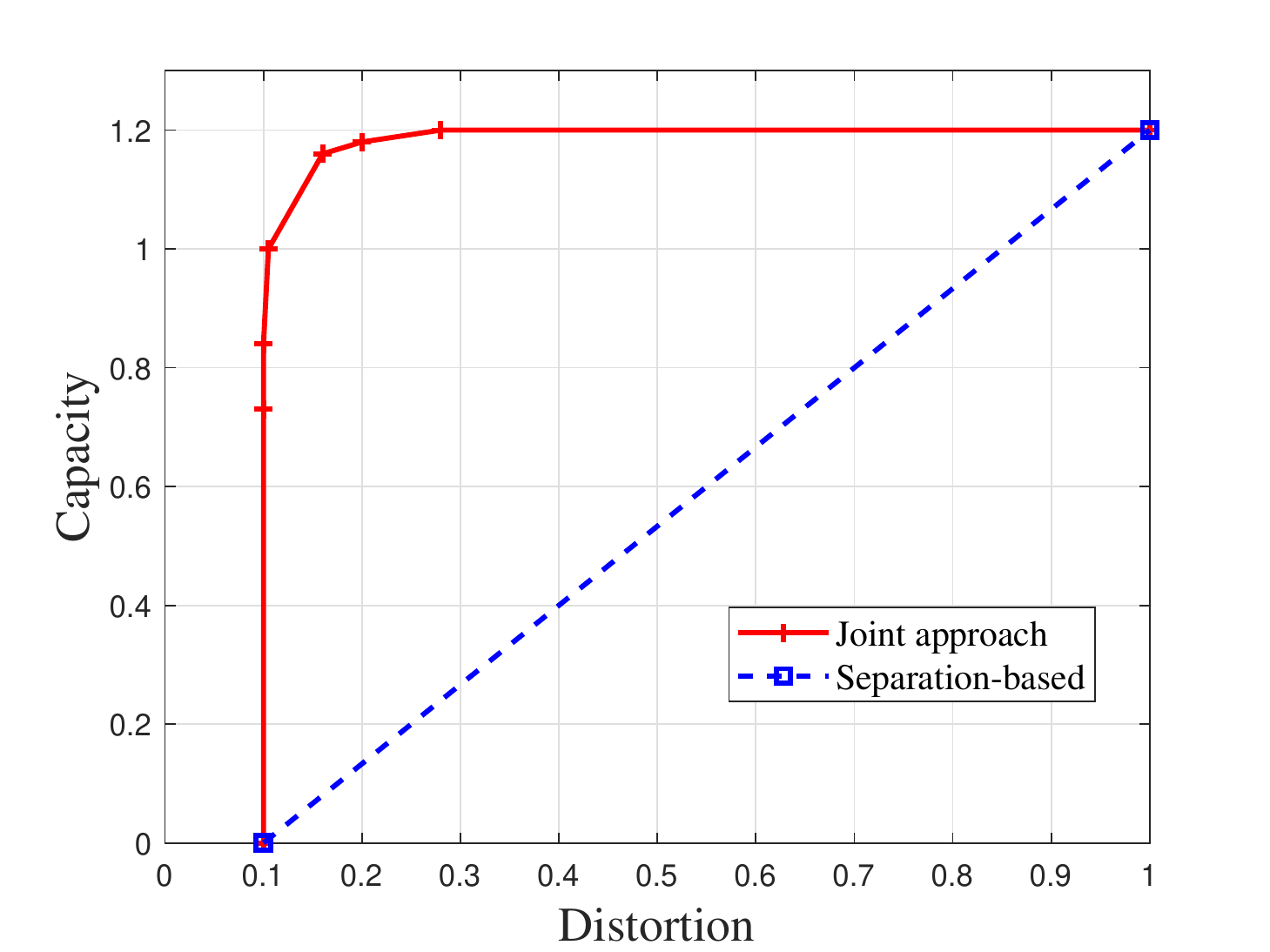}
\par\end{centering}
\caption{\label{fig:P2P channel}The capacity-distortion tradeoff when $P=10$
dB \cite{P2P}.}
\end{figure}

Fig. \ref{fig:P2P channel} plots the capacity-distortion tradeoff
when $P=10$ dB. On one extreme, when $D=1$, $C(D=1)$ reduces to
the unconstrained capacity (note that the distortion $D=1$ can be
achieved by setting $\hat{S}=0$ no matter what the input distribution
is), while on the other extreme, when $D=0$, positive capacity $C(D=0)=0.733$
bcu is still achievable. In general the joint transmission design
as proposed in \cite{P2P} outperforms a communication and sensing
separation-based approach.

\subsubsection{Device-free ISAC over Memoryless Multiple-Access Channels with Mono-static
Sensing}

\begin{figure}[tbh]
\begin{centering}
\includegraphics[width=85mm]{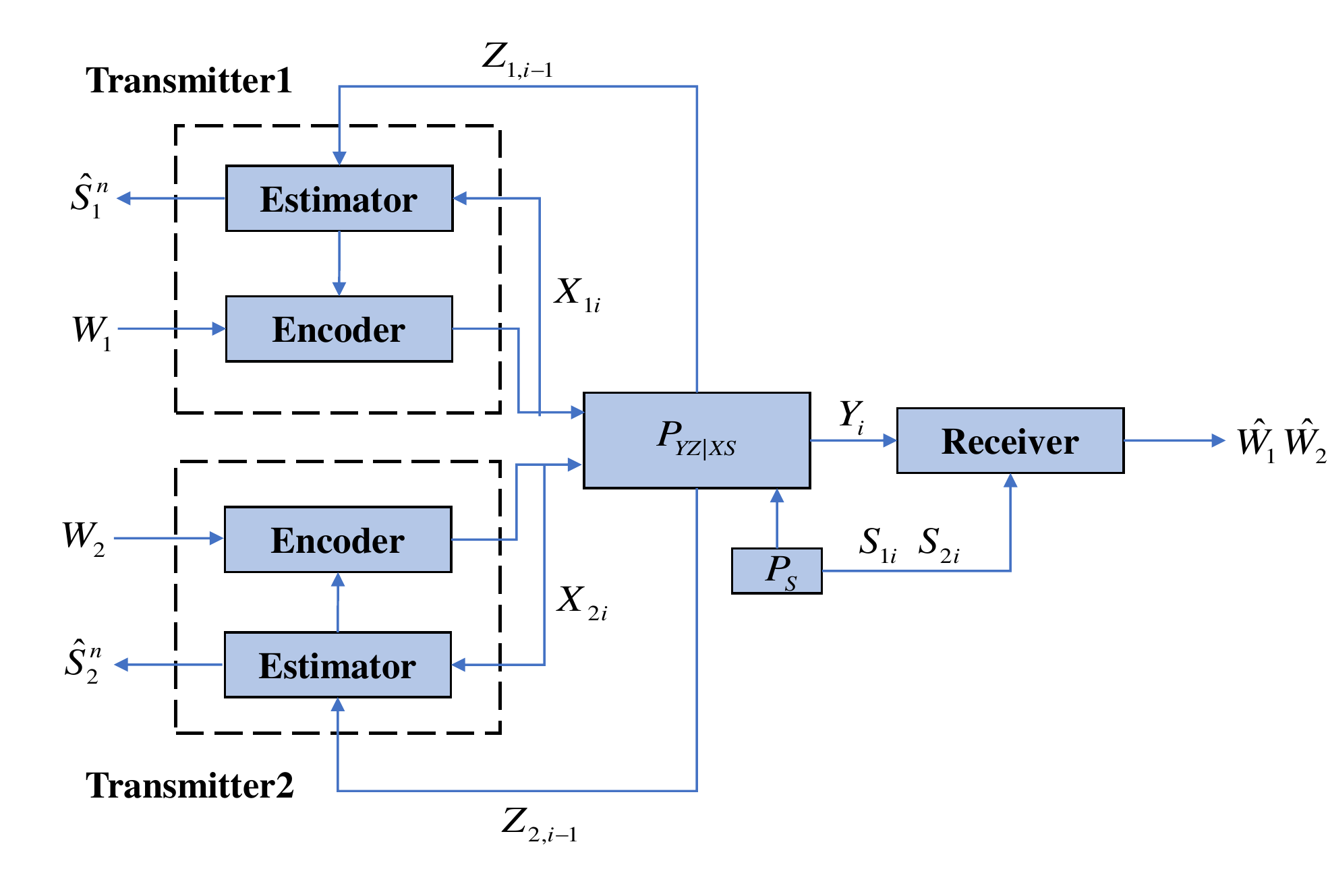}
\par\end{centering}
\caption{\label{fig:MAC_device_free}The system model of multiple-access channel
with mono-static sensing.}
\end{figure}

Reference \cite{bistatic_MAC} instead considered a two-user multiple-access
channel where the $k$-th ($k=1,2$) ISAC transmitter wishes to send
message $W_{k}$ to the receiver while simultaneously estimating its
channel state via output feedback, as illustrated in Fig. \ref{fig:MAC_device_free}.
The channel is memoryless with i.i.d. state sequence $s_{k}^{n}$
($s_{k,i}\in\mathcal{S}_{k}$) in a coding block, and the receiver
knows both $s_{1}^{n}$ and $s_{2}^{n}$. With input/output alphabet
$\mathcal{X}_{k},Y$ and $\mathcal{Z}_{k}$, given input $X_{k,i}=x\in\mathcal{X}_{k}$,
the channel produces outputs $(Y_{i},Z_{1,i},Z_{2,i})\in\mathcal{Y}\times\mathcal{Z}_{1}\times\mathcal{Z}_{2}$
according to a given transition law $P_{YZ_{1}Z_{2}|X_{1}X_{2}S_{1}S_{2}}(.,.,.|x_{1},x_{2},s_{1},s_{2})$
for each time instance $i$.

For this ISAC MAC model, outer and inner bounds on the capacity-distortion
region have been established, see Theorem 1 and Theorem 2 of \cite{bistatic_MAC}.
The inner bound exploits the feedback-induced cooperation between
the transmitters and is achieved by block Markov encoding and backward
decoding, in which three auxiliary random variables $U,V_{1},V_{2}$
are introduced. Specifically, $U$ is the common decoded message from
the previous coding blocks through feedback $Z_{1},Z_{2}$, $V_{1}$
is the partial message transmitted by the current coding block of
user 1 which can be decoded by user $2$ through feedback $Z_{2}$,
and $V_{2}$ is the partial message transmitted by the current coding
block of user 2 which can be decoded by user $1$ through feedback
$Z_{1}$. Consequently, the estimation of $s_{1}$ at Transmitter
$1$ is based on $x_{1}$, $v_{2}$, $z_{1}$ and the optimal estimator
$\psi_{1}^{*}\left(x_{1},v_{2}\text{,}z_{1}\right)$ for $S_{1}$
is given by
\begin{align}
\psi_{1}^{*}\left(x_{1},v_{2}\text{,}z_{1}\right) & =\arg\underset{\psi_{1}}{\min}\underset{s_{1}\in\mathcal{S}_{1}}{\sum}P_{S_{1}\mid X_{1}V_{2}Z_{1}}\left(s_{1}\mid x_{1}v_{2}z_{1}\right)\nonumber \\
\times & d_{1}\left(s_{1},\psi_{1}\left(x_{1},v_{2},z_{1}\right)\right)
\end{align}
where $d_{1}$ is the distortion function at Transmitter $1$. Given
$X_{1}=x_{1},V_{2}=v_{2}$, the estimation cost for $S_{1}$ is
\begin{equation}
c_{1}(x_{1},v_{2})=\mathbb{E}\left[d_{1}\left(s_{1},\psi_{1}^{*}\left(x_{1},v_{2},z_{1}\right)\right)\mid X_{1}=x_{1},V_{2}=v_{2}\right]
\end{equation}
The optimal estimator $\psi_{2}^{*}\left(x_{2},v_{1}\text{,}z_{2}\right)$
for $S_{2}$ and the corresponding estimation cost $c_{2}(x_{2},v_{1})$
can be obtained similarly.

For given average distortion constraints $\mathbb{E}[c_{1}(X_{1},V_{2})]\le D_{1},\mathbb{E}[c_{2}(X_{2},V_{1})]\le D_{2}$,
the inner bound $\mathcal{R}(D_{1},D_{2})$ achieved by the above
block Markov encoding and backward decoding scheme consists of all
rate pairs $(R_{1},R_{2})$ satisfying
\begin{align}
R_{1} & \leq I\left(X_{1};Y\mid X_{2}V_{1}US\right)+I\left(V_{1};Z_{2}\mid X_{2}U\right)\nonumber \\
R_{2} & \leq I\left(X_{2};Y\mid X_{1}V_{2}US\right)+I\left(V_{2};Z_{1}\mid X_{1}U\right)\nonumber \\
R_{1}+R_{2} & \leq\begin{array}{l}
\min\{I\left(X_{1}X_{2};Y\mid S\right),I\left(X_{1}X_{2};Y\mid SV_{1}V_{2}U\right)\\
+I\left(V_{1};Z_{2}\mid X_{2}U\right)+I\left(V_{2};Z_{1}\mid X_{1}U\right)\}
\end{array},
\end{align}
where $V_{1}X_{1}-U-V_{2}X_{2}$ and $UV_{1}V_{2}-X_{1}X_{2}-YZ_{1}Z_{2}$
form Markov chains.

To gain insights on the resultant rate-distortion tradeoff, consider
the following state-dependent MAC channel:
\begin{equation}
Y=S_{1}X_{1}+S_{2}X_{2},
\end{equation}
with binary input $X_{1}$ and $X_{2}$, i.i.d. states $S_{1}$ and
$S_{2}$ Bernoulli distributed with parameter $p(s_{k}=1)=p_{s}$
and output feedback $Z_{1}=Z_{2}=Y$.

\begin{figure}[tbh]
\begin{centering}
\includegraphics[width=85mm]{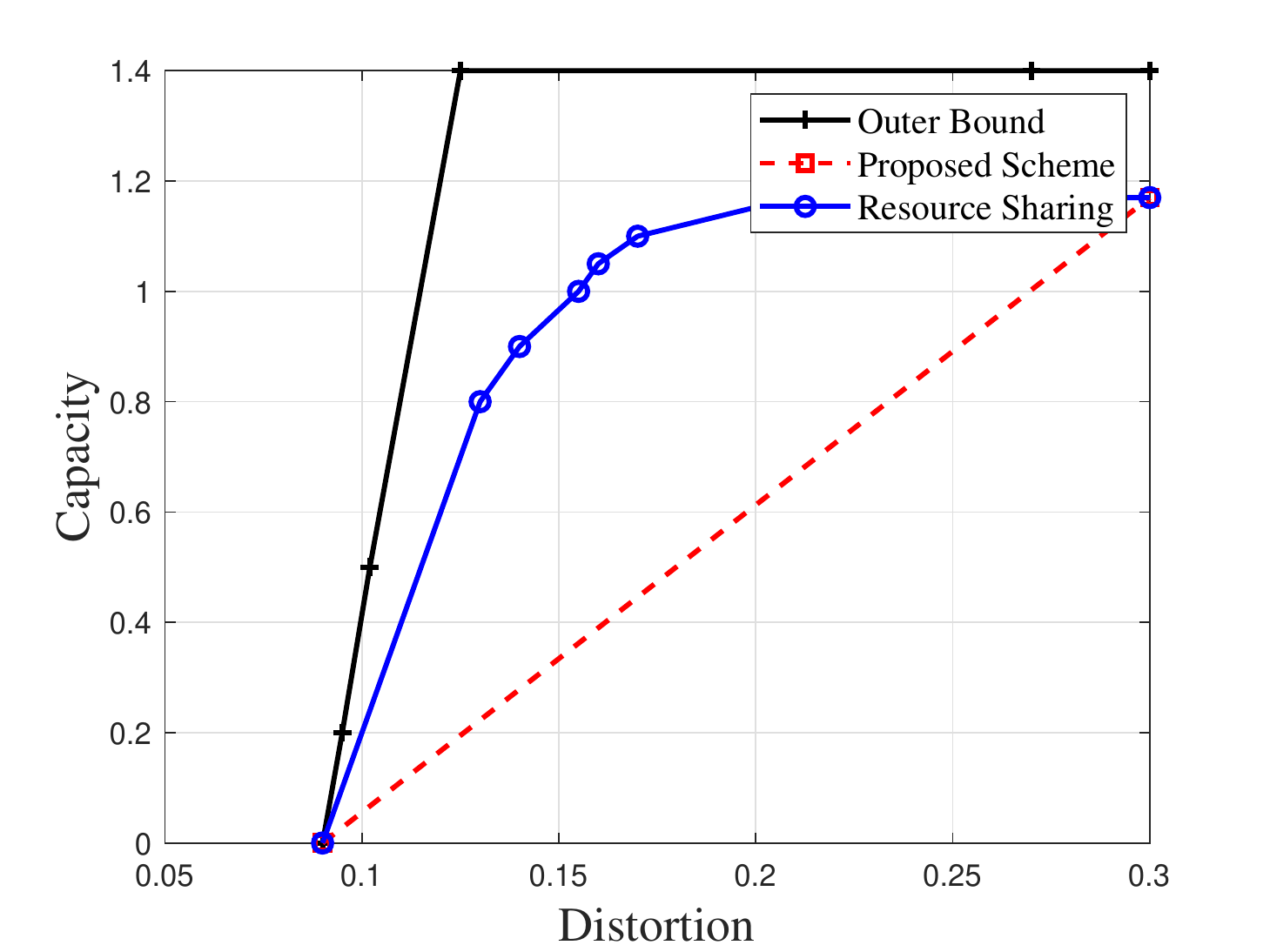}
\par\end{centering}
\caption{\label{fig:Tradeoff of ISAC2}Tradeoff between sum rate and distortion
for $p_{s}=0.7$ \cite{bistatic_MAC}.}
\end{figure}

Fig. \ref{fig:Tradeoff of ISAC2} plots the sum rate-distortion tradeoff
when $D_{1}=D_{2}=D$ and $p_{s}=0.7$, where the x-axis denotes the
distortion and the y-axis denotes the sum capacity. There exists gap
between the inner bound and the outer bound, however, when the distortion
is small, the proposed scheme achieves near-optimal performance. In
general the joint transmission design as proposed outperforms a communication
and sensing resource sharing approach.

\subsubsection{Device-free ISAC over Memoryless Broadcast Channels}

\begin{figure}[tbh]
\begin{centering}
\includegraphics[width=85mm]{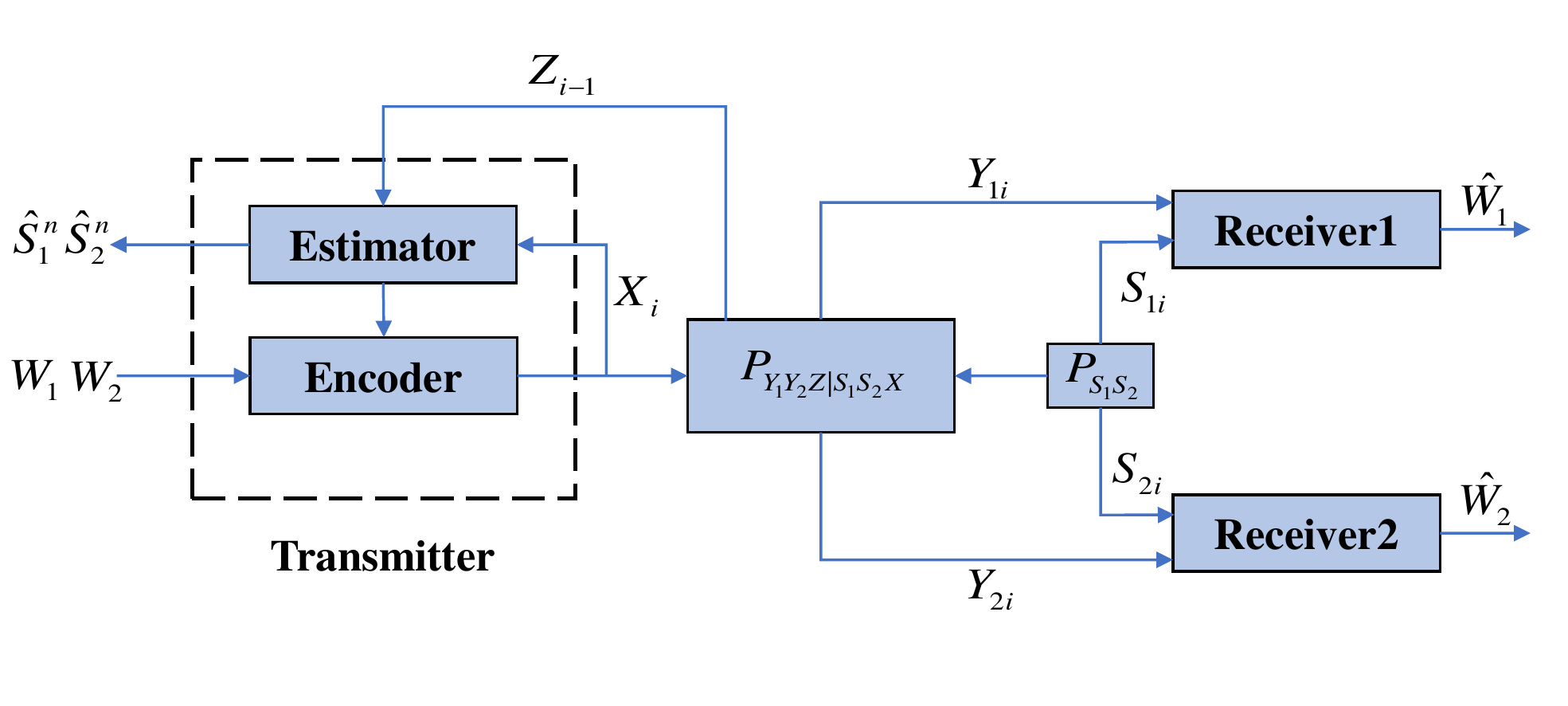}
\par\end{centering}
\caption{\label{fig:ISACDFBC}The system model of broadcast channel with mono-static
sensing.}
\end{figure}

A more recent work \cite{Ahmadipour2020JointSA} studied a two-user
broadcast channel where the ISAC transmitter wishes to send message
$W_{k}$ to the $k$-th ($k=1,2$) receiver while simultaneously estimating
its channel state via output feedback, as illustrated in Fig. \ref{fig:ISACDFBC}.
The channel is memoryless with i.i.d. state sequence $s_{k}^{n}$
($s_{k,i}\in\mathcal{S}_{k}$) in a coding block, and each receiver
knows its own state sequence $s_{k}^{n}$. With input/output alphabet
$\mathcal{X},\mathcal{Y}_{k}$ and $\mathcal{Z}_{k}$, given input
$X_{i}=x\in\mathcal{X}$, the channel produces outputs $(Y_{1,i},Y_{2,i},Z_{1,i},Z_{2,i})\in\mathcal{Y}_{1}\times\mathcal{Y}_{2}\times\mathcal{Z}_{1}\times\mathcal{Z}_{2}$
according to a given transition law $P_{Y_{1}Y_{2}Z_{1}Z_{2}|XS_{1}S_{2}}(.,.,.,.|x,s_{1},s_{2})$
for each time instance $i$.

In \cite{Ahmadipour2020JointSA}, the capacity-distortion tradeoff
region was fully characterized for the physically degraded ISAC broadcast
channel. In addition, outer and inner bounds on the tradeoff region
was established for the general ISAC broadcast channel. As an example
of the physically degraded case, consider the following broadcast
channel with multiplicative binary states:
\begin{equation}
Y_{k}=XS_{k},~~k=1,2,
\end{equation}
where the joint state probability mass function $P(S_{1}=0,S_{2}=0)=1-q$,
$P(S_{1}=0,S_{2}=1)=0$, $P(S_{1}=1,S_{2}=1)=q\gamma$ and $P(S_{1}=1,S_{2}=0)=q(1-\gamma)$
with $q,\gamma\in[0,1]$, and output feedback $Z=(Y_{1},Y_{2})$.
The capacity-distortion tradeoff region is depicted in \cite[Fig. 2]{Ahmadipour2020JointSA}
for $\gamma=0.5$ and $q=0.6$. Again in general the joint transmission
design as proposed outperforms a resource-sharing (or separation-based)
approach that splits the resource either for sensing or communication.

\subsection{Capacity-Distortion Tradeoff for Device-based ISAC Memoryless Channels\label{subsec:Memoryless channel device-based}}

For the device-based ISAC system, the receiver aims to simultaneously
decode the message and estimate some random parameters of interest
from its received signal. By modeling the parameter as a random state,
\cite{zhang} presented an information-theoretic framework of joint
communication and state estimation.

Consider a point-to-point memoryless channel. With input/output alphabet
$\mathcal{X}$ and $\mathcal{Y}$, given input $X_{i}=x\in\mathcal{X}$,
the channel produces output $Y_{i}\in\mathcal{Y}$ according to a
given transition law $P_{Y|XS}(.|x,s)$ for each time instance $i$.
Assume that the state sequence $s^{n}$ to be estimated is i.i.d with
$P(s^{n})=\prod\nolimits _{i=1}^{n}p_{S}\left(s_{i}\right)$ and is
unknown to the transmitter. For the model considered, the capacity-distortion
function is established in Theorem 1 of \cite{zhang}.
\begin{align}
C\left(D\right)=\underset{P_{X}\in\mathcal{P}_{D}}{max}I\left(X;Y\right),
\end{align}
where $\mathcal{P}_{D}=\left\{ P_{X}:\underset{x\in\mathcal{X}}{\sum}P_{X}\left(x\right)d^{*}\left(x\right)\leq D\right\} $.
Here, $d^{*}(x)$ is the estimation cost function due to signaling
with $x\in\mathcal{X}$. In other words, $d^{*}(x)$ is the minimum
distortion that can be achieved for a given signaling $x\in\mathcal{X}$.
$\mathcal{P}_{D}$ regulates the input distribution so that the signaling
is estimation-efficient. When $D=\infty$, $C(D)$ reduces to the
classic unconstrained channel capacity.

To gain insights on the resultant capacity-distortion tradeoff here,
consider the following state-dependent Gaussian channel \cite{zhang}:
\begin{align}
Y_{i}=X_{i}+S_{i}+Z_{i},
\end{align}
where $S_{i}\sim\mathcal{\mathcal{C}N}(0,Q)$, $Z_{i}\sim\mathcal{C}N(0,N)$
and $X_{i}$ is subject to an average power constraint $P$. The system
achieves the following tradeoff
\begin{align}
C\left(D\right)=\begin{cases}
\begin{array}{c}
\log\left(1+\frac{P}{Q+N}\right)\\
0
\end{array} & \begin{array}{c}
D>\frac{QN}{Q+N}\\
D\leq\frac{QN}{Q+N}
\end{array}\end{cases}.
\end{align}
It can be seen that a non-zero communication rate is achieved only
when the estimation distortion required is not very stringent and
above the threshold $QN/(Q+N)$. This demonstrates the cost of achieving
finer estimation at the receiver.

The study is also extended to a two-user multiple access channel with
device-based sensing and the corresponding capacity-distortion region
is the union of all rate pairs $(R_{1},R_{2})$ satisfying
\begin{align}
R_{1}\leq I\left(X_{1};Y\mid X_{2},Q\right),\nonumber \\
R_{2}\leq I\left(X_{2};Y\mid X_{1},Q\right),\nonumber \\
R_{1}+R_{2}\leq I\left(X_{1},X_{2};Y\mid Q\right),
\end{align}
over $P_{Q}(q)P_{X_{1}|Q}(x_{1}|q)P_{X_{2}|Q}(x_{2}|q)P_{Y|X_{1},X_{2}}(y|x_{1},x_{2})$
that satisfies
\begin{equation}
D\ge\underset{\left(q,x_{1},x_{2}\right)}{\sum}P_{Q}\left(q\right)P_{X_{1}\mid Q}\left(x_{1}\mid q\right)P_{X_{2}\mid Q}\left(x_{2}\mid q\right)d^{*}\left(x_{1},x_{2}\right).
\end{equation}

\subsection{Summary}

Information-theoretic state-dependent channels with generalized output
feedback have been shown to be useful in modeling and assessing the
performance of device-free ISAC systems. The capacity-distortion tradeoff
has been fully characterized for a point-to-point channel under some
simplified assumptions, while inner and outer bounds on the capacity-distortion
region have been proposed for multiple access and broadcast ISAC channels.
The benefit of joint-design approach over separation-based (or resource-sharing)
approach is clearly evident. However, the results are derived under
some restrictive assumptions, such as the state sequence is i.i.d.
or/and channel state and sensing state are the same. New modeling
and bounding techniques shall be developed for memoryless device-free
ISAC channels under more realistic assumptions.

Similar to device-free ISAC, device-based ISAC also embraces performance
tradeoff between communication and sensing. Information-theoretic
state-dependent channels with receiver state estimation have been
proposed to establish the fundamental capacity-distortion tradeoff
in device-based ISAC. In particular, the optimal tradeoff has been
fully characterized for a point-to-point channel and a two-user MAC
channel with i.i.d. state sequence. This approach can also be generalized
and applied to more complicated channel topologies, such as broadcast
channels, which are worth further studies.

\section{Design and Performance Analysis of Specific ISAC Systems \label{sec:device-based isac}}

In this section, we discuss some designs and performance analysis
of ISAC systems tailored to different application scenarios. Similar
to the previous section, we first focus on the device-free ISAC and
then go on to the device-based ISAC.

\subsection{Design and Performance Analysis of Some Device-Free ISAC Applications}

\subsubsection{Application of MAC with Mono-Static BS Sensing}

\begin{figure}[tbh]
\begin{centering}
\includegraphics[width=75mm]{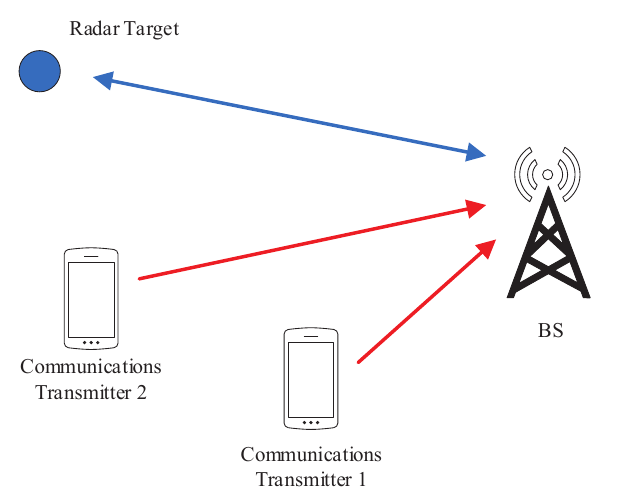}
\par\end{centering}
\centering{}\caption{\label{fig:ISAC-MAC}An uplink communication system with mono-static
BS sensing.}
\end{figure}

The first application considered is a joint radar-communication system
as shown in Fig. \ref{fig:ISAC-MAC}. The base station is sensing
$K$ targets of interest while serving an uplink communication user.
Therefore, the echo signals reflected by the targets will be superimposed
on the uplink communication signal. This is a MAC with mono-static
BS sensing. For $K$ targets, the observed complex baseband signal
at the BS is given by
\begin{equation}
y\left(t\right)=\sqrt{P_{com}}\alpha_{cu}s_{c}\left(t\right)+\sqrt{P_{rad}}\sum\limits _{k=1}^{K}\alpha_{rk}s_{r}\left(t-\tau_{k}\right)+z\left(t\right),
\end{equation}
where $P_{com}$ is the communication transmit power, $P_{rad}$ is
the radar transmit power, $\alpha_{cu}$ is the channel gain for the
$u$-th user, $\alpha_{rk}$ is the RCS of the $k$-th target, $\tau_{k}$
is the delay of the $k$-th target and $z\left(t\right)$ is the additive
Gaussian noise.

\begin{figure}[tbh]
\begin{centering}
\includegraphics[width=75mm]{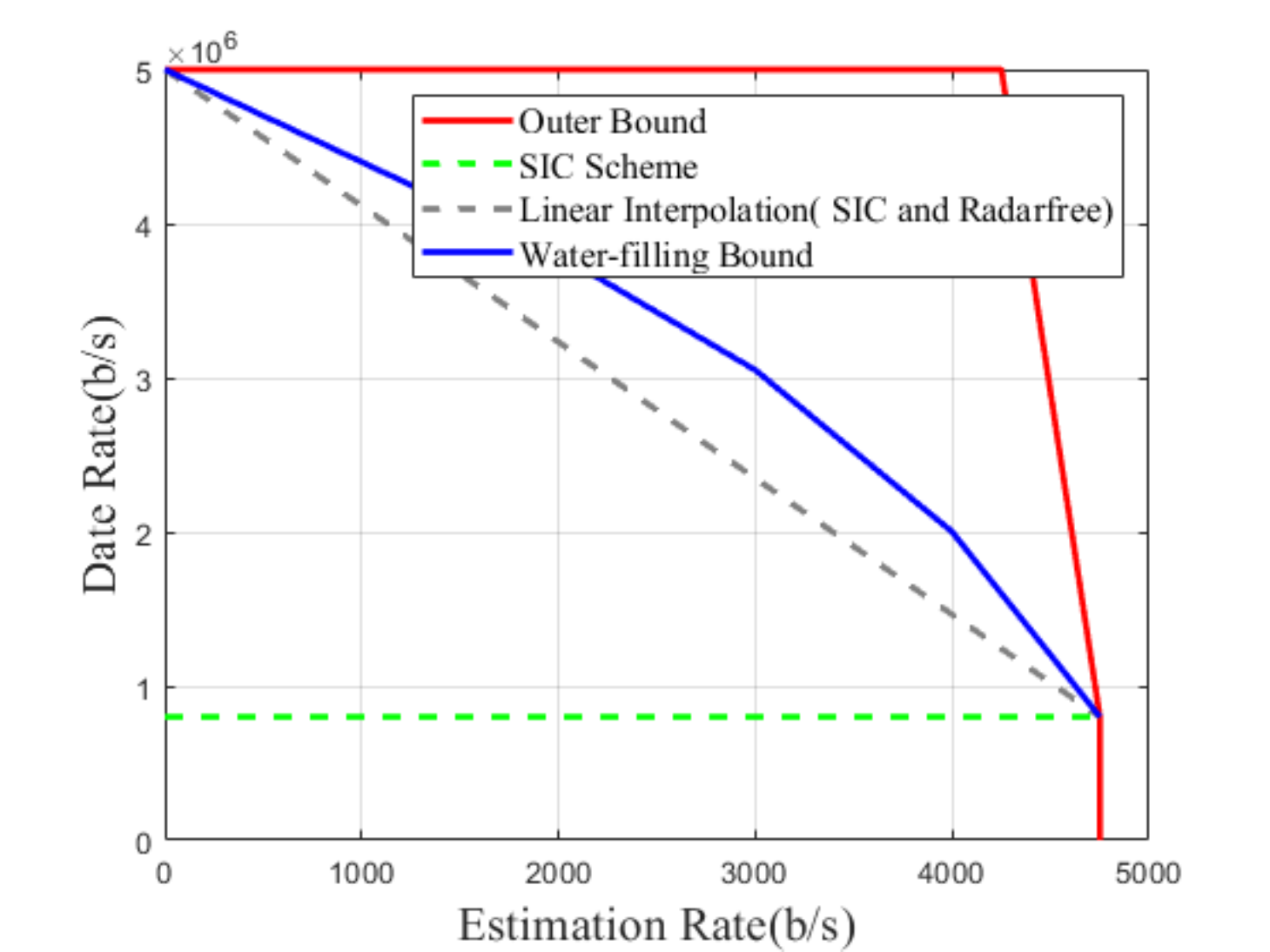}
\par\end{centering}
\caption{\label{fig:MAC_tradeoff2}The tradeoffs between communication information
rate and estimation information rate under different schemes \cite{estimation_rate}.}
\end{figure}

For this application scenario, various resource-sharing schemes that
involve bandwidth partitioning and power allocation have been proposed
and their associated performance bounds have been established for
each scheme, see Section IV. of \cite{estimation_rate}. Therein,
the performance of estimating $\tau_{k}$ is quantified by the notion
of estimation information rate as discussed in Section III in this
paper. Fig. \ref{fig:MAC_tradeoff2} compares the tradeoffs between
communication information rate and estimation information rate under
different schemes. Outer bounds on communications and radar are indicated
by the red lines. Successive interference cancellation (SIC) bound
for the communications rate is indicated by the green dashed line.
The linear interpolation between SIC vertex and the radarfree data
rate bound is indicated by the gray dashed line. The water-filling
inner bound is indicated by the blue line. It can be seen that the
water-filling bound generally outperforms the other inner bounds,
because it advocates flexible sub-band partitioning (one for communications
only and the other for both radar and communications) and optimized
power allocation between the sub-bands.

\subsubsection{Application of Point-to-Point Channel with Mono-Static Sensing}

\begin{figure}[tbh]
\begin{centering}
\includegraphics[width=85mm]{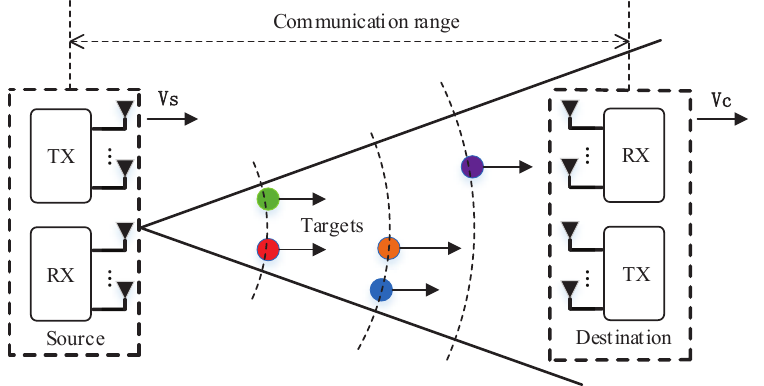}
\par\end{centering}
\centering{}\caption{\label{fig:DMMSE system model}A point-to-point communication system
with mono-static sensing.}
\end{figure}

The second application considered is a point-to-point channel with
mono-static sensing in vehicular networks as shown in Fig. \ref{fig:DMMSE system model}.
Specifically, a source vehicle sends an adaptive IEEE 802.11ad single-carrier
physical layer frame to a target vehicle and uses the reflections
from the target vehicle to derive its range and velocity. Each frame
has $K$ symbols in total, with $\alpha=\frac{K_{c}}{K}$ fraction
of them for data and the rest for preamble.

For the system considered, the effective maximum achievable communication
spectral efficiency depends on $\alpha$ and is expressed as
\begin{equation}
r_{\text{eff}}=\alpha\log(1+\text{SNR}_{c})=\log(1+\text{SNR}_{c})^{\alpha},\label{eq:rate}
\end{equation}
where $\text{SNR}_{c}$ represents the communication SNR that accounts
for the path-loss.

As for sensing, in case of velocity estimation using the preamble
of the IEEE 802.11ad frame, the CRB is given by \cite{DMMSE}
\begin{equation}
\text{CRB}_{v}=\frac{6\lambda^{2}}{16\pi^{2}\left(1-\alpha\right)^{3}K^{3}T_{s}^{2}\text{SNR}_{r}},\label{eq:CRBv-1}
\end{equation}
where $\text{SNR}_{r}$ is the radar SNR, $\lambda$ is the wave-length
and $T_{s}$ is the symbol duration. On the other hand, the CRB for
the range estimation of a target vehicle is given by
\begin{equation}
\text{CRB}_{d}=\frac{c^{2}}{32\pi^{2}B_{\text{rms}}^{2}\left(1-\alpha\right)K\text{SNR}_{r}},\label{eq:CRBd}
\end{equation}
where $B_{\text{rms}}$ is the root-mean square bandwidth of the Fourier
transform of the preamble and $c$ is the speed of the light.

It can be seen from (\ref{eq:rate}) - (\ref{eq:CRBd}) that both
radar and communication performance metrics are dependent on $\alpha$.
In addition, the communication rate $r_{\text{eff}}$ can be derived
to its equivalent MSE metric as $\text{MMSE}_{\text{eff}}=2^{-r_{\text{eff}}}$
as discussed in Section III. Therefore, one can optimize $\alpha$
through the following weighted optimization problem:
\begin{align}
 & \mathop{\min}\limits _{\alpha}\omega_{d}\log\left(\text{CRB}_{d}\right)+\omega_{v}\log\left(\text{CRB}_{v}\right)-\omega_{c}\log\left(\text{MMSE}_{\text{eff}}\right)\nonumber \\
 & ~~~~~~~~~~~~~~~\text{s.t.}~0\leq\alpha\leq1,
\end{align}
to achieve the optimal tradeoff between communication and radar sensing
performance.

\subsection{Design and Performance Analysis of Some Device-Based ISAC Applications}

\subsubsection{Performance Analysis for Point to Point Communication with Localization}

In \cite{ILAC_waveform9}, Armin introduced a parametric positioning
waveform design which provides a scalar parameter for controlling
the distribution of the PSD. When the single-carrier transmission
scheme is considered, the signal which has more power concentrated
at the edges of the spectrum leads to a larger equivalent signal bandwidth,
resulting a lower CRB in positioning. However, for communication,
the optimal signal PSD scheme is to concentrate the signal power at
the central of the mainlobes. Consequently, there is a trade-off for
waveform design between localization and communication. Besides the
study of waveform design, the power-partitioning scheme for satisfying
different localization and data-rate requirements is also researched
in the millimeter wave network \cite{ILAC_pow12}. Ghatak adopted
the Bayesian CRLB and Rate Coverage Probability as the performance
metrics of localization and communication respectively to be satisfied.
Apparently, as more transmit power allocated for data services, the
rate coverage probability will improve while the localization accuracy
will degrade. In \cite{ILAC_10}, Destino also considered a millimeter
wave wireless network, which utilized a beam training period for localization.
Different from the RSS measurements used by \cite{ILAC_pow12} for
localization, the TOA-based beam alignment scheme is considered, which
reveals a trade-off between localization accuracy and effective communication
rate. Nevertheless, only an exhaustive search strategy is considered
and without the consideration of beam misalignment error.

\subsubsection{Performance Analysis and Optimization for Multiple Access Communication
with Localization}

\begin{figure}[tbh]
\centering{}\includegraphics[width=75mm]{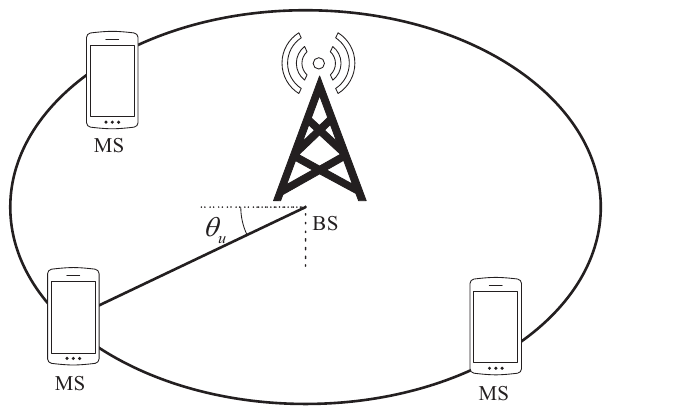}\caption{\label{fig:Uplink-system-model}An uplink communication system with
localization.}
\end{figure}

\begin{figure}[tbh]
\begin{centering}
\includegraphics[width=75mm]{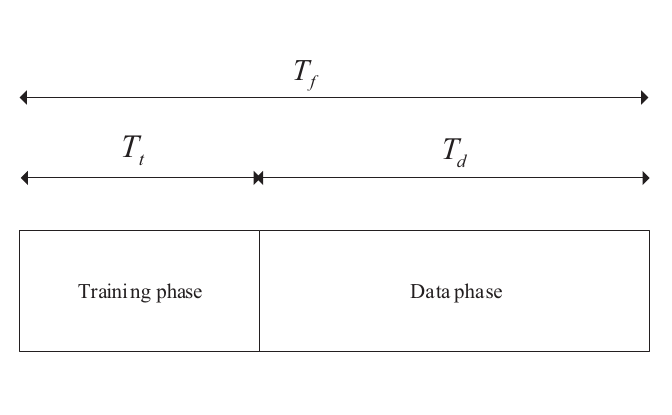}
\par\end{centering}
\centering{}\caption{\label{fig:Joint-localization-and}Joint localization and data transmission
scheme.}
\end{figure}

In \cite{ILAC_11}, the prior work \cite{ILAC_10} is extended to
multi-user scenario. The author considered a millimeter wave based
multi-user single-input-multiple-output (SIMO) wireless uplink system
with an $N$-antenna BS and $U$ single-antenna mobile stations (MSs)
as shown in Fig. \ref{fig:Uplink-system-model}, and orthogonal resource
allocation for different MSs is assumed, e.g., by using time-division-multiple-access
(TDMA) or frequency-division-multiple-access (FDMA). In this system,
a joint localization and data transmission scheme was proposed. In
each transmission block of fixed duration $T_{f}$, the training phase
of duration $T_{t}$ is used for beam alignment and localization while
the data phase of duration $T_{d}$ is used for data transmission
as shown in Fig. \ref{fig:Joint-localization-and}.

During the training phase, an exhaustive beam alignment strategy is
used. Specifically, for each user, the BS sequentially trains each
of the beam in the codebook set $\mathcal{W}$ and find the best one
that maximizes the beamforming gain of this user. Assuming LOS channels
for all users, the received signal at the BS for training the $k$-th
MS can be written as
\begin{equation}
y_{k}\left(t\right)=\alpha_{k}\mathbf{w}^{H}\mathscr{\mathbf{a}}_{R}\left(\theta_{k}\right)x_{k}\left(t-\tau_{k}\right)+\mathbf{w}^{H}\mathbf{n}\left(t\right),
\end{equation}
where $\alpha_{k}$ and $\theta_{k}$ are the complex gain and AOA
associated with the LOS path of the $k$-th MS, $\mathbf{w}\in\mathcal{W}$
is the receive beamforming vector, $\mathscr{\mathbf{a}}_{R}\left(\theta_{k}\right)$
is the receive array response vector with AOA $\theta_{k}$, $x_{k}\left(t\right)$
is the reference signal, $\tau_{k}$ is the delay of the LOS path,
and $\mathbf{n}\left(t\right)$ is the additive white Gaussian noise.

For a fixed frame duration $T_{f}$, one can expect that there is
a trade-off between communication QoS and localization accuracy, which
are quantified by effective data-rate $R_{k}$ and Position Error
Bound (PEB) $Q_{k}$
\begin{equation}
R_{k}=B\frac{T_{d}}{T_{f}}f_{k}\log_{2}\left(1+\frac{\mathrm{SNR}_{k}}{f_{k}}\right),\label{eq:effective data rate}
\end{equation}
\begin{equation}
Q_{k}=\mathrm{tr\left(\sum_{\mathbf{w}\in\mathcal{W}}\mathbf{J}_{k,\boldsymbol{w}}\right)},
\end{equation}
where $f_{k}\in(0,1)$ denotes the fraction of $T_{d}$ resources
allocated to the $k$-th MS, $\mathrm{SNR}_{k}$ denotes the receive
SNR, $B$ denotes the signal bandwidth, and $\mathbf{J}_{k,\boldsymbol{w}}$
denotes the FIM associated with a single beam $\mathbf{w}$. The PEB
is calculated by the summation of the FIM over all beams in codebook
set $\mathcal{W}$ used in the exhaustive beam alignment strategy.
Since the receive beamforming vector remain fixed for the complete
data reception phase, a compromise receive beamforming vector is adopted
by the superposition of best receive beam of each MS. Then an optimization
problem is formulated to find the optimal resources allocation scheme
based on the optimization variables $f_{k}$ and $T_{t}$ with the
objective of maximizing the minimum data rate of MSs subject to the
PEB constraints.

The simulation results indicate that more time spent for beam training
(i.e., larger $T_{t}$) leads to better beam alignment and localization
accuracy. However, this would reduce the time left for data transmission
(i.e., smaller $T_{d}$). Hence, there exists an optimal training
overhead that strikes the balance between the effective achievable
sum rate and the localization performance.

\subsubsection{Other Performance Analysis}

There are other applications that involve broadcasting, relaying or
D2D communication, as discussed in Section \ref{sec:Classifications-of-Integrated}.
For relay channel with cooperative localization, the relay can assist
communication and localization, the typical infrastructure of which
is the unmanned aerial vehicles (UAV). The UAV can be located by the
ground BSs and then can be used as a new anchor node to assist the
terrestrial localization. Meanwhile, the aerial mobile networks can
also provide communication services via UAV-aided relaying. Hence,
there are two links between the BS and the user, therein the BS-User
direct link and the BS-Relay-User link, both of which can provide
data communication and positioning functionalities. Apparently, the
performance bound for relay channel will be more sophisticated than
the single link channel topology. Moreover, the ISAC problem is studied
under RIS-aided wireless network, where a beamforming design was studied
for integrated localization and communication \cite{ILAC_survey14}.
In D2D channel topology, the user receives waveforms both from the
BS and other neighboring users, which can be either communication
signals or localization reference signals. The user needs to achieve
accurate positioning by cooperation with other users and D2D communication
simultaneously. Different from the relay or RIS aided channel, the
indirect link between the BS and the user is connected by another
user.

\subsection{Summary}

The typical application of device-free ISAC is for joint radar and
communication. Two specific designs that involve mono-static sensing
have been reviewed in this section. Estimation-information-rate induced
approach and equivalent-MSE induced approach as presented in Section
\ref{sec:Performance-Metrics} have been shown to be useful in establishing
bounds on the sensing-communication performance tradoff. On the other
hand, the typical application of device-based ISAC is for joint communication
and localization. A number of past studies have focused on optimizing
the resource allocation in either power, time or spatial domain to
strike a good balance between the achievable data rate and localization
accuracy. To further improve the performance, the concept of relay-aided
(static relay, or mobile relay such as UAV) cooperative communication
and localization has also been explored yet in a very preliminary
manner. Very few performance limits were reported in this case, which
deserves further study.

\section{Open Problems and Future Research Directions}

The research on the fundamental limits of device-free and device-based
sensing (especially those based on the EFIM and CRB analysis) is relatively
mature. However, the fundamental limits of many ISAC scenarios remain
open. For example, even for the simplest scenario of point-to-point
channel with mono-static sensing, a complete characterization of the
communication capacity and sensing distortion region (\textit{capacity-distortion
region}) is still unknown for non-i.i.d. channel/sensing states. There
are many other ISAC network topologies that have not been studied
before. It is also important to study the fundamental limits under
more practical considerations such as the imperfect CSI, frequency
offset and timing synchronization error, different mobility models,
etc., and optimize the system performance based on the fundamental
limits analysis. In the following, we will discuss some of these open
problems and future research directions in details.

\subsection{Tighter Information-Theoretical Bounds for Capacity-Distortion Region
of ISAC}

In classic sensing scenarios, the parameter estimation is performed
based on a known waveform send by the transmitter, and the prior distribution
of the parameters is given and cannot be controlled by the sensing
scheme. In this case, the CRB provides a tight lower bound for the
MSE performance of unbiased estimators in the high SNR regime. There
also exist other sensing bounds such as Weiss-Weinstein bound (WWB)
and Zik-Zakai bound (ZZB), which can provide a tighter lower bound
of MSE for moderate and low SNR regimes. However, in ISAC scenarios,
the estimation is usually performed based on communication signals,
which can be encoded signals. The receiver needs to recover the communication
message and the sensor needs to estimate the parameter from the encoded
communication signals. In this case, we need to characterize the capacity-distortion
region, which is fundamentally different from the classic sensing
or communication scenarios. In general, we cannot separately analyze
the communication capacity and the sensing performance using the classic
bounds such as CRB. We have to derive new inner and outer bounds for
the capacity-distortion region, based on novel information-theoretical
bounds.

\subsubsection{Tight Bounds for Memoryless ISAC Channels}

In Section \ref{subsec:memoryless channel device-free}, we have presented
some existing inner and outer bounds for a few simple memoryless ISAC
channels, where both the channel state and sensing state are assumed
to be i.i.d. and ergodic over one codeword. However, only in some
special cases, the inner and outer bounds coincide with each other
and part of the capacity-distortion region can be determined. The
optimal achievable scheme and the associated capacity-distortion region
remain unknown for most cases. In addition, the current information-theoretical
bounds for the memoryless ISAC channels are obtained under some restrictive
assumptions. For example, in \cite{P2P,bistatic_MAC,Ahmadipour2020JointSA},
the channel state and sensing state are assumed to be the same. In
\cite{zhang}, the receiver is assumed to know the perfect channel/sensing
state. However, in practice, the channel and sensing states are usually
different but correlated with each other, and the receiver usually
does not have the channel state information to begin with. Therefore,
an important future research direction is to derive tighter bounds
for memoryless ISAC channels under more realistic assumptions. To
achieve this, we need to develop joint sensing and channel coding
schemes as well as new bounding techniques that can work with more
realistic assumptions (e.g., with different channel and sensing states,
and without perfect channel state information at the receiver) and
can close the gap between the inner and outer bounds. For example,
it is possible to exploit the sensed state at the transmitter to improve
the joint sensing and channel coding scheme and improve the achievable
region (i.e., inner bounds of the capacity-distortion region).

\subsubsection{Information-Theoretical Bounds for Block-Varying ISAC Channels}

In many practical applications, the channel and sensing states are
not i.i.d. but block-varying, i.e., the channel/sensing state (approximately)
remains constant over one codeword. There still lacks studies on the
fundamental limits of such block-varying ISAC channels. One major
challenge of characterizing the capacity-distortion region for the
block-varying ISAC channel is as follows. In traditional communication
scenario, the block-varying ISAC channel reduces to the block fading
channel. In this case, the Shannon capacity region is well defined
under the assumption of perfect CSI at the transmitter (CSIT). However,
in block-varying ISAC channels, it may not make sense to assume perfect
CSIT, especially when the communication channel is also part of the
sensing channel. In this case, the transmitter may learn some imperfect
CSIT via self-sensing or CSI feedback from the receiver. As such,
we may need to incorporate the overhead of CSIT acquisition/state
sensing and the effect of imperfect CSIT in the analysis of the capacity-distortion
region, which is very challenging. In fact, the Shannon capacity may
not be well defined under imperfect CSIT. In this case, how to properly
define and characterize the capacity-distortion region of block-varying
ISAC channels is still an open problem and deserves further study.

\subsection{Fundamental Limits of Emerging ISAC Scenarios}

The study of the fundamental limits of ISAC is still at an early stage,
and many ISAC scenarios have not been investigated. In the following,
we discuss several important ISAC scenarios that have not been considered
before.

\subsubsection{More Complicated ISAC Network Topologies}

One interesting research direction is to study the fundamental limits
for other important ISAC network topologies obtained by merging the
sensing network topologies with communication network topologies.
For example, we may consider mono-static interference networks where
there are multiple communication transmitter-receiver pairs interfering
with each other and each communication transmitter also serves as
a radar transmitter to detect some moving targets. Furthermore, we
may introduce cooperation between the transmitters to enhance both
the communication and sensing performance, which is a useful ISAC
scenario for 6G cellular networks where the BSs can perform cooperative
communication and sensing via backhaul/fronthaul connections.

\subsubsection{Intelligent reflecting surface (IRS) aided ISAC }

IRS-aided ISAC is another ISAC scenario deserving further study. The
IRS can be used to change the communication/sensing channel and thus
has the potential to enhance the communication and sensing performance.
For example, the IRS may create NLOS paths with known scatter locations
(the IRS serves as a scatter with known location). In this case, the
NLOS paths created by the IRS can provide useful information to both
localization and communications, and thus enhance the coverage and
performance of communication and localization services. In some device-free
sensing scenarios, it is possible to equip an IRS at the target surface
(e.g., when the target is an autonomous vehicle) to enhance the target
estimation performance via passive beamforming at the target IRS.
Since the IRS-aided ISAC systems have the ability to adjust the communication/sensing
channel through passive beamforming, the analysis of fundamental limits
of IRS-aided ISAC systems is completely different from the conventional
ISAC systems.

\subsubsection{Environment Side Information aided ISAC}

When environment side information such as map information is available,
we can exploit this prior information to enhance the performance of
ISAC systems. A new information-theoretical framework is needed to
incorporate the map information into the fundamental limits analysis. 

\subsection{Fundamental Limits of ISAC under Practical Considerations}

Most existing works on the fundamental limits of ISAC have ignored
some important practical issues, such as the channel estimation error,
the frequency offset and timing synchronization error, the mobility,
etc. In the following, we shall point out several important practical
issues that should be taken into account in future studies.

\subsubsection{Channel Estimation Error}

In practice, the channel state information is never perfect due to
the channel estimation error, CSI feedback delay and CSI quantization
error. As already mentioned before, it is important to study how to
incorporate the overhead of CSI acquisition and the effect of imperfect
CSI in the analysis of fundamental limits of ISAC.

\subsubsection{Frequency Offset and Timing Synchronization Error}

Due to the hardware impairments, there always exist frequency offset
and timing synchronization error between different sensing or communication
transceivers. Unlike the traditional communication systems which have
relatively low requirement on the frequency offset and timing synchronization
error, the sensing performance of ISAC is very sensitive to the frequency
offset and timing synchronization error especially for future ISAC
systems with a high requirement on the sensing accuracy. For example,
6G communication systems are expected to achieve a positioning accuracy
at the subcentimeter level \cite{you2021towards}. In this case, a
small timing synchronization error of 0.1 nanosecond will lead to
a localization error of several centimeters. Therefore, the future
ISAC systems must take this problem into account. We notice that several
works have studied the fundamental limits of radar sensing/localization
under the consideration of frequency offset and timing synchronization
error \cite{spital_temp}, which is however, not the case for ISAC.

\subsubsection{Tracking Performance Analysis under Different Mobility Models}

The channel and sensing states usually change smoothly over time following
certain dynamics induced by the mobility pattern. Therefore, it is
of great importance to study the tracking performance of channel/sensing
state under different mobility models. Some initial tracking performance
analysis has been conducted in \cite{VLP5} for visible light-based
positioning, where the conditions under which the tracking process
is stable (i.e., the state tracking error is bounded as time goes
to infinity) is derived, and the converged state error is also analyzed.
We may leverage the tools therein and study the fundamental tracking
performance limits in more general ISAC systems.

\section{Conclusions}

In this work, a survey of recent studies on the fundamental limits
of integrated sensing and communication has been provided. According
to whether the sensing targets participating the sensing procedure
by transmitting and/or receiving, we first classify the ISAC related
technologies into four major categories: device-free sensing, device-based
sensing, device-free ISAC and device-based ISAC, and then each category
is further divided into different cases. For each case, we highlight
several important works, and present the system model, performance
bounds and key insights learned from the analysis of the fundamental
limits. In particular, we propose several typical ISAC channel topologies
as abstracted models for various ISAC systems, and present the current
research progress on the fundamental limits for each ISAC channel.
We show that the fundamental limits of ISAC channels cannot be obtained
by a trivial combinations of existing performance bounding techniques
in separate sensing and communication systems. Finally, we present
a list of important open challenges and potential research directions
on ISAC, many of them have not been mentioned in the previous works.


\end{document}